\documentclass[final,3p,11pt,pdflatex]{elsarticle}
\usepackage{xcolor, xspace, colortbl}
\usepackage[T1]{fontenc}
\usepackage{lmodern}
\usepackage{graphicx}
\usepackage{amsmath,amssymb,amsthm,nicefrac,mathtools}
\usepackage{bbold}
\usepackage{array}
\usepackage{subcaption}
\usepackage[force]{feynmp-auto}
\usepackage{slashed}
\usepackage{tikz}
\usetikzlibrary{shadings}
\usetikzlibrary{fit,backgrounds}
\usetikzlibrary{positioning}
\usepackage{tabularx}
\usepackage{multicol, multirow}
\usepackage{hyperref}
\hypersetup{
	colorlinks,
	linkcolor={red!50!black},
	citecolor={blue!50!black},
	urlcolor={blue!80!black}
}
\usepackage{orcidlink}
\usepackage[inline]{enumitem}
\setlist[description]{font=\normalfont\itshape}

\newcommand{\code}[1]{\texttt{#1}}
\newcommand{\str}[1]{\code{\color{color-listings-string}#1}}
\newcommand{\wl}{\code{Wolfram Language}\xspace}
\newcommand{\fortran}{\code{Fortran}\xspace}
\newcommand{\sh}{\code{Bourne shell}\xspace}
\newcommand{\cpp}{\code{C++}\xspace}
\newcommand{\fs}{\code{FlexibleSUSY}\xspace}
\newcommand{\fa}{\code{FeynArts}\xspace}
\newcommand{\fc}{\code{FormCalc}\xspace}
\newcommand{\colormath}{\code{ColorMath}\xspace}
\newcommand{\spheno}{\code{SPheno}\xspace}
\newcommand{\flavorkit}{\code{FlavorKit}\xspace}
\newcommand{\flexibledecay}{\code{FlexibleDecay}\xspace}
\newcommand{\npf}{\code{NPointFunctions}\xspace}
\newcommand{\sarah}{\code{SARAH}\xspace}
\newcommand{\observable}{\code{O$_{\code i}$}\xspace}
\newcommand{\model}{\code{M$_{\code a}$}\xspace}
\newcommand{\dir}[1]{\code{#1/}\xspace}
\newcommand{\file}[1]{\code{#1}\xspace}

\newcommand{\mec}{\ensuremath{\mu}--\ensuremath{e}~conversion\xspace}

\newcommand{\meee}{\ensuremath{\mu\to 3e}\xspace}
\newcommand{\opertyp}[1]{\mathcal{#1}}
\newcommand{\norm}[1]{\left|#1\right|^2}

\newcommand{\hc}{\ensuremath{\text{h.c.}}}
\newcommand{\bra}[1]{\langle#1\rvert}
\newcommand{\ket}[1]{\lvert#1\rangle}
\newcommand{\coefficientUV}[3][]{\ensuremath{F^{\opertyp{#2}#1}_{#3}}\xspace}
\newcommand{\twodecays}{\ensuremath{\ell_i\to\ell_j\gamma}\xspace}
\newcommand{\threedecays}{\ensuremath{\ell_i^{\phantom{c}}\to\ell_j^{\phantom{c}}\ell_k^{\phantom{c}}\ell_k^c}\xspace}
\newcommand{\Lagr}{\mathcal{L}}
\newcommand{\eftLagrangian}{\ensuremath{\Lagr_{\text{EFT}}}\xspace}
\newcommand{\fullLagrangian}{\ensuremath{\Lagr_{\model}}\xspace}
\newcommand{\muFermiFactor}{m_\mu G_F}
\newcommand{\coefficientEFT}[3][]{\ensuremath{C^{\opertyp{#2}#1}_{#3}}\xspace}
\newcommand{\operatorEFT}[3][]{\ensuremath{O^{\opertyp{#2}#1}_{#3}}\xspace}
\newcommand{\chain}[1]{[#1]}

\newcommand{\Lnucl}{\ensuremath{\Lagr_{\text{N}}^{\text{coh}}}\xspace}
\newcommand{\gluonNew}[2]{{\tilde{C}^{\opertyp{#1}}_{#2}}}
\newcommand{\cohFactor}[3][]{\ensuremath{{g^{\opertyp{#2}#1}_{#3}}}\xspace}
\newcommand{\nuclFactor}[3][]{\ensuremath{G^{\opertyp{#2}#1}_{#3}}\xspace}
\newcommand{\tensorRecoil}{}
 
\newcommand\exampleonename{output a given number\xspace}
\newcommand\exampletwoname{show fermion masses\xspace}
\newcommand\examplethreename[1][]{lepton self-energy with \npf #1\xspace}
\newlength\stextwidth
\newcommand\makesamewidth[3][c]{%
	\settowidth{\stextwidth}{#2}%
	\makebox[\stextwidth][#1]{#3}%
}
\newcommand\helperintex[1]{\text{\makesamewidth{1}{\tiny #1}}}
\newcommand{\figref}[1]{Figure~\ref{#1}}
\newcommand{\lstref}[1]{Listing~\ref{#1}}
\newcommand{\secref}[1]{Section~\ref{#1}}
\newcommand{\secsref}[2]{Sections~\ref{#1}--\ref{#2}}
\newcommand{\tabref}[1]{Table~\ref{#1}}
\newcommand{\tabsref}[2]{Tables~\ref{#1}--\ref{#2}}
\newcommand{\lineref}[1]{line~\ref{#1}}
\newcommand{\linesref}[2]{lines~\ref{#1}--\ref{#2}}

\usepackage[acronym, nohypertypes={acronym}]{glossaries-extra}
\usepackage{xcolor}
\setabbreviationstyle[acronym]{long-short}
\glssetcategoryattribute{acronym}{nohyperfirst}{true}
\newacronym{EDM}{EDM}{Electric Dipole Moment}
\newacronym{BSM}{BSM}{Beyond the Standard Model}
\newacronym{SM}{SM}{Standard Model}
\newacronym{MSSM}{MSSM}{Minimal Supersymmetric Standard Model}
\newacronym{2HDM}{2HDM}{Two-Higgs-Doublet Model}
\newacronym{SUSY}{SUSY}{supersymmetric}
\newacronym{RGE}{RGE}{Renormalization Group Equation}
\newacronym{SLHA}{SLHA}{SUSY Les Houches Accord}
\newacronym{FLHA}{FLHA}{Flavour Les Houches Accord}
\newacronym{WCxf}{WCxf}{Exchange Format for Wilson Coefficients}
\newacronym{MRSSM}{MRSSM}{Minimal R-Symmetric Supersymmetric Standard Model~\cite{Kribs:2007ac}}
\newacronym{MSBAR}{\ensuremath{\overline{\text{MS}}}}{Modified Minimal Subtraction Scheme}
\newacronym{DRBAR}{\ensuremath{\overline{\text{DR}}}}{Dimensional Reduction Scheme}
\newacronym{CLFV}{CLFV}{Charged Lepton Flavor Violation}
\newacronym{EFT}{EFT}{Effective Field Theory}
\newacronym{GNM}{GNM}{Grimus-Neufeld Model}
\newacronym{LQ}{LQ}{Leptoquark}
\newacronym{FFS}{FFS}{Fermion-Fermion-Scalar}
\newacronym{SSF}{SSF}{Scalar-Scalar-Fermion}
\newacronym{QCD}{QCD}{Quantum Chromodynamics}
\newacronym{QED}{QED}{Quantum Electrodynamics}
\newacronym{JSON}{JSON}{\code{JavaScript} Object Notation}

\usepackage{listings}
\usepackage[scaled]{beramono}
\definecolor{color-listings-background}{RGB}{253,246,227}
\definecolor{color-listings-string}{RGB}{182,141,180}
\definecolor{color-listings-comments}{RGB}{53,102,164}
\definecolor{color-listings-break}{RGB}{255,0,0}
\definecolor{color-listings-context}{RGB}{170,170,170}
\definecolor{color-listings-local}{RGB}{109,164,30}
\definecolor{color-listings-syntax}{RGB}{182,100,9}
\definecolor{color-cpp-token}{RGB}{135, 118, 65}
\definecolor{color-cpp-keyword}{RGB}{109,164,30}
\definecolor{color-cpp-preprocessor}{RGB}{148,89,131}
\definecolor{color-cpp-template-arguments}{RGB}{200,70,0}
\makeatletter
\let\orig@lstnumber=\thelstnumber

\newcommand\lstresetnumber{\global\let\thelstnumber=\orig@lstnumber}
\makeatother

\lstnewenvironment{meta}[1]{%
	\lstset{
		rulecolor=\color{color-listings-background},
		mathescape=true,
		language=Mathematica, 
		caption={#1},
		emph={Option, Value, LoopNumber, Contribution, TopologyName, WhenToApply, Description, Command, Momenta, Rule, ChainNumber, Entry}, 
		emphstyle=\itshape,
		emph={[2]Private,FlexibleSUSY,Utils,TextFormatting,WriteOut,FlexibleSUSYObservable,Observables, CConversion,FeynArts,NPointFunctions,SARAH,FormCalc,WilsonCoeffs},
		emphstyle={[2]\color{color-listings-context}},
		emph={[3]lep,lep_, iI_, iJ_, iK_, contr_, loopN_, iO_, nucl_, parA_, parB_, num_, fermion_, gen_, field_, obs, gen, slha_, files_, _obs, allObs_, expr, expr_, allObs,
		observables, prototypes, definitions, npfHeaders, npfDefinitions, slha, res_, res, npfOutput, npfVertices, f_, f, field, contr, npf, basis, code, vertices, higgs, l_, cxxVertices, name},
		emphstyle={[3]\color{color-listings-local}},
		literate={`}{{\textcolor{color-listings-context}\textasciigrave}}{1},
		backgroundcolor=\color{color-listings-background}
	}%
}{}
\newcommand\placeintable{
\lstset{
	aboveskip = -6pt,
	belowskip= -10pt,
	xleftmargin=0pt,  
	xrightmargin=0pt, 
	frame=none
}}
\lstnewenvironment{cppcode}[1]{
	\lstset{
		mathescape=true,
		caption={#1},
		emph={namespace,using,class,template,auto,int,const,double,switch,case,break,return,complex,typename, default}, 
		emphstyle={\color{color-cpp-keyword}},
		emph={[2]FIELD,RTYPE},
		emphstyle={[2]\color{color-cpp-template-arguments}},
		morestring=*[b][\color{color-listings-string}]{"},
		morecomment=[s][\color{color-cpp-preprocessor}]{\#}{\ },
		morecomment=[s][\color{color-listings-comments}]{/*}{*/},
		morecomment=*[l][\color{color-listings-comments}]{//},
		morecomment=[s][\color{color-cpp-token}]{@}{@},
		backgroundcolor=\color{color-listings-background}
	}
}{}
\newcommand\cpptoken[1]{\code{\color{color-cpp-token}#1}}
\lstdefinelanguage{LesHouches}{
	morecomment=[l][\color{color-listings-comments}]{\#}
}
\newcommand\lsyntax[1]{\code{\color{color-listings-syntax}#1}}
\newcommand{\commnamespace}{{\color{color-listings-comments}\code{namespace}}}
\newcommand{\option}[2][]{\textit{\code{#2}}$_{\textit{\code{#1}}}$}
\newcommand{\optionindex}[1]{\textit{\code{#1}}}
\newcommand{\stringindex}[1]{\code{\color{color-listings-string}#1}}
\newcommand\context[1]{\code{{\color{color-listings-context}#1\textasciigrave}\xspace}}
\newcommand\local[1]{\code{{\color{color-listings-local}#1}\xspace}}

\newcommand\fntoken{\cpptoken{@\observable{}\_filename@}\xspace}
\newcommand{\task}[1]{\code{\color{color-listings-comments}Task~#1}\xspace}
\begin{document}
\journal{Computer Physics Communications}

\begin{frontmatter}
	\title{FlexibleSUSY extended to automatically compute physical quantities in any Beyond the Standard Model theory: Charged Lepton Flavor Violation processes, Higgs decays, and user-defined observables}
	
	\author[a]{Uladzimir Khasianevich\,\orcidlink{0000-0003-0255-0674}\corref{author}}
	\author[b]{Wojciech Kotlarski\,\orcidlink{0000-0002-1191-6343}}
	\author[a]{Dominik St\"ockinger}
	\author[c]{Alexander Voigt\,\orcidlink{0000-0001-8963-6512}}
	
	\cortext[author]{Corresponding author.\\\textit{E-mail address:} uladzimir.khasianevich@tu-dresden.de}
	\address[a]{Institut f\"ur Kern- und Teilchenphysik, TU Dresden, Zellescher Weg 19, 01069 Dresden, Germany}
	\address[b]{National Centre for Nuclear Research Pasteura 7, 02-093 Warsaw, Poland}
	\address[c]{Institute for Theoretical Solid State Physics, RWTH Aachen University, Sommerfeldstraße 16, 52074 Aachen, Germany}
	
	\begin{abstract}

\fs is a framework for the automated computation of physical quantities (observables)
in models beyond the Standard Model (BSM). This paper describes an
extension of \fs which allows to define and add new observables that can be
enabled and computed in applicable user-defined BSM models. The extension has
already been used to include Charged Lepton Flavor Violation (CLFV)
observables, but further observables 
can now be added straightforwardly.
The paper is split
into two parts. The first part is non-technical and describes from the
user's perspective how to enable the calculation of predefined
observables, in particular CLFV observables. The second part of the
paper explains how to define new observables such that their automatic
computation in any applicable BSM model becomes possible. A key ingredient is the
new \texttt{NPointFunctions} extension which allows to use tree-level
and loop calculations in the 
model-independent setup of observables. Three examples of increasing
complexity are fully worked out. This illustrates the features and
provides code snippets that may be used as a starting point for implementation of further observables.
\end{abstract}
	
	
\end{frontmatter}

{\bf\noindent Program summary}

\begin{small}
	\noindent
	{\em Program Title:} \npf\\
	{\em CPC Library link to program files:} (to be added by Technical Editor) \\
	{\em Developer's repository link:} \href{https://github.com/FlexibleSUSY/FlexibleSUSY}{https://github.com/FlexibleSUSY/FlexibleSUSY}\\
	{\em Code Ocean capsule:} (to be added by Technical Editor)\\
	{\em Licensing provisions:} GPLv3\\
	{\em Programming language:} \cpp, \wl, \fortran, \sh\\
	{\em Journal reference of previous version:} Comput. Phys. Commun.
	230 (2018) 145–217; PoS CompTools2021 (2022) 036 \\
	{\em Does the new version supersede the previous version?:} Yes\\
	{\em Reasons for the new version:} Program extension including new observables and file structures\\
	{\em Nature of problem: }Determining observables for an arbitrary extension of the Standard Model supported by \fs, input by the user.\\
	{\em Solution method: }Generation of the code from automated algebraic manipulations. Automatic filling and compiling of predefined template files.\\
	{\em Additional comments including restrictions and unusual features: }
	Vertices with a direct product of Lorentz and color structures are supported. Settings of the advanced \npf mode rely on explicit specification of topologies.\\
\end{small}

\tableofcontents
\section{Introduction}
Exploring the parameter space of \gls{BSM} theories, researchers frequently employ software packages to automate both intricate calculations and parameter scans. 
Nevertheless, these software tools are often restricted to a limited
range of models, closely related to the  \gls{SM}, the \gls{MSSM}, the \gls{2HDM}, and their extensions involving higher-dimensional operators.
This is only a slice of all intriguing possibilities that include a broad set of models.

To address this limitation, \fs~\cite{Athron:2014yba, Athron:2017fvs} was created
to be used for a broad class of \gls{SUSY} or non-\gls{SUSY} models, providing a tool for the comprehensive investigation of diverse theoretical scenarios.%
\footnote{A software with similar capabilities is \sarah/\spheno/\flavorkit~\cite{Staub:2013tta, Porod:2003um, Porod:2011nf, Porod:2014xia}.}
It is a software application primarily implemented in the \wl \cite{Mathematica} and
\cpp based on \sarah \cite{Staub:2009bi,Staub:2010jh,Staub:2012pb,Staub:2013tta} and components from \code{SoftSUSY} \cite{Allanach:2001kg,Allanach:2013kza}, designed to produce an efficient and accurate \cpp spectrum
generator (a program searching for a consistent set of model
parameters and calculating the pole mass spectrum and a set of observables) for physical models specified by the user.

The produced \cpp program applies user-defined boundary conditions at
up to three distinct energy scales within the model, incorporating \gls{RGE} evolution between these scales. 
It further generates a collection of mixing matrices, pole masses, and auxiliary quantities.
Recent versions of \fs have introduced the computation of several
important observables that are suitable for phenomenological
investigations and comparisons with experimental data.
In particular, we highlight the extensions \code{FlexibleAMU} and \code{FlexibleCPV} introduced in Ref.~\cite{Athron:2017fvs} (they are responsible for the calculations of the anomalous magnetic moment and \gls{EDM} of leptons), an update for \code{FlexibleMW} on precise calculation of the $W$-boson pole mass from Ref.~\cite{Athron:2022isz}, and
\flexibledecay~\cite{Athron:2021kve} (a tool to calculate decays of scalars in a broad class of \gls{BSM} models).

A key point of these observables is that they are integrated in \fs
such that they are ready to be computed for any desired \gls{BSM}
scenario \fs is applied to. In order to achieve this, the mentioned
extensions store information about the observables in suitable \wl and
\cpp meta code which is then automatically converted into actual \cpp code
specifically for each \gls{BSM} scenario.

So far, new observables were added by individually modifying the
internals of \fs, hence users could not add new observables in such a
model-independent way.
In the present paper, we explain a new \fs~\code{2.8} design structure which
solves this problem. It allows to integrate new observables on the
meta code level, and it provides
powerful options to define and finetune the computation of new
observables without having to touch internals of \fs.

A number of new observables has already been integrated by using this
new structure (they correspond to various \gls{CLFV} processes), and in the future, further additional observables may be integrated either by \fs developers or by users of \fs.

To streamline and modularize the integration of new observables into \fs, an extension named \npf~\cite{Khasianevich:2022ess} was developed.
This extension serves to automate the calculation of amplitudes and other quantities that rely on them for any high-energy model supported by \fs.
In essence, \npf incorporates a well-defined approach of widely used packages \fa~\cite{Hahn:2000kx}, \fc~\cite{Hahn:1998yk}, and \colormath~\cite{Sjodahl:2012nk} into \fs (up to technical implementation details to be mentioned later in appropriate sections).

The article is separated into two parts depending on the readers' interests: 
\begin{enumerate}
	\item 
	\secref{sec:all-obs} describes how to install and use \fs to
        calculate any of the available observables. This section is of
        interest for all users of \fs who may want to switch on the
        computation of observables. Reading it does not require
        knowledge of the internal structure of \fs.
	To get physical insights about the new \gls{CLFV} observables, one is invited to read~\ref{sec:physical-details}.
	\item 
	 \secref{sec:new-observables} presents details on how to
         implement new observables. It provides a general outline and background information relevant for all observables,
         and it covers three specific examples of increasing
         complexity. It thus illustrates the range of possibilities
         and equips users with code snippets which can be used as a
         basis for further developments.
	Interesting features improving the functionality of \fs are mentioned separately in \ref{sec:additional-features}.
\end{enumerate}
\section{Available observables and how to calculate them with \fs}\label{sec:all-obs}
\begin{table}[t!]
	\renewcommand*{\arraystretch}{1.2}
\setlength{\tabcolsep}{4pt}
\centering
\begin{tabularx}{\textwidth}{>{\centering}p{3cm}>{\centering}p{3cm}>{\centering}p{3.2cm}X}
	\hline
	Observable & \fs name & Loop level & \multicolumn{1}{c}{Hints and comments}
	\\\hline
	$\Delta a_\ell$ & \code{AMM}, \code{AMMUncertainty} & & Anomalous magnetic moment of a lepton~\cite{Athron:2017fvs}
	\\
	$d_\ell$ & \code{EDM} & & Electric dipole moment of a lepton~\cite{Athron:2017fvs}
	\\
	\twodecays & \code{BrLToLGamma} & 1 & See~\ref{sec:two-decay-physics}
	\\
	\threedecays & \code{BrLTo3L} & 0--1 & See \tabref{tab_l3l-simple} and~\ref{sec:three-decay-physics}
	\\
	\mec & \code{LToLConversion} & 0--1 & See \tabref{tab_llc-simple} and~\ref{sec:mec-physics}
	\\
	$b\to s\gamma$ & \code{bsgamma} & 1 & ---
	\\
   --- & \code{FlexibleDecay} & known \gls{SM} (0--4), LO (0--1) for \gls{BSM} & Decays of scalars~\cite{Athron:2021kve}
	\\\hline
\end{tabularx}%

	\caption{All observables currently supported by \fs.}
	\label{tab_all-fs-observables}
\end{table}
\begin{table}[t!]
	\renewcommand*{\arraystretch}{1.2}
\setlength{\tabcolsep}{4pt}
\centering
\begin{tabular}{>{\centering\arraybackslash}p{\textwidth-2\tabcolsep}}
\hline
	Usage
\\\hline
\cellcolor{color-listings-background}
\placeintable
\begin{meta}{}
FlexibleSUSYObservable`BrLTo3L[lep_, iI_ -> {iJ_, iK_, iK_}, contr_, loopN_]
\end{meta} 
\\\hline
\end{tabular}
\begin{tabularx}{\textwidth}{>{\centering}p{2.2cm}>{\centering}p{2cm}X}
	\hline
	Abbreviation & Values & \multicolumn{1}{c}{Hints}
	\\\hline
	\local{lep} & symbol & Leptons $\ell$ in \threedecays in \sarah model (\code{Fe} in {SM} or {MRSSM})
	\\
	\local{iI}, \local{iJ}, \local{iK} & integer & Generations $i$, $j$, $k$ in \threedecays, starting from 1
	\\
	\local{contr} &
	\code{Vectors}, \code{Scalars}, \code{Boxes} & 
	Contributions (a synonym or a subset is allowed), see  
	\file{FlexibleSUSY.m},
	\file{NPointFunctions.m} in the directory \dir{meta/Observables/BrLTo3L}
	\\
	\local{loopN} & 0--1 & Loop level, \fc limitation
	\\\hline
\end{tabularx}%
	\caption{Options for \code{BrLTo3L}, see \file{meta/Observables/BrLTo3L/Observables.m}.}
	\label{tab_l3l-simple}
\end{table}
\begin{table}[t!]
	\renewcommand*{\arraystretch}{1.2}
\setlength{\tabcolsep}{4pt}
\centering
\begin{tabular}{>{\centering\arraybackslash}p{\textwidth-2\tabcolsep}}
\hline
Usage
\\\hline
\cellcolor{color-listings-background}
\placeintable
\begin{meta}{}
FlexibleSUSYObservable`LToLConversion[lep_, iI_ -> iO_, nucl_, contr_, loopN_]
\end{meta}
\\\hline
\end{tabular}
\begin{tabularx}{\textwidth}{>{\centering}p{2.2cm}>{\centering}p{2cm}X}
	\hline
	Abbreviation & Values & \multicolumn{1}{c}{Hints}
	\\\hline
	\local{lep} & symbol & Leptons in \sarah model (\code{Fe} in {SM} or {MRSSM})
	\\
	\local{iI}, \local{iO} & integer & Muon and electron generations, starting  from 1
	\\
	\local{nucl} & \code{Al} or \code{Au} & Nucleus, see \file{l\_to\_l\_conversion.cpp.in} in the directory \dir{templates/observables}
	\\
	\local{contr} & \code{Vectors}, \code{Scalars}, \code{Boxes} & 
	Contributions (a synonym or list of a subset is allowed), see  
	\file{FlexibleSUSY.m},
	\file{NPointFunctions.m} in the directory \dir{meta/Observables/LToLConversion}
	\\
	\local{loopN} & 0--1 & Loop level, \fc limitation
	\\\hline
\end{tabularx}%
	\caption{Options for \code{LToLConversion}, see \file{meta/Observables/LToLConversion/Observables.m}.}
	\label{tab_llc-simple}
\end{table}

Observables that are currently available in \fs are shown in \tabref{tab_all-fs-observables}. This section shows the reader how to calculate them with \fs. 

\subsection{Installation}
There are no changes to the previous \fs version of Refs.~\cite{Athron:2017fvs, Athron:2021kve} with respect to mandatory installation steps. 
All missing dependencies will be highlighted by \fs during the execution of the \file{configure} script (their list and hints about the installation can be found at  \href{https://github.com/FlexibleSUSY/FlexibleSUSY}{developer's repository}, see the program summary).
To use observables that rely on \npf module (in particular, \gls{CLFV} ones), \fa and \fc must be installed.

\subsection{Output defined observable with \fs}

To switch on the calculation of a  desired, predefined  observable \observable (for example, \code{bsgamma} or \code{LToLConversion} from \tabref{tab_all-fs-observables}) for a physical model \model (like \gls{SM} or \gls{MSSM}), one needs to modify either \file{model\_files/\model/\fs.m} or \file{models/\model/\fs.m}.
These files define the \cpp spectrum generator output in the \gls{SLHA}~\cite{Skands:2003cj, Allanach:2008qq} or
the \gls{FLHA}~\cite{Mahmoudi:2010iz} formats.

The first file mentioned above is used by the script \file{createmodel} (see \secsref{sec:ex1-final}{sec:lepton-se-exe}) to create both the directory \dir{models/\model} and the second file, while the latter is executed by commands \file{configure} and \code{make} that create \cpp spectrum generator itself.
This means, that to always include the desired observables after the directory \dir{models/\model} is purged or cleaned, one modifies the first file.

The required modification of the file \file{\fs.m} is in the list \code{ExtraSLHAOutputBlocks}. Each observable that should be computed and appear in the spectrum generator output needs a corresponding entry which specifies the details of the observable and how it should appear in the output. For example,  \mec is switched on in the \gls{MRSSM} (for its particular \fs configuration called \code{MRSSM2}) as follows:
\newpage
\begin{meta}{Adding \mec observable (example for \file{models/MRSSM2/\fs.m}).}
ExtraSLHAOutputBlocks = {
	{
		FlexibleSUSYLowEnergy, 
		{
			{41, FlexibleSUSYObservable`LToLConversion[Fe, 2 -> 1, Al, All, 1]}, 
			...
		}
	},
	...
};
\end{meta}
The numerical value of \mec after the execution of the spectrum generator will be stored in \gls{SLHA} format under the user-chosen number \code{41} in the block \code{FlexibleSUSYLowEnergy} (see \file{meta/WriteOut.m}).
More details for arguments of \code{LToLConversion} are provided in \tabref{tab_llc-simple}.

To numerically calculate all added observables (apart from \flexibledecay), the observable calculation must be enabled in the \gls{SLHA} input file for \cpp spectrum generator:\footnote{%
	Here, the number \code{15} corresponds to the \fs conventions for the \gls{SLHA} block \code{Block \fs}, introduced in Appendix~B of Ref.~\cite{Athron:2017fvs}.
	Calculation of observables using \flexibledecay is controlled by a separate flag as explained in Ref.~\cite{Athron:2021kve}.
}
{
\lstset{backgroundcolor=\color{color-listings-background}}
\begin{lstlisting}[language=LesHouches,caption={In \file{models/\model/LesHouches.in.\model}. All added observables are enabled.}]
Block FlexibleSUSY
	...
	15   1                    # calculate all observables
\end{lstlisting}
}%

\subsection{Output Wilson coefficients with \fs}
One can also output Wilson coefficients used in derivation of \code{LToLConversion} or \code{BrLTo3L}.
To do that, one places the corresponding observable into the \code{FWCOEF} (\code{IMFWCOEF}) block to output their real (imaginary) part, for example:
			
\begin{meta}{Showing Wilson coefficients (example for \file{models/MRSSM2/\fs.m}).}
ExtraSLHAOutputBlocks = {
	{
		FWCOEF, 
		{
			{1, FlexibleSUSYObservable`LToLConversion[Fe, 2 -> 1, Al, All, 1]}, 
			...
		}
	}, 
	...
};
\end{meta}
The coefficients are defined in files \file{meta/Observables/\observable/WriteOut.m}.
\section{File structure of new observables \observable}\label{sec:new-observables}
\begin{figure}[t]
\centering
\newcommand{\tikzvdist}{0.2cm}
\usetikzlibrary{positioning,calc,decorations.pathreplacing}
\tikzset{node distance = \tikzvdist}
{%
\begin{tikzpicture}
	\node (FSdir) {In \dir{meta}\vphantom{\dir{/O/\observable}}};
	\node[below=of FSdir] (mO) {\file{Observables.m}};
	\node[below=of mO] (mW) {\file{WriteOut.m}};
	\node[below=of mW] (mF) {\file{FlexibleSUSY.m}};
	\node[below=of mF] (mN) {\file{\vphantom{[]}NPointFunctions.m}};
	\node[below=of mN] (mL) {\file{\vphantom{[]}FSMathLink.m}};
	
	\node[right=1.5cm of FSdir] (OBSdir) {In \dir{meta/Observables/\observable}};
	\node[below=of OBSdir] (mOO) {\file{Observables.m}};
	\node[below=of mOO] (mOW) {\file{WriteOut.m}};
	\node[below=of mOW] (mOF) {\file{FlexibleSUSY.m}};
	\node[below=of mOF] (mON) {\file{[NPointFunctions.m]}};
	\node[below=of mON] (mOL) {\file{[FSMathLink.m]}};
	
	\node[right=0.5cm of OBSdir] (Tdir) {In \dir{templates/observables}};
	\draw let \p1=(mOW.east), \p2=(mOF.east), \p3=(Tdir) in
	node (thin) at ($(\x3,0)+(0,\y1)!0.5!(0,\y2)$) {\file{\fntoken.hpp.in}};
	\draw let \p1=(mON.east), \p2=(mOF.east), \p3=(Tdir) in
	node (tcin) at ($(\x3,0)+(0,\y1)!0.5!(0,\y2)$) {\file{\fntoken.cpp.in}};
	
	\draw let \p1=(OBSdir), \p2=(Tdir), \p3=(mOL) in
	node (MTdir) at ($(\x1,0)!0.5!(\x2,0)+(0,\y3-1.2cm)$) {In \dir{models/\model/observables}};
	\node[below=of MTdir] (th) {\file{\model{}\_\fntoken.hpp}};
	\node[below=of th] (tc) {\file{\model{}\_\fntoken.cpp}};
	
	\draw let \p1=(FSdir), \p2=(OBSdir), \p3=(MTdir) in
	node (Mdir) at ($(\x1,0)!0.5!(\x2,0)+(0,\y3)$) {In \dir{models/\model}};
	\draw let \p1=(Mdir), \p2=(th), \p3=(tc) in
	node (run) at ($(\x1,0)+(0,\y2)!0.5!(0,\y3)$) {\file{run\_\model.x}};
	
	\begin{pgfonlayer}{background}
		\node[inner sep=0cm, draw=lightgray, fit=(FSdir)(mO)(mW)(mF)(mN)(mL)] {};
		\node[inner sep=0cm, draw=lightgray, fit=(OBSdir)(mOO)(mOW)(mOF)(mON)(mOL)(Tdir)(thin)(tcin)] (OBSfiles) {};
		\node[inner sep=0cm, draw=lightgray, fit=(MTdir)(th)(tc)(Mdir)(run)] {};
	\end{pgfonlayer}  

	\node[above=of OBSfiles] {New files, corresponding to an observable \observable};
	
	\draw[decorate,decoration={brace}] let \p1=(FSdir.north), \p2=(mL.south), \p3=(mN.west) in
	($(\x3-\tikzvdist,\y2)$) -- node [above, rotate=90, text width=5cm,align=center] {\footnotesize meta phase controlled by \file{models/\model/FlexibleSUSY.m}} ($(\x3-\tikzvdist,\y1)$) ;
	
	\draw[decorate,decoration={brace}] let \p1=(Mdir.north), \p2=(tc.south), \p3=(mN.west) in
	($(\x3-\tikzvdist,\y2)$) -- node [above, rotate=90, text width=2cm,align=center] {\footnotesize \cpp spectrum generator} ($(\x3-\tikzvdist,\y1)$);
\end{tikzpicture}}%
\caption{The place and role of new files, required by new observables. The files in square brackets are optional as described in the text.
The name of \cpp files contains \fntoken, replaced by the rule \code{GetObservableFileName}, see Listing~\ref{code:general-observables.m}.
}
\label{fig_code-structure}
\end{figure}
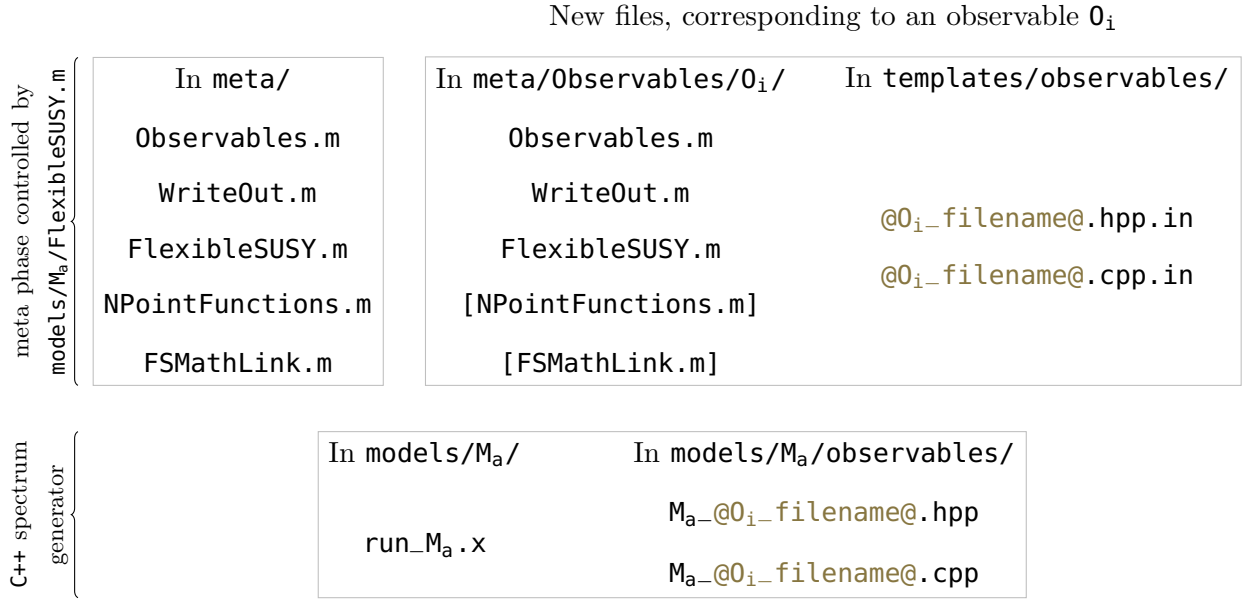

Two new \fs features are described in this paper: a way to add new observables to \fs (implying new auxiliary files), and a way to generate \cpp code to numerically calculate amplitudes that are required by these observables (\npf extension).
This section addresses both extensions and explains how to implement new observables with the help of toy examples and code snippets of already implemented observables.

To create a spectrum generator, \fs usually first runs \wl code (located in \dir{meta}) to obtain model-specific expressions for mass matrices, self-energies, amplitudes, etc. The obtained expressions are converted to \cpp form and are filled into \cpp template files (located in \dir{templates}) to generate model-specific \cpp code (located in \dir{model/\model}).
\figref{fig_code-structure} shows the files relevant for the calculation of observables. 
The top left block shows \wl files in \dir{meta} that contain general
routines and observables implemented in previous versions of \fs. 
The top right block shows the corresponding observable-specific \wl files (located in \dir{meta/Observables/\observable}) and the necessary \cpp template files (located in \dir{templates/observables}). 
The bottom block shows the generated relevant \cpp files that are eventually compiled and combined with other \cpp files to the final spectrum generator executable (\file{models/\model/run\_\model.x}).

In the following, by creating a new observable we mean the generation of a specific \cpp function that calculates the observable within the final spectrum generator.
That is, one needs to create several \cpp code blocks in appropriate places of the generated \cpp spectrum generator, as the following template-listing 
with observable- and model-specific blocks shows:\footnote{%
	These \cpp expressions might seem to be trivial at this point and the natural question arises whether it is really necessary to make use of the file structure described in \secsref{sec:observables-general}{sec:content-cpp}.
	One needs to keep in mind, that the mentioned expressions depend both on the physical model \model and the observable \observable settings, which change the parts generated during the meta phase.
	These modifications can, in general, be very complex, and \fs meta-phase routines help to apply them.
}
{
\lstset{label={code:all-cpp-blocks}}
\begin{cppcode}{\cpp template code to calculate an observable for a specific model \model.}
// 1. Function definition in $\color{color-listings-comments}\model$_$\fntoken$.cpp:
namespace $\model$_@$\cpptoken\observable$_namespace@ {
@$\cpptoken\observable$_output_type@ @$\cpptoken\observable$_calculate(...)@ {
	// Complicated function body generated during the meta phase and specific for ${\color{color-listings-comments}\observable}$ and ${\color{color-listings-comments}\model}$ 
}
}
	
// 2. Internal calculation of the observable:
observables.@$\cpptoken\observable$_name@ = $\model$_@$\cpptoken\observable$_namespace@::@$\cpptoken\observable$_calculate(...)@;

// 3. Writing the observable in Les Houches format to an output stream:
block << FORMAT_*(observables.@$\cpptoken\observable$_name@, ...);
\end{cppcode}
}
\noindent In the \cpp source code snippet above, \cpptoken{@\observable{}\_name@} is replaced by the name of the observable and \cpptoken{@\observable{}\_output\_type@} is its \cpp type. The \cpptoken{@\cpptoken\observable\_calculate(...)@} pattern will be replaced by the name of function that performs the calculation, which is defined in the $\model$\_\cpptoken{@\observable{}\_namespace@} namespace. The body of the function will depend on the observable and the model under consideration.

In the following subsections it is described how the expressions are determined that replace the different patterns in the above \cpp template code snippet. The files where these expressions are determined are the \wl files located in the directory \dir{meta/Observables/\observable}, which will be described in the following subsections.
In general, to define a new observable one creates five new files filled with the content described in the following steps:
\begin{enumerate}
\item In \secref{sec:observables-general} we show how to define the
  \cpp name of the observable, the observable's \cpp type, etc.~(done
  in the file \file{meta/Observables/\observable/Observables.m}).
\item In \secref{sec:write-general} we show how to connect the
  calculation of the observable to the spectrum generator output (done
  in the file \code{meta/Observables/\observable/WriteOut.m}).
\item In \secsref{sec:content-fs-file}{sec:content-cpp} we
  describe how to fill the body of the function that calculates the
  numeric value of the observable for a given parameter point (done
  in the meta-phase file \code{meta/Observables/\observable/FlexibleSUSY.m} in \secref{sec:content-fs-file} and
  two \cpp template files in \secref{sec:content-cpp}).
\item As a last step, we show how to modify the model-specific \fs model file named
  \file{models/\model/\fs.m} to register the observable to the model
  under consideration.
\end{enumerate}

Each subsection will begin with general explanations and definitions. Then three recurring examples will be used for	illustration. The first example illustrates the definition of an observable with minimal complexity --- the output of a single numerical constant, where the numerical constant is not hard-coded but can be specified separately for each model \fs is applied to. The second example outputs the value of fermion masses, where	the decision of which fermions to select can be specified separately for each model. The third example outputs the value of one-loop self-energies and thus illustrates how to use loop calculations in the definition of observables.

\subsection{Content of the file \file{\observable/Observables.m}}\label{sec:observables-general}
\subsubsection*{General case}

The creation of a new observable \observable starts with the file
\file{meta/Observables/\observable/Observables.m}. This file is
supposed to define the \cpp tokens such as the observable name
(\cpptoken{@\observable{}\_name@}) or the observable's \cpp type
(\cpptoken{@\observable{}\_output\_type@}), etc. The file contains
only one call to the function \code{DefineObservables}, which defines all \cpp tokens.
The general structure of this function call is shown in the following \wl source code snippet:
{
\lstset{label={code:general-observables.m}}
\begin{meta}{General content of \file{meta/Observables/\observable/Observables.m}.}
Observables`DefineObservable[
	FlexibleSUSYObservable`$\observable$[parA_, parB_, ...], (* Wolfram Language symbol for observable *)
	GetObservableType        -> {1},                        (* $\cpptoken{@\observable{}\_output\_type@}$ *)
	GetObservablePrototype   -> "calc_parA(int parB, ...)", (* $\cpptoken{@\cpptoken\observable\_calculate(...)@}$ prototype *)
	
	(* Optional: to override defaults, use commands below *)
	GetObservableName        -> "name_with_parAparB...",    (* $\cpptoken{@\observable{}\_name@}$ *)
	CalculateObservable      -> "calc_parA(parB+1, ...)",   (* $\cpptoken{@\cpptoken\observable\_calculate(...)@}$ call *)
	GetObservableNamespace   -> "observable_namespace",     (* $\cpptoken{@\cpptoken\observable\_namespace@}$ *)
	GetObservableFileName    -> "observable_file",          (* $\fntoken$ *)
	GetObservableDescription -> "observable: parA, parB"    (* description *)
];
\end{meta}
}
\noindent
The first argument of the function \code{DefineObservables}  defines the \wl symbol of the observable \observable, which is used in the file \file{models/\model/\fs.m} to register the observable to the model-specific \fs model file, see \tabsref{tab_l3l-simple}{tab_llc-simple} for examples.
The optional parameters (like \local{parA\_}) can be freely chosen and may be used to specify anything that might be changed from one model \model to another (like the \sarah name of a particle multiplet, relevant contributions, etc.), or may be specific for the observable itself (like a particle generation index, the loop level, etc.). 
The parameters may be wrapped by other \wl functions (like \code{Rule}) to improve readability.
Note that by convention, the names of all \fs observables belong to \context{FlexibleSUSYObservable} context.

All further arguments of the function \code{DefineObservables} define how \cpp tokens (for example, \cpptoken{@\observable{}\_output\_type@}) for the observable \observable should be replaced in code from \lstref{code:all-cpp-blocks}. 
Note that the names of the parameters defined for the observable (like \str{parA}, \str{parB}, etc.)\@ are replaced in the strings by their values given in the model-specific \fs model file (the only exception: arguments in the prototype). For example, if one writes in the \fs model file \code{\context{FlexibleSUSYObservable}\observable{}[Fe, 3]}, then the strings \str{parA} and \str{parB} are replaced by \code{Fe} and \code{3}, respectively, in all strings on the r.h.s.~in the function \code{DefineObservable}, so that for example the execution of the function \code{GetObservableName} produces \str{"name\_with\_Fe3"}.%
\footnote{%
More advanced ways to name the \cpp code parts are supported:
\begin{enumerate*}
\item 
	One may use expression like \str{"\$(parA+1)"}, which evaluates the content inside the parenthesis after substitution of observable pattern values (\local{parA\_}).
\item
	The description of observable in the \gls{SLHA} output can be generated from its name automatically by replacing underscores with spaces. 
	This can be redefined by \code{GetObservableDescription}.
\item
	Both names for \cpp templates and \cpp namespace can be automatically generated from the \wl name of the observable by inserting an underscore before capital letters (or before numbers in front of them) and lowering the letter case.
	If the generated replacement for \cpptoken{@\cpptoken\observable\_namespace@}  (\fntoken) leads to undesired side effects, then one may use \code{GetObservableNamespace} (\code{GetObservableFileName}) to override the default behavior.
\item 
	\code{GetObservableName}
	uniquely defines the \cpp name of the observable, which replaces \cpptoken{@\observable{}\_name@}.
	This name is used to store the numerical value of the observable internally and is generated automatically but can be overridden.
\end{enumerate*}
}

The meaning of two other mandatory arguments of the function \code{DefineObservables} is the following:
\begin{description}
\item[\code{GetObservableType}]
	defines the \cpp type of the observable and replaces \cpptoken{@\observable{}\_output\_type@}.
	In general, it can be any (real or complex) scalar, array or matrix \cpp type.
        The syntax \code{\{1\}} in \lstref{code:general-observables.m} defines a complex-valued array of length~1 (corresponds to the \cpp type \code{Eigen::Array<std::complex<double>,1,1>}). See also \file{BrLToLGamma/Observables.m} for another example. Note that array, vector or matrix types are convenient for storing Wilson coefficients, debugging information, etc.
	As described later, the connection between the observable type and the \fs Les Houches output is specified in the file named \code{meta/Observables/\observable/WriteOut.m}.
\item[\code{GetObservablePrototype}]
	defines the \cpp prototype for the function that calculates the observable (\cpptoken{@\cpptoken\observable\_calculate(...)@}).
	Note that the function name is modified with the values of the parameters of the first argument of the function \code{DefineObservable} (\str{parA} is replaced with value of \local{parA\_} in \lstref{code:general-observables.m}), while the function arguments are kept intact (\str{parB} is kept).\footnote{%
The following two additional function arguments may be added: \str{auto model} and \str{auto qedqcd}. The first one allows to access all numeric quantities of the model (model parameters, masses, etc.). The second one provides access to known low-energy input observables.}
\end{description}

Let us now illustrate how to create new observables in \fs starting from the usage of the function \code{DefineObservable} in \file{Observables.m} file. 
The examples are of increasing complexity and they will be continued in the next subsections. 

\subsubsection{Example 1: \exampleonename}\label{sec:constant-observable.m}

Let us create an observable, called \code{ExampleConstantObservable}, that outputs a numerical constant specified by the user in the \file{\fs.m} file, so that the value of the numerical constant is not hard-coded in the body of the observable \cpp code but rather configured with the  \wl files.
This example will demonstrate the basic usage of the function that calculates an observable and allows us to familiarize ourselves with the workflow, as no physical calculation is required.
First, one creates the \dir{meta/Observables/ExampleConstantObservable} directory. Afterwards one creates the file \file{Observables.m} inside this directory with the following content:
{
\lstset{label={code:ex1-obs}, numbers=left}
\begin{meta}{Content of \file{ExampleConstantObservable/Observables.m}.}
Observables`DefineObservable[
	FlexibleSUSYObservable`ExampleConstantObservable[num_],
	GetObservableType      -> {1},
	GetObservablePrototype -> "calculate_example_constant_observable(double num)"
];
\end{meta}
}
\noindent As in the general case discussed above, the function
call of \code{DefineObservables}  defines the \wl symbol of the
observable to be \code{ExampleConstantObservable}, and it specifies
that the observable depends on one parameter \local{num\_}; the meaning
of the parameter will be the value of the numerical constant to be
output. Accordingly, as \cpp return type for the function that calculates the observable we use an array of size one in \code{GetObservableType}.
\code{GetObservablePrototype} defines the prototype of the function that performs the calculation and we use \str{double num} as function argument.

\subsubsection{Example 2: \exampletwoname}\label{sec:ex1-observables}

Let us turn to a little more involved example to illustrate the usage of model-specific quantities in the calculation of an observable and the possibility of filling the \cpp template files with model-specific content.
As an example, we define an ``observable'' which can output several
masses whose values are determined in \fs. Specifically we choose to
allow the output the \gls{MSBAR}/\gls{DRBAR} mass of a user-selected fermion at a given renormalization scale and/or the pole mass of a \gls{SM} lepton.
Again, similar to the example above, the fermion field and its
generation number will be specified by the user in the \file{\fs.m}
file, so that these values are not hard-coded in the body of the
observable \cpp code but are configured with the \wl files.

To set up the observable, we create the directory \dir{meta/Observables/ExampleFermionMass} and the file \file{Observables.m} within it, with the following content:
{
\lstset{label={code:ex2-obs}, numbers=left}%
\begin{meta}{Content of \file{ExampleFermionMass/Observables.m}.}
Observables`DefineObservable[
	FlexibleSUSYObservable`ExampleFermionMass[fermion_[gen_]],
	GetObservableType        -> {2},$\label{line:ex2-obs-type}$
	GetObservablePrototype   -> "ex_fermion_mass(int gen, auto model, auto qedqcd)",
	GetObservableDescription -> "fermion[gen] (lepton[gen] if in Block ExampleLeptonMass) mass"
];
\end{meta}
}
\noindent Again, the function
call of \code{DefineObservables} first  defines the \wl symbol of the
observable to be \code{ExampleFermionMass}, and it specifies
that the observable depends on two arguments, merged into one \wl function. We assume that the concrete name of the fermion to
output will be defined in the \fs model file
\file{models/\model/\fs.m}, and we assume that it has the form
\code{\local{fermion}[\local{gen}]}, where \local{fermion} is the name
of the fermion multiplet and \local{gen} is its index in the
multiplet. Therefore, the observable \code{ExampleFermionMass} has the form of a function with the \code{\local{fermion\_}[\local{gen\_}]} pattern as argument, which matches the multiplet name with \local{fermion\_} and the index in the multiplet with \local{gen\_}.%
\footnote{%
	Note that in the generated \cpp code the multiplet index is decreased by one in order to match the \cpp index convention. I.e.~if one writes \code{Fe[1]} in the \wl model file \file{\fs.m} and refers in the generated \cpp code to the argument of \code{Fe}, then its value will be \code{0}, not \code{1}.
}
We would like to return two numerical values, corresponding to
the  \gls{MSBAR}/\gls{DRBAR} mass of the fermion
\code{\local{fermion}[\local{gen}]} at a given renormalization scale
and the pole mass of a \gls{SM} lepton of generation \code{\local{gen}}. Hence we define the observable type to be an array of length~2 in \code{GetObservableType}.
The prototype of the function that calculates the observable is
defined in \code{GetObservablePrototype}. The function takes the index
of the fermion in the multiplet as first argument. The remaining two
arguments \str{model} and \str{qedqcd} allow the access to the model
parameters, masses, mixing matrices, etc.
In the name of the function
  prototype \str{ex\_fermion\_mass}, the string \str{fermion} is replaced by the
  value of the variable \code{\local{fermion}}, which may be \code{Fe},  \code{Fd}, etc.~depending on the user-selected model \model.
As we plan to calculate different masses depending on the \gls{SLHA} output block (see \secref{sec:ex2-writeout}), let us implement the description of the observable as specified with \code{GetObservableDescription} above to have an explicit indication of the generated numerical values.
There, the strings \str{fermion} and \str{gen} are replaced by the
user-specified values of \local{fermion} and \local{gen}, respectively.

\subsubsection{Example 3: \examplethreename}
\label{sec:ex3-observable.m}
As a third example, let us implement an observable to calculate the self-energy of a user-specified lepton at the one-loop level.
This example will illustrate how to use the \npf extension in both simple and advanced ways, as well as how to output quantities in the \gls{FLHA} format.
As in the previous examples, we start by defining the observable and a suitable \cpp function prototype that allow us to select the lepton (from a multiplet) whose self-energy shall be calculated and the contributions that should be taken into account:
\begin{meta}{Content of \file{ExampleLeptonSE/Observables.m}.}
Observables`DefineObservable[
	FlexibleSUSYObservable`ExampleLeptonSE[field_[gen_], contr_],
	GetObservableType      -> {2},
	GetObservablePrototype -> "ex_lepton_se_contr(int gen, auto model)"
];
\end{meta}
We call the observable \code{ExampleLeptonSE}, which takes the multiplet name (\local{field\_}) and the index of the particle in the multiplet (\local{gen\_}) as first argument, as in the previous example. Furthermore, we'd like to be able to select a subset of the full one-loop contribution. To achieve this, we provide a second parameter (\local{contr\_}) which we will use later to only calculate a certain part of the full one-loop self-energy.
In general, a fermion self-energy can be split into different covariants involving left-handed or right-handed projection
  operators $P_{L,R}$ and possibly the covariant $\slashed{p}$, where
  $p^\mu$ is the external fermion momentum.
Let us in this example for simplicity output only the coefficients of
the two covariants  $\slashed{p}P_L$ and $\slashed{p}P_R$, so we choose an array of length~2.
The \cpp function needs the index of the particle in the multiplet (\str{gen}) and the model object (\str{model}) as arguments.
 
\subsection{Content of the file \file{\observable/WriteOut.m}}\label{sec:write-general}
\subsubsection*{General case}

After the definition of the observable name, type, etc.~in the file \file{\observable/Observables.m}, one can now connect the numerical value(s) of the observable to the Les Houches output of the \cpp spectrum generator in the file \file{meta/Observables/\observable/WriteOut.m}.
Currently, there are two automated ways to output the results of calculations in \fs: via the \gls{SLHA} (exists since \fs~\code{1.0}) and the \gls{FLHA} (the usage is explained in this article) formats.
Both are defined in the \file{meta/WriteOut.m} file, which provides functions to write numbers, arrays or matrices to specified output blocks.

In the simplest case the value of an observable \code{\observable{}[\local{parA}, \local{parB}, ...]} is a single complex number (see Examples 1--2 below or the \twodecays decay; their Les Houches output is shown in \secsref{sec:ex1-final}{sec:ex2-final}). If we want to write its real part to the \code{FlexibleSUSYLowEnergy} \gls{SLHA} output block, we define a function called \code{WriteObservable} in \file{meta/Observables/\observable/WriteOut.m} as follows:
\begin{meta}{Content of \file{meta/Observables/\observable/WriteOut.m}.}
WriteOut`WriteObservable[
	"FlexibleSUSYLowEnergy",
	obs:FlexibleSUSYObservable`$\observable$[parA_, parB_, ...]
] := "Re(observables." <> Observables`GetObservableName[obs] <> "(0))";
\end{meta}

\noindent
The first argument to the \code{WriteObservable} function is a string with the \gls{SLHA} output block name (here: \str{"FlexibleSUSYLowEnergy"}), which reflects the \gls{SLHA} block name in the \fs model file \code{models/\model/\fs.m}.
Since the observable type is a complex-valued array of length~1 in this example, we have to output the 0-th element of the array (in \cpp index convention). We obtain the real part of the array element by applying the \code{Re} function.

As another example we consider the output of the real parts of several Wilson coefficients (see also Example~3 below and its Les Houches output in \secref{sec:lepton-se-exe}). This requires a more involved definition of the function \code{WriteObservable}:
\begin{meta}{Content of \file{meta/Observables/\observable/WriteOut.m} to output the real parts of several Wilson coefficients.}
WriteOut`WriteObservable[
	"FWCOEF", (* Or "IMFWCOEF" with Re below replaced with Im *)
	obs:FlexibleSUSYObservable`$\observable$[parA_, parB_, ...]
] := 
StringReplace[
	{
		"fermions$_\stringindex{1}$, operator$_\stringindex{1}$, Oalpha$_\stringindex{1}$, Oalphas$_\stringindex{1}$, contributions$_\stringindex{1}$, num_value, \ $\texttt{\kern-0.6em\color{color-listings-string}"}$comment$_\stringindex{1}$\ $\texttt{\color{color-listings-string}\kern-0.6em"}$", 
		"fermions$_\stringindex{2}$, operator$_\stringindex{2}$, Oalpha$_\stringindex{2}$, Oalphas$_\stringindex{2}$, contributions$_\stringindex{2}$, num_value, \ $\texttt{\kern-0.6em\color{color-listings-string}"}$comment$_\stringindex{2}$\ $\texttt{\color{color-listings-string}\kern-0.6em"}$", 
		...
	}, 
	{
		"num_value" -> "Re(observables." <> Observables`GetObservableName[obs] <> ")", 
		...
	}
];
\end{meta}
We define the function \code{WriteObservable} to write the real parts of the Wilson coefficients to the \code{FWCOEF} Les Houches output block.\footnote{%
If the imaginary parts of the Wilson coefficients shall be written to the output, one should use \str{"IMFWCOEF"} as block name and replace  \code{Re} with \code{Im}.
} The return value of \code{WriteObservable} is a list of strings, with each string consisting of a comma-separated tuple of a fermion name (\str{fermions$_\stringindex{1}$}), an operator name (\str{operator$_\stringindex{1}$}), etc., which should be replaced by appropriate values.  See Ref.~\cite{Mahmoudi:2010iz} for a description of the \gls{FLHA} format.
The numbering convention for the \gls{FLHA} format in \fs is the following: Wilson coefficients must occupy the positions from the end of the \cpp output array.
In the definition of the function \code{WriteObservable} the substring \str{"num\_value"} is replaced by the real parts of the Wilson coefficients (accessed via \code{GetObservableName}) via the function \code{StringReplace}.  Note that there is no need to explicitly specify the numbering of the Wilson coefficients in the replacement rule for \str{"num\_value"}, as it is done automatically.

\subsubsection{Example 1: \exampleonename}\label{sec:constant-writeout.m}

We continue our minimal Example~1 from \secref{sec:constant-observable.m}, where the observable is defined to just be a user-defined numeric constant specified in the \fs model file. The second step is to connect the numeric value of the observable to the \gls{SLHA} (or \gls{FLHA}) format, which is the output of the spectrum generator.
This is done by the following definition placed in the file \file{meta/Observables/ExampleConstantObservable/WriteOut.m}:
\begin{meta}{Content of \file{ExampleConstantObservable/WriteOut.m}.}
WriteOut`WriteObservable[
	"FlexibleSUSYLowEnergy",
	obs:FlexibleSUSYObservable`ExampleConstantObservable[_]
] := "Re(observables." <> Observables`GetObservableName[obs] <> "(0))";
\end{meta}
As the first argument of the function \code{WriteObservable} shows, the numeric value of the observable is written to the \gls{SLHA} output block \code{FlexibleSUSYLowEnergy}, which matches a corresponding definition in the model file \file{models/\model/\fs.m}.
The function returns a string (which must be valid \cpp syntax), where we take the real part of the first entry from the array of length~1, where the observable is stored (with the zero-based index convention in \cpp).
Since for the output no information about the observable is needed, one uses the \code{\_} pattern in the specification of \code{ExampleConstantObservable}.
The code listing above leads to the usage of the \cpp parser for the \gls{SLHA} format named \code{FORMAT\_ELEMENT}, see Listing~\ref{code:all-cpp-blocks}.

\subsubsection{Example 2: \exampletwoname}\label{sec:ex2-writeout}

Now we turn back to Example~2 from above and connect our observable (two given fermion masses) to the output of the spectrum generator. Similarly to Example~1, one creates the file \file{meta/Observables/ExampleFermionMass/WriteOut.m} and places the definitions of the function \code{WriteObservable} there.
In this example we would like to output two different masses at the same time. To store them internally, we have defined the observable type to be an array of length~2, see the specification of \code{GetObservableType} in \lineref{line:ex2-obs-type} of \lstref{code:ex2-obs}. To write the two fermion masses to the output, one could do the following:
\begin{meta}{Content of \file{ExampleFermionMass/WriteOut.m}.}
WriteOut`WriteObservable[
	"FlexibleSUSYLowEnergy",
	obs:FlexibleSUSYObservable`ExampleFermionMass[_]
] := "Re(observables." <> Observables`GetObservableName[obs] <> "(0))";

WriteOut`WriteObservable[
	"ExampleLeptonMass",
	obs:FlexibleSUSYObservable`ExampleFermionMass[_]
] := "Re(observables." <> Observables`GetObservableName[obs] <> "(1))";
\end{meta}
Here, we define two distinct behaviours of our observable with two different Les Houches output blocks (which must be reflected by appropriately named lists in \file{models/\model/\fs.m}): \str{"FlexibleSUSYLowEnergy"} and \str{"ExampleLeptonMass"}. 
In both cases, definitions lead to the \cpp parser named \code{FORMAT\_ELEMENT}, see Listing~\ref{code:all-cpp-blocks}.
The fermion mass in the first entry of our two-component observable array is written to the block \code{FlexibleSUSYLowEnergy}, while the second entry is written to the block \code{ExampleLeptonMass}.
Note that the name of this second block is not a part of the official \gls{SLHA} standard, but is a non-standard addition we use for this example.

\subsubsection{Example 3: \examplethreename}

In Example~3, we aim to output the two components of a lepton self-energy.
We can use the automatic \gls{FLHA} output format to achieve this:
\begin{meta}{Content of \file{ExampleLeptonSE/WriteOut.m}.}
WriteOut`WriteObservable[
	"FWCOEF",
	obs:FlexibleSUSYObservable`ExampleLeptonSE[_[gen_], _]
] :=
StringReplace[
	{
		"leptons, 31, 0, 0, 2, num_value, \ $\texttt{\kern-0.6em\color{color-listings-string}"}$left\ $\texttt{\kern-0.6em\color{color-listings-string}"}$", (* P_L *)
		"leptons, 32, 0, 0, 2, num_value, \ $\texttt{\kern-0.6em\color{color-listings-string}"}$right\ $\texttt{\kern-0.6em\color{color-listings-string}"}$" (* P_R *)
	},
	{
		"num_value" -> "Re(observables." <> Observables`GetObservableName[obs] <> ")",
		"leptons"   -> Switch[gen, 0, "1111", 1, "1313", 2, "1515"]
	}
];
\end{meta}
In the definition of \code{WriteObservable} we provided a pattern for the index of the lepton in the lepton multiplet (\local{gen\_}) explicitly in the first argument of the observable name \code{ExampleLeptonSE}, because the concrete value of this index must be known to select the appropriate self-energy for the output.

\subsection{Content of the file \file{\observable/FlexibleSUSY.m}, simple settings for \npf}\label{sec:content-fs-file}
\subsubsection*{General case}
In the previous sections it was described how to define a new observable (done in the file \file{Observables.m}) and how to write the numeric value of the observable to the Les Houches output of the spectrum generator (defined in the file \file{WriteOut.m}).
In this section we describe how to generate the content of the \cpp function, that calculates the numeric value of the observable. This \cpp function is defined in the \cpp template files located in \dir{templates/observables}, which contains placeholders that are replaced by expressions to calculate the observable. The rules that specify how to replace the placeholders by appropriate expressions are defined in the observable-specific file \file{meta/Observables/\observable/\fs.m}. This file contains at least the function \code{WriteClass}, which performs the following tasks:
\begin{enumerate}
	\item
	Fill the \cpp templates from \dir{templates/observables} with appropriate \cpp code.
	\item 
	Move the filled \cpp templates into \dir{models/\model/observables}.
\end{enumerate} 

The following source code listing shows an example for the function \code{WriteClass} that does some basic replacements that we will explain in the following:
{
\lstset{label={code:general-fs},numbers=left, breaklines=false}
\begin{meta}{General content of \file{meta/Observables/\observable/FlexibleSUSY.m}.}
FlexibleSUSY`WriteClass[obs:FlexibleSUSYObservable`$\observable$, allObs_, files_] :=
Module[
	{
		observables = DeleteDuplicates[Cases[Observables`GetRequestedObservables[allObs], _obs]],
		prototypes = {}, definitions = {}, 
		npfHeaders = "", npfDefinitions = {}, cxxVertices = {}
	},

	If[observables =!= {}, 
		Utils`PrintHeadline["Creating " <> SymbolName[obs] <> " class ..."];
		
		prototypes = TextFormatting`ReplaceCXXTokens[
			"@type@ @prototype@;",
			{
				"@type@"      -> CConversion`CreateCType[Observables`GetObservableType[$\lsyntax{\#}$]],
				"@prototype@" -> Observables`GetObservablePrototype[$\lsyntax{\#}$]
			}
		] $\lsyntax{\&/@}$ observables;
		
		(* Task 1: filling definitions *)$\label{line:general-fs-task1}$
	];

	(* Task 2: filling templates and moving them into models/${\color{color-listings-comments}\model}$/observables/ *)
	WriteOut`ReplaceInFiles[
		files,
		{
			"@npf_headers@"           -> npfHeaders,$\label{line:general-fs-cpp-replacements-start}$
			"@npf_definitions@"       -> StringRiffle[DeleteDuplicates[npfDefinitions], "\n\n"],
			"@calculate_prototypes@"  -> StringRiffle[DeleteDuplicates[prototypes],     "\n\n"],
			"@calculate_definitions@" -> StringRiffle[DeleteDuplicates[definitions],    "\n\n"],
			"@include_guard@"         -> SymbolName[obs],
			"@namespace@"             -> Observables`GetObservableNamespace[obs],
			"@filename@"              -> Observables`GetObservableFileName[obs],
			Sequence $\lsyntax{@@}$ $\context{FlexibleSUSY\textasciigrave Private}$GeneralReplacementRules[]$\label{line:general-fs-cpp-replacements-end}$
		}
	];
	
	(* Task 3: returning something to the outside world *)$\label{line:general-fs-task3}$
	{
		"for_outside_usage$_\stringindex{1}$" -> ...,$\label{line:general-fs-outside1}$
		"for_outside_usage$_\stringindex{2}$" -> ...,
		...
	}
];
\end{meta}
}
\noindent
The function \code{WriteClass} has three parameters: the explicit name of the observable (\local{obs}), the list of all observables that are requested --- by the variable \code{ExtraSLHAOutputBlocks} from the file \file{models/\model/\fs.m} --- to be calculated (\local{allObs\_}), and a list of the \cpp template file names, where the replacements should be performed, and the corresponding output file names (\local{files\_}).

The body of the function \code{WriteClass} consists of three parts:
\begin{enumerate}
\item The first part is the \code{If} statement, where all \cpp expressions
are created (\task{1} as indicated in the listing above). 
In general, we are interested in defining one observable \observable unified by similar calculations, e.g.~we define one observable \observable for the set of processes \threedecays instead of multiple observables $\mu\to 3e$, $\tau\to 3\mu$, etc. Then, we enable specific realizations of \observable in \code{ExtraSLHAOutputBlocks}, see also explicit examples in Listing~\ref{code:ex1-multiple-observables} and \ref{code:ex2-multiple-observables}. 
So, we start with selecting unique realizations of chosen \observable from all \local{allObs} and storing them into \local{observables}.
Then, inside \code{If} statement we make changes to the \cpp code generation based on the possible realization-specific features (e.g.~the process $\tau\to 3\mu$ might require additional contributions, compared to $\mu\to 3e$).
Changes in \cpp prototypes stored in the variable \local{prototypes} are handled automatically due to the function \code{WriteObservable}, see Listing~\ref{code:general-observables.m}, while the \cpp definitions in the variable \local{definitions} usually require manual coding, see Listings~\ref{code:ex1-fs}--\ref{code:ex3-simple-fs}.
In the above example source code listing \local{prototypes} strings are created by replacing the \str{"@type@"} and \str{"@prototype@"} tokens by the observable's \cpp type and the prototype of the function that calculates the numeric value of the observable, respectively.
\item In the second part (\task{2} indicated in the listing) the \cpp tokens (\str{"@npf\_headers@"}, \str{"@npf\_definitions@"}, etc.\@) are replaced in the \cpp template files (\local{files}) that are passed to the function with the help of the \code{ReplaceInFiles} function.
\item The third part (\task{3}) is to specify the function's returned expression. In the example above the function returns a list of replacement rules, whose use is described below.

Note that in \task{2} the tokens in the \cpp template files can be replaced by strings that can in principle be arbitrarily large. However, we recommend to put as much generic information as possible into the \cpp template files and replace the tokens in the template files only by model-specific information. 
For the latter we recommend to use the full power of \fs's helper routines and functions located in \file{meta/TextFormatting.m}, \file{meta/CConversion.m} and \file{meta/Utils.m}.
We refer the reader also to \secref{sec:advanced-npf-usage}, where we describe the \npf extension and how one can use it to generate \cpp code for amplitudes.
\end{enumerate}

The expression returned by the \code{WriteClass} function (\task{3})
should be a list of replacement rules, which gets stored internally in
the variable \code{ObservablesExtraOutput[\str{"\observable{}"}]}. The
replacement rules stored in the returned list can be accessed and
re-used later, if needed. For example, one can access expression stored in the rule defined in \lineref{line:general-fs-outside1} from \lstref{code:general-fs} as follows:
\begin{meta}{Accessing an expression returned by the \code{WriteClass} function.}
expr = Cases[ObservablesExtraOutput["$\str{\observable}$"], ("for_outside_usage$_\stringindex{1}$" -> res_) :> res];
\end{meta}
Besides the possible manual re-use of the returned expressions, \fs automatically performs the following two tasks with the returned list of replacement rules:
\begin{enumerate}
\item If there exists a replacement rule of the form \code{\str{"C++ vertices"} -> \local{list$_{\code{1}}$}}, then the expression \local{list$_{\code{1}}$} must be a list of vertices that are required to calculate the observable. Each vertex is represented by a list of \sarah fields, e.g.~\code{\{bar[Chi], Chi, VP\}} in the \gls{MRSSM} (represents the $\bar\chi_a^0\chi_b^0 \gamma$ vertex with two neutralinos and a photon). For each required vertex \fs creates a corresponding \cpp function for its numerical evaluation when calculating the numeric value of the observable. 
\item If there exists a replacement rule of the form \code{\str{"C++ replacements"} -> \local{list$_{\code{2}}$}}, then the expression \local{list$_{\code{2}}$} must be a list of replacement rules that should be applied to all \cpp template files, see \ref{sec:extra-mec-input} for an example.
\end{enumerate}

\subsubsection*{Enabling optional \npf extension, simple settings}
If the observable relies on tree-level or one-loop level Feynman
diagrams then there is an automated way to generate them in \fs
with the help of the \npf extension already announced in
Ref.~\cite{Khasianevich:2022ess}. \npf is typically used from within
the \code{WriteClass} function. Internally, the extension calls
\fa~\cite{Hahn:2000kx}, \fc~\cite{Hahn:1998yk}, and \colormath~\cite{Sjodahl:2012nk} to generate analytic expressions and converts them into \cpp
form for their numeric evaluation in \fs. It thus allows access to
Feynman diagrammatic computations in the definition of observables.

\npf can be used in two modes which we will refer to as simple and
advanced modes. The modes differ by the accessible settings. 	Both types of settings serve to modify the calls of the \fa and \fc routines: In the simple mode only the
settings listed in \tabref{tab_simple-options} are accessible,
which allow topology-indepedent modifications. The advanced mode
allows many more settings, enabling in particular to select specific
option values for selected topologies. 
	In the present section, we focus on the simple settings from
        \tabref{tab_simple-options}. 
	These simple settings are also illustrated with the help of
        Example~3 in \secref{sec:lepton-se-simple}. Later, the
        advanced settings are explained in detail in
        \secref{sec:advanced-npf-usage} and exemplified in an extra \secref{sec:lepton-se-npf}.
\begin{table}[t!]
\renewcommand*{\arraystretch}{1.2}
\setlength{\tabcolsep}{4pt}
\centering
\begin{tabular}{>{\centering\arraybackslash}p{\textwidth-2\tabcolsep}}
	\hline
	Usage
	\\\hline
	\cellcolor{color-listings-background}
	\placeintable
\begin{meta}{}
NPointFunction[..., Option$_\optionindex{i}$ -> Value$_\optionindex{j}$]
\end{meta}
\\\hline
\end{tabular}
\begin{tabularx}{\textwidth}{>{\centering}p{4cm}>{\centering}p{3cm}X}
	\hline
	Option & Values & \multicolumn{1}{c}{Hints}
	\\\hline
	\code{Regularize} & \code{MSbar}, \code{DRbar} & Corresponds to \code{\context{FormCalc}Dimension}: \code{D}, \code{4}; can be overriden by \code{regularization} setting
	\\
	\code{ZeroExternalMomenta}
	&
	\code{True}, \code{False}, \code{OperatorsOnly}, \code{ExceptLoops}
	&
	External momenta and mass treatment
	\\
	\code{OnShellFlag}
	&
	\code{True}, \code{False}
	&
	Usage of on-shell external particles
	\\
	\code{UseCache}
	&
	\code{True}, \code{False}
	&
	Cache usage 
	\\
	\code{Observable}
	&
	\code{None} or \code{\observable{}[]}
	&
	Mode of \npf: simple or advanced, that enables \file{\observable/NPointFunctions.m}
	\\
	\code{KeepProcesses} & & For \code{Observable -> None}, see allowed values in the function \code{GetExcludeTopologies} in the file \file{meta/\npf/Topologies.m}
	\\
	\code{LoopLevel} & \code{0} or \code{1} & Loop level, \fc limitation
	\\\hline
\end{tabularx}%
\caption{
		Mandatory options for the function \code{NPointFunction}, see the function \code{CheckOptionValues} in \file{meta/NPointFunctions.m} file for allowed values, and \file{meta/Observables/\observable/FlexibleSUSY.m} files for examples.
	}
\label{tab_simple-options}
\end{table}

\begin{description}
\item[\code{ZeroExternalMomenta}] 
	can currently be \code{True}, \code{False}, \code{OperatorsOnly}, and \code{ExceptLoops}.
	This option also specifies the way external masses should be treated for specific topologies, fermions bispinors, and in loop integrals.
	If set to \code{True}, all external momenta are set to zero everywhere, while if set to \code{ExceptLoops}, the scalar products in loop integrals are kept.
	This allows one to correctly implement the expressions for self-energy-like diagrams relevant for \gls{CLFV} processes in particular.
\item[\code{UseCache}] 
	stores the \fa and \npf output for future usage, if enabled.
	This speeds up the code generation if \dir{model/\model} is purged or parts of the meta code are modified.
\item[\code{KeepProcesses}] 
	specifies the mode of \npf.
	There are two modes to calculate an amplitude: using simple settings only (\code{Observable -> None}) and with the help of advanced ones defined in \file{meta/Observables/\observable/NPointFunctions.m} files (\code{Observable -> \observable{}[]}), see \secref{sec:advanced-npf-usage}.%
	\footnote{%
		The simple mode with the simple settings discussed
                here was developed mainly to allow users a simple starting
                point. It is also used in unit testing of \fs.
		Currently, \fs performs unit tests of self-energy expressions in the \gls{SM}/\gls{MSSM}  and $Z$-boson penguins in the \gls{MRSSM}.
		It is also used in Example 3 to demonstrate the basic
                usage of the \npf package and can serve to write the
                first iterations of the code that relies on
                \npf. Typically, the code for actual physics
                observables such as the ones discussed in
                \secref{sec:all-obs} will use the advanced settings of \npf.
	}	
\end{description}

\subsubsection{Example 1: \exampleonename}\label{sec:constant-flexiblesusy.m}

Each observable undergoes various calculational steps before being
evaluated into an actual numerical result. In general, the definitions
of the calculational steps are distributed between the \wl file
\file{meta/Observables/\observable/FlexibleSUSY.m} and the \cpp
template files discussed.  Specifically, the \code{WriteClass}
function in the \file{\fs.m} file is supposed to generate
model-specific expressions to be placed into \cpp template files from
\secref{sec:constant-cpp}. As stated before, in general it is
recommended to put as much information as possible into the \cpp
template files and keep the \code{WriteClass} function minimal. In
case of Example~1, however, the example is so simple that we would
like \code{WriteClass} to generate the complete \cpp code for all
calculations (just output a number in this case), thus the \cpp
template files will not contain much code.  This code generation is
done with the definition of \task{1} and \task{3} in the \code{WriteClass}
function (by replacing \lineref{line:general-fs-task1} and \lineref{line:general-fs-task3} in \lstref{code:general-fs}) as
follows: {%
  \lstset{label={code:ex1-fs},numbers=left}
\begin{meta}{Content of \file{ExampleConstantObservable/FlexibleSUSY.m}.}
(* Task 1: generate function definition by replacing tokens by concrete C++ code *)
definitions = TextFormatting`ReplaceCXXTokens["
	@type@ @prototype@ {$\label{line:ex1-fs-cpp-start}$
		@type@ res {num};
		return res;
	}",$\label{line:ex1-fs-cpp-end}$
	{
		"@type@"      -> CConversion`CreateCType[Observables`GetObservableType[$\lsyntax{\#}$]],
		"@prototype@" -> Observables`GetObservablePrototype[$\lsyntax{\#}$]
	}
] $\lsyntax{\&/@}$ observables;

...

(* Task 3: return an empty list *)
{}
\end{meta}
}%
\noindent In the above source code listing the variable
\local{definitions} is defined to contain the entire definition of the
function that calculates the observable (including function body). It
is generated by replacing the \str{@type@} token by the concrete \cpp
type \code{Eigen::Array<std::complex<double>,1,1>} and the function
name \str{@prototype@} (including its argument \str{double num}) as
specified by the definitions of \secref{sec:constant-observable.m} in
a generic string. In the body of the function, we initialize a local
variable \str{res} of type \str{@type@} and set its value to the value
of \str{num}. The value of \str{res} is then returned to
  further fill the \gls{SLHA} block entries. The final function
definition in the \local{definitions} variable will be used later in
\secref{sec:constant-cpp} together with the \cpp template files to
construct the complete \cpp code to output the numerical value of the
observable.

The \code{WriteClass} function finally returns an empty list (\task{3}),
because we do not need any of the defined expressions anywhere else in
this example.

Note that the function definition generated in \task{1} is later put
into the \cpp template files in \dir{templates/observables} an is thus
closely connected to their content, as shown in
\secref{sec:constant-cpp}.

\subsubsection{Example 2: \exampletwoname}\label{sec:ex2-fs}

Now we return to Example~2 where we output fermion masses.  In this
example we follow the general recommendation to put as much \cpp code
as possible into the \cpp template files and as little as possible
into \file{\fs.m}. Thus, in this example we fill the following code
into the \code{WriteClass} function in the \file{\fs.m} file of
\lstref{code:general-fs}:
{\lstset{label={code:ex2-fs}}%
\begin{meta}{Content of \file{ExampleFermionMass/FlexibleSUSY.m}.}
(* Task 1: generate function definition by replacing tokens by concrete C++ code *)
definitions = TextFormatting`ReplaceCXXTokens["
	@type@ @prototype@ {
		return forge<@type@, fields::@fermion@>(gen, model, qedqcd);
	}",
	{
		"@fermion@"   -> SymbolName[Head[First[$\lsyntax{\#}$]]],
		"@type@"      -> CConversion`CreateCType[Observables`GetObservableType[$\lsyntax{\#}$]],
		"@prototype@" -> Observables`GetObservablePrototype[$\lsyntax{\#}$]
	}
] $\lsyntax{\&/@}$ observables;

...

(* Task 3: return an empty list *)
{}
\end{meta}
}
\noindent
Again, the variable \local{definitions} contains the definition
for each \cpp function for the observable calculation of the
function that calculates the numerical value of the observable. This
function has the name and type that were specified via
\code{GetObservablePrototype} and \code{GetObservableType},
respectively. In this example the function body contains only a single
line, which contains a call of the template function \str{forge}. By
this call we delegate the computation of the observable to the
\str{forge} function, which is defined entirely at the \cpp level in
\secref{sec:fermion-mass-cpp}. The same strategy is applied in several
of the predefined observables discussed in \secref{sec:all-obs} such
as \threedecays or \mec.  The template parameter
\str{fields::@fermion@} of the \str{forge} function above is an
internal \cpp type that represents the fermion whose mass shall be
output (in the \gls{MRSSM}, for example, the token \str{@fermion@} is
replaced by \code{Fe}).

Again, we do not want to use any expressions defined in the
\code{WriteClass} function anywhere else, so we return an empty list
from the function (\task{3}).

\subsubsection{Example~3: \examplethreename[(\code{Observable -> None})]}\label{sec:lepton-se-simple}

This example illustrates the use of the \npf extension of \fs. We will
use the \npf extension in the simplified mode, where we do not make
use of the advanced settings in \file{\npf.m} and just specify the
option \code{Observable -> None}. The following code snippet shows the
content of the \code{WriteClass} function:
{%
\lstset{label={code:ex3-simple-fs},numbers=left}
\begin{meta}{Content of \file{ExampleLeptonSE/\fs.m} with the option \code{Observable -> None}.}
(* Task 1: generate function definition by replacing tokens by concrete C++ code *)

(* Task 1a: delete duplicates and ignore fermion generation index *)
observables = DeleteDuplicates[observables /. f_[_Integer] :> f[_]];

(* Task 1b: run NPointFunction *)
Module[{field, contr, npf, basis, name},
	field = Head[First[$\lsyntax{\#}$]];
	contr = Last[$\lsyntax{\#}$];$\label{line:ex3-simple-fs-contr-definition}$
	npf = NPointFunctions`NPointFunction[
		{field}, (* Incoming particles *)
		{field}, (* Outgoing particles *)
		NPointFunctions`UseCache            -> False,
		NPointFunctions`OnShellFlag         -> True,
		NPointFunctions`ZeroExternalMomenta -> True,
		NPointFunctions`LoopLevel           -> 1,
		NPointFunctions`Regularize          -> FlexibleSUSY`FSRenormalizationScheme,
		NPointFunctions`Observable          -> None,$\label{line:ex3-simple-fs-observable}$
		NPointFunctions`KeepProcesses       -> {Irreducible}$\label{line:ex3-simple-fs-keep}$
	];

	basis =
	{
		"left_wilson"  -> NPointFunctions`DiracChain[SARAH`DiracSpinor[field[{SARAH`gt2}], 0, 0], 7, SARAH`DiracSpinor[field[{SARAH`gt1}], 0, 0]],
		"right_wilson" -> NPointFunctions`DiracChain[SARAH`DiracSpinor[field[{SARAH`gt2}], 0, 0], 6, SARAH`DiracSpinor[field[{SARAH`gt1}], 0, 0]]
	};

	npf = WilsonCoeffs`InterfaceToMatching[npf, basis];$\label{line:ex3-simple-fs-interface}$
	name = "se_irr";

	AppendTo[cxxVertices, NPointFunctions`VerticesForNPointFunction[npf]];
	AppendTo[npfDefinitions, NPointFunctions`CreateCXXFunctions[npf, name, Identity, basis][[2]]];$\label{line:ex3-simple-fs-create}$
	AppendTo[definitions,
		TextFormatting`ReplaceCXXTokens["
			@type@ @prototype@ {
				const auto npf = npointfunctions::@name@(model, {gen, gen}, {});
				return {npf[0], npf[1]};
			}",
			{
				"@type@"      -> CConversion`CreateCType[Observables`GetObservableType[$\lsyntax{\#}$]],
				"@prototype@" -> Observables`GetObservablePrototype[$\lsyntax{\#}$],
				"@name@"      -> name
			}
		]
	];
] $\lsyntax{\&/@}$ observables;

(* Task 1c: obtain list of C++ header files *)
npfHeaders = NPointFunctions`CreateCXXHeaders[];

...

(* Task 3: returning something to the outside world *)
{"C++ vertices" -> Flatten[cxxVertices, 1]}
\end{meta}
}%
\noindent First of all, as \npf returns the same code for different generations of external particles, we remove potential duplicates by replacing integer generation with the \code{\_} pattern for simplicity (it could have been any symbol or number), see \task{1a}.

Afterwards, we run \npf (\task{1b}) to generate the relevant \cpp code,
and all required \cpp vertices. For this we create a local function
(defined as a \code{Module}), which we map to all observables in the
\local{observables} list. Note, however, that in more involved
calculations it may be better to define a non-local, separate function
to process all \local{observables}, instead of a local one via a
\code{Module}, as done here. The local function to obtain the \cpp
code to numerically evaluate the observable(s) stores its results in
the following local variables:
\begin{description}
\item[\local{field}] represents the lepton particle that we specify in
  the definition of the observable (like \code{Fe} for example in the
  \gls{SM} or \gls{MRSSM}).
	
\item[\local{contr}] contains an expression that allows one to select
  certain contributions from the self-energy that should be output. In
  this example we do not make use of this selection feature. See
  \secref{sec:lepton-se-npf} for an advanced example.
	
\item[\local{npf}] contains the main result of \npf: an object with
  generic amplitudes and, sometimes, replacement rules. See
  \tabref{tab_simple-options} for a list of all possible options.  In
  this example, we calculate the \code{Fe -> Fe} amplitude, from which
  we will extract the self-energy (note that the outgoing particle is
  typed as \code{Fe} and not \code{bar[Fe]}).  We do not store the
  results in the cache.  \code{OnShellFlag} is used internally by the
  option \code{\context{FormCalc}OnShell} for the function
  \code{\context{FormCalc}CalcFeynAmp}.  \code{ZeroExternalMomenta} is
  passed to the function \code{\context{FormCalc}OffShell}.  The loop
  level is specified explicitly to be \code{1}.  The renormalization
  scheme is decided to be set by \fs.

  We chose a simple operation mode of the \npf extension by setting
  \code{Observable -> None}.  This implies, that no optional file \file{\observable/\npf.m} with advanced settings is used, see \secref{sec:advanced-npf-usage} for more details and the usage example in \secref{sec:lepton-se-npf}. The option \code{KeepProcesses} allows selecting certain topologies, provided by the list of default values within the function
  \code{GetExcludeTopologies} in \file{meta/\npf/Topologies.m}.  To
  extend this list, we refer to the usage of
  \code{\context{FeynArts}\$ExcludeTopologies}.
	
\item[\local{name}] contains the \cpp name of \npf-generated function
  that calculates the numerical value of the observable.

\item[\local{basis}] is a list of replacement rules to extract certain
  sub-expressions of the amplitude and store them in local \cpp
  variables. The rhs.~of each replacement rule defines the
  expression, whose prefactor should be extracted. The lhs.~defines
  the name of the \cpp variable in which this prefactor shall be
  stored. Note that patterns are currently not supported to select
  the sub-expression.
  The replacement rules stored in the \local{basis} variable should be
  passed to the \code{InterfaceToMatching} function, which performs
  the extraction of the prefactors.\footnote{%
Sometimes the question arises how to know
the exact form of the expressions in \local{basis}? As a practical recommendation, one can place the \code{Print[\local{npf}];} statement right before the definition of \local{basis} and then iteratively run \fs several times to figure out the exact form of the basis elements. 
In general, they have to explicitly represent the structures, no patterns are supported here at this moment.  
}

\item[\local{cxxVertices}] contains the list of all vertices that are
  required to numerically calculate the observable. They have to be
  returned from the \code{WriteClass} function (see \task{3}), because
  \fs stores the \cpp code for the vertices in other files.
	
\item[\local{npfDefinitions}] contains strings with all generated
  amplitude-related \cpp functions required for the numerical
  evaluation of the amplitude. The function \code{CreateCXXFunctions}
  converts the \npf object into \cpp code. The user provides the
  \local{name} of the \cpp function and specifies which color
  structures are expected (\code{Identity} or
  \code{\context{SARAH}Delta}).
	
\item[\local{definitions}] contains the \cpp function that calculates
  the numerical value of the observable with the help of \cpp code
  created by the \npf extension. The function body consists of a call
  to the generated \str{irr\_se} function (the function name is stored
  in the \local{name} variable), to which the model parameters
  (\str{model} object) and the indices of the incoming and outgoing
  particles (both are \str{gen} here) are passed. The last argument
  contains the set of momenta of the external particles. In the
  current implementation external momenta are assumed to be zero and
  so the last function argument should be an empty initializer list
  \code{\{\}}.

  Since \local{basis} contains two replacement rules that define
  prefactors of certain Lorentz structures, \str{irr\_se} will return
  a two-valued array. The generated function eventually returns the
  two components of this array.
\end{description}
In \secref{sec:lepton-se-npf} the example described here is extended to use advanced \npf options.

\subsection{Content of the \cpp template files}\label{sec:content-cpp}
\subsubsection*{General case}

The function \code{WriteClass} defined in the file \file{meta/FlexibleSUSY.m} in the previous section fills \cpp templates with model-specific information, and places the resulting files into the directory \dir{models/\model}.
The content of the \cpp template files is discussed in this section.

For each observable one has to create two \cpp template files in the \file{templates/observables} directory: a \file{.hpp.in} and \code{.cpp.in} file.
These files may contain the tokens that are replaced by the \code{WriteClass} function. The following two source code listings show generic examples of these template files:
\newpage 
{%
\lstset{label={code:general-hpp-in}}
\begin{cppcode}{General content of \cpp header template \file{templates/observables/\fntoken.hpp.in}.}
#ifndef @ModelName@_@include_guard@_H
#define @ModelName@_@include_guard@_H
#include "lowe.h"
// Auxiliary $\code{\color{color-listings-comments}\#include}$ directives could come here

namespace flexiblesusy::@namespace@ {

// Auxiliary observable-specific expressions could come here

// declaration of the function that calculates the observable
@calculate_prototypes@

} // $\commnamespace$ flexiblesusy::@namespace@

#endif
\end{cppcode}
}%
{%
\lstset{label={code:general-cpp-in}}	
\begin{cppcode}{General content of \cpp definitions template \file{templates/observables/\fntoken.cpp.in}.}
#include "@ModelName@_mass_eigenstates.hpp"
#include "cxx_qft/@ModelName@_qft.hpp"
#include "@ModelName@_@filename@.hpp"
@npf_headers@
// Auxiliary $\code{\color{color-listings-comments}\#include}$ directives could come here

namespace flexiblesusy {
namespace @ModelName@_cxx_diagrams::npointfunctions {	
	
@npf_definitions@

} // $\commnamespace$ @ModelName@_cxx_diagrams::npointfunctions

using namespace @ModelName@_cxx_diagrams;
namespace @namespace@ {	
	
// Auxiliary observable-specific expressions could come here

// Definition of the function that calculates the observable
@calculate_definitions@

} // $\commnamespace$ @namespace@
} // $\commnamespace$ flexiblesusy
\end{cppcode}
}%
Both template files contain \cpp tokens (e.g.~\code{\color{color-cpp-token}@calculate\_definitions@}) to be replaced by the \code{WriteClass} function, defined in \file{\fs.m} during \fs's meta phase. For example, \cpptoken{@npf\_headers@} will be replaced with the content of \local{npfHeaders} and similar for other tokens, see \linesref{line:general-fs-cpp-replacements-start}{line:general-fs-cpp-replacements-end} of \lstref{code:general-fs}.
If no replacement rule for a token is found, the token is replaced by an empty strings.

The numerical calculation of the observable might require more \cpp auxiliary functions or classes. To use extra functions/classes one could add appropriate preprocessor \code{\color{color-cpp-preprocessor}\#include} directives.

Note again, that there is some freedom to move code between the
\file{\observable/\fs.m} and \file{\fntoken.cpp.in} files. However, as
stated before, we recommend to move as much \cpp code as possible into
the dedicated directory \dir{templates/observables}, as \cpp source
code is often more robust compared to \wl scripts due to static type
checking.

\subsubsection{Example 1: \exampleonename}\label{sec:constant-cpp}

To output a single number we create two \cpp template files in \dir{templates/observables}, as stated above.  They will
be filled by the \code{WriteClass} function defined in \file{\fs.m}
created in \secref{sec:constant-flexiblesusy.m}.  The content
of the header template file
\file{example\_constant\_observable.hpp.in} is the same as in general
\lstref{code:general-hpp-in}. In principle, we can fill the
\cpp definitions template file with the default content provided in
\lstref{code:general-cpp-in}. Nevertheless, as we do not use the \npf
module for the observable calculation (realized with \linesref{line:ex1-fs-cpp-start}{line:ex1-fs-cpp-end} in
\lstref{code:ex1-fs}), we can simplify the file as follows:
\begin{cppcode}{Content of \cpp definitions template \file{example\_constant\_observable.cpp.in}.}
#include "@ModelName@_@filename@.hpp"

namespace flexiblesusy::@namespace@ {
		
@calculate_definitions@
	
} // $\commnamespace$ flexiblesusy::@namespace@
\end{cppcode}
In this file, the template \cpp token
\cpptoken{@calculate\_definitions@} will be replaced with the content of the
variable \local{definitions} defined in the file
\file{\observable/\fs.m} described in 
\secref{sec:constant-flexiblesusy.m}. In this way, during the meta phase of \fs the 
 complete \cpp code will be generated for the function
\str{calculate\_example\_constant\_observable} (defined in Listing~\ref{code:ex1-obs}) which
outputs the number \str{num}.
We do not need to provide any additional observable-specific code, as
everything is already provided by the \code{WriteClass} function in
\file{\observable/\fs.m} during the \wl meta phase.

\subsubsection{Example 2: \exampletwoname}\label{sec:fermion-mass-cpp}

As stated above, we need to provide two \cpp template files in \dir{templates/observables}.  They will be filled during the
\fs meta phase and embedded into the rest of the \cpp spectrum
generator. In this example the content of the template header file
\file{example\_fermion\_mass.hpp.in} is the same as in the general
\lstref{code:general-hpp-in}.
The template file \file{example\_fermion\_mass.cpp.in}, however,
contains two changes compared to the generic example from
\lstref{code:general-cpp-in}. First of all, all commands related to
\npf can be omitted, since we do not use Feynman diagrammatic
calculations in this example (like in Example~1). Apart from this, our
goal with this example is to demonstrate how one can move as many
calculations as possible to the \cpp template file, while keeping the
flexibility to insert model-specific information. As discussed in
\secref{sec:ex2-fs}, we prepared for this by having a minimal function
body defined on the \wl level in the file
\file{ExampleFermionMass/FlexibleSUSY.m}, where the function body only
calls another \cpp function named \code{forge}. This function shall now
be defined. The following source code listing shows the content of the
\cpp template file that contains this definition:\footnote{%
  Instead of the \code{switch} statement one can also use
  \code{qedqcd.displayLeptonPoleMass(idx);}.}%
\begin{cppcode}{Content of \cpp definitions template \file{example\_fermion\_mass.cpp.in}.}
#include "@ModelName@_mass_eigenstates.hpp"
#include "cxx_qft/@ModelName@_qft.hpp"
#include "@ModelName@_@filename@.hpp"
#include "error.hpp"

namespace flexiblesusy {
using namespace @ModelName@_cxx_diagrams;
namespace @namespace@ {
		
template <typename RTYPE, typename FIELD>
auto forge(int idx, const @ModelName@_mass_eigenstates& model, const softsusy::QedQcd& qedqcd) 
{
	context_base context {model};
	auto context_mass = context.mass<FIELD>({idx});
			
	std::complex<double> lepton_mass;
	switch (idx) {
		case 0:  lepton_mass = qedqcd.displayPoleMel();
		         break;
		case 1:  lepton_mass = qedqcd.displayPoleMmuon();
		         break;
		case 2:  lepton_mass = qedqcd.displayPoleMtau();
		         break;
		default: throw OutOfBoundsError("fermion index out of bounds");
	}
	RTYPE res {context_mass, lepton_mass};
	return res;
}
		
@calculate_definitions@
		
} // $\commnamespace$ @namespace@
} // $\commnamespace$ flexiblesusy
\end{cppcode}
In the code listing above, the template \cpp token
\cpptoken{@calculate\_definitions@} will be replaced by the content of
the variable \local{definitions}, defined in the
\file{ExampleFermionMass/FlexibleSUSY.m} file in \secref{sec:ex2-fs},
which calls the \code{forge} function.  The \code{forge} function
defined above illustrates how to access two different types of
particle masses.  First, the running \gls{MSBAR}/\gls{DRBAR} mass of
the fermion specified by
\code{\color{color-cpp-template-arguments}FIELD} and \code{idx} (which
can correspond to, e.g.~\code{Fe[2]} or \code{Fd[3]} at the \wl level in the selected \gls{BSM} model)
is obtained by calling the \code{mass} function template. Afterwards,
the pole mass of the \gls{SM} lepton of generation \code{idx} is obtained from
the \code{qedqcd} object.
The \code{forge} function finally returns an array of length~2
containing these two masses.

\subsubsection{Example 3: \examplethreename}\label{sec:ex3-cpp}

As in the examples above, we have to create two \cpp template files
that are responsible for the numerical calculation of the
observable. This example uses \npf to generate analytical expressions
for the self energies, but it is otherwise a structurally simple
example. Hence, like in Example~1 (but unlike Example~2) we do not
delegate computations to a \code{forge} function; rather, the main
definition of the observable is done by the function \code{WriteClass}
in \file{\fs.m} created in \secref{sec:lepton-se-simple}.  For this
reason we use the \cpp template files given in
\lstref{code:general-hpp-in}--\ref{code:general-cpp-in} without any
changes in this example. Note that these general template files are
generally appropriate for observables that use the \npf extension,
hence we can keep the present subsection very short.

We recall that in this example we use \npf in the simple mode
specified by the option \code{Observable -> None}; the example will be
continued in \secref{sec:lepton-se-exe}, which shows how to enable and
use the computation of the self-energy in a concrete spectrum
generator. An alternative version of the example is provided in
\secref{sec:lepton-se-npf}, where the example is modified to
illustrate the use of the advanced settings of \npf.

\subsection{Content of the optional file \file{\observable/NPointFunctions.m}, advanced settings}\label{sec:advanced-npf-usage}

The \npf extension allows to generate Feynman diagrammatic
calculations to obtain analytical expressions for observables in any
specific model \model when the \fs meta phase is executed.  The simple
usage of \npf was demonstrated in \secref{sec:lepton-se-simple}, but
frequently it is necessary to fine-tune calculations done with
\npf. Such a fine-tuning is possible via advanced settings of \npf
explained in this section. It may involve the selection of subsets of
Feynman diagrams, remove contributions, selecting the regularization
scheme, or specifying the fermion order in e.g.\ four-fermion
amplitudes. Parts of the fine-tuning may be accessible to users of the
observable (such as the arguments \code{Vector}, \code{Scalars}, etc.\
of observables in \tabsref{tab_l3l-simple}{tab_llc-simple}), other
parts may be found only in the definition of the advanced settings.

In the following we begin by explaining how to enable the advanced
settings and giving an overview, then we will explain all advanced
settings and finally provide an example.

\begin{table}[t!]
\renewcommand*{\arraystretch}{1.2}
\setlength{\tabcolsep}{4pt}
\centering
\begin{tabularx}{\textwidth}{c>{\centering}p{3cm}X}
	\hline
	Option & Usage & \multicolumn{1}{c}{Purpose and hints}
	\\\hline
	\code{topologies} & See \tabref{tab_settings-topologies} & Select topologies based on adjacency matrices \\
	\code{diagrams} & See \tabref{tab_settings-diagrams} & Exclude contribution based on generic fields \\
	\code{amplitudes} & See \tabref{tab_settings-diagrams} & Remove contribution based on class insertions \\
	\code{chains} & See \tabref{tab_settings-chains} & Specify operators to neglect \\
	\code{order} & See \secref{sec:chains} & Select fermion order \\
	\code{sum} & See \secref{sec:other-modifications} & Define particle generations to skip on the \cpp level \\
	\code{regularization} & See \secref{sec:other-modifications} & Change scheme for a specified topologies \\
	\code{momenta} & See \secref{sec:other-modifications} & Eliminate given external momenta for a given topology \\
	\code{mass} & See \secref{sec:other-modifications} & Treat masses in an selected way
	\\\hline
\end{tabularx}%
\caption{Purpose of all available advanced settings and their usage.}
\label{tab_advanced-options}
\end{table}

\subsubsection{Enabling advanced settings, overview}
\label{sec:advancedoverview}

We begin by explaining how to enable and access the advanced settings
when setting up the definition of an
observable. The \npf extension is called from the observable-specific file
\file{\observable/FlexibleSUSY.m} discussed in
\secref{sec:content-fs-file}. \lstref{code:ex3-simple-fs} combined
with \lstref{code:general-fs} provides an example for using \npf in
its simple mode without advanced settings. To use the advanced mode, this file must be
modified as follows: 
\newpage 
{
\lstset{label={code:enableNPF},numbers=left, breaklines=false}
\begin{meta}{Modifications in \file{\observable/\fs.m} for the option \code{Observable -> \observable{}[]}.}
(* Instead of lines $\ref{line:ex3-simple-fs-observable}$ and $\ref{line:ex3-simple-fs-keep}$ in Listing $\ref{code:ex3-simple-fs}$ *)
NPointFunctions`Observable    -> obs[],
NPointFunctions`KeepProcesses -> If[Head[contr] === List, contr, {contr}]$\label{line:enableNPF-keep}$
\end{meta}
}
\noindent Here the line \code{Observable -> \observable{}[]} enables the advanced
settings (note that the variable \local{obs} was defined in the first line in Listing~\ref{code:general-fs}). This option requires the existence of the corresponding file
\code{meta/Observables/\observable/\npf.m} which should contain the
detailed advanced settings described further below.

The second line defines the option \code{KeepProcesses} via the
variable \local{contr} (instead of the hard-coded value
\code{{Irreducible}} in \lstref{code:ex3-simple-fs}) and opens up an
important way to access advanced settings. To explain the
meaning of this line we briefly recall the overall structure of
setting up an observable. From the user's perspective, an observable
ultimately is called as described in \secref{sec:all-obs}, and
observables may have options \local{contr}, see
e.g.~\tabsref{tab_l3l-simple}{tab_llc-simple}, with
possible values being a keyword or a list of keywords such as
\code{Vectors}, \code{Scalars}, \code{Boxes}, etc. 
Using these options influences how the observable is evaluated. 
The definition of such
keywords and how they influence the observable is part of the advanced
settings of \npf.

Using such a \local{contr} option has to be prepared in the file
\file{\observable/Observables.m} as exemplified in 
\secref{sec:ex3-observable.m} by specifying the appropriate argument
in the definition of the observable. Via \lineref{line:ex3-simple-fs-contr-definition} of
\lstref{code:ex3-simple-fs} and \lineref{line:enableNPF-keep} of \lstref{code:enableNPF} the variable
\local{contr} propagates into the option of \code{KeepProcesses}. The
construction in the code makes sure that the option is always a
list. 

The advanced settings are listed in \tabref{tab_advanced-options} and
explained in detail in the following \secsref{sec:option-topo}{sec:other-modifications}.
As an overview, \npf relies on \fa and \fc to produce \wl expressions for model-specific class-level amplitudes, and the advanced settings
fine-tune the calls to \fa and \fc routines, so familiarity with these tools is required to some extent.\footnote{%
Nevertheless, due to \cpp \code{template} capabilities, we use these packages in a non-standard way.
For example, the particle content of the model, vertices, and masses are stored as \cpp structures.
This allows the \cpp compiler to substitute all particle classes and sum over generation indices.
Thus, the amplitudes are modified before the usage of \fc to avoid particle-level computations.
In addition, all color factors for amplitudes are computed with \colormath.
}
Among the advanced settings, \code{topologies}, \code{diagrams} and
\code{amplitudes} can be used to define the keywords which can be
set by users as values of \local{contr} as explained above; these keywords then
influence 
 the execution of the functions \code{\context{FeynArts}CreateTopologies}, \code{\context{FeynArts}InsertFields}, and \code{\context{FormCalc}CalcFeynAmp}.
The further advanced settings influence the calls to \fa and \fc and
manipulate their output in ways which are fixed for the observable and
can be influenced only by directly modifying the file \file{\observable/\npf.m}.
 
\subsubsection{\code{topologies}}\label{sec:option-topo}

The observable calculation by the \npf extension starts from the
generation of topologies. It is possible to select particular
topologies in the calculation by using the option \code{KeepProcesses} in \lineref{line:enableNPF-keep} of \lstref{code:enableNPF} as explained above. 
The possible values of \local{contr} are (lists of) keywords: \code{Vectors}, \code{Scalars},
\code{TreeLevelSChannel}, etc.

\begin{figure}[t]
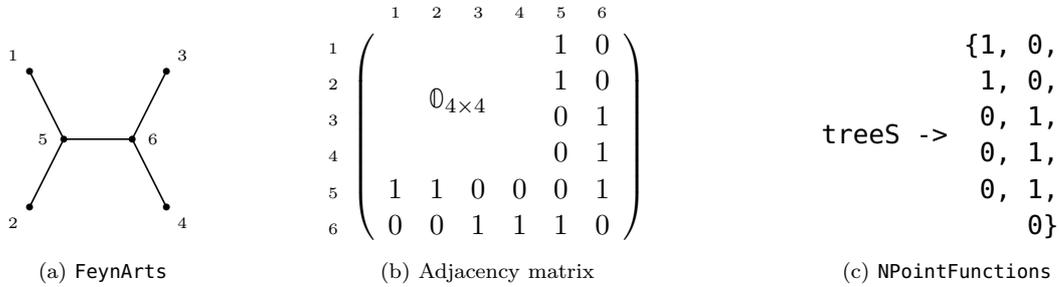

\include{fig_topology-explanation}
\caption{
		The connection between \fa topology, adjacency matrix, and \npf name \code{treeS}.
}
\label{fig_topology-explanation}
\end{figure}

In this section we focus on how to define such keywords, generally now
called \option[i]{Contribution}, and to associate them with topologies
of Feynman diagrams. 
\begin{table}[t]
\centering
\renewcommand*{\arraystretch}{1.2}
\begin{tabular}{>{\centering\arraybackslash}p{\textwidth-2\tabcolsep}}
	\hline
	Usage
	\\\hline
	\cellcolor{color-listings-background}
	\placeintable
	\begin{meta}{}
topologies[LoopNumber] = {
	Contribution$_\optionindex{i}$ -> TopologyName$_\optionindex{j}$,	
	...
};
	\end{meta}
	\\\hline
\end{tabular}
\begin{tabularx}{\textwidth}{>{\centering}p{3cm}>{\centering}p{2cm}X}
		\hline
		Abbreviation & Values & \multicolumn{1}{c}{Hints}
		\\\hline
		\option{LoopNumber} & \code{0} or \code{1} & \code{FormCalc} limitation
		\\
		\multirow{3}{*}{\option[i]{Contribution}} & 
		\multirow{3}{*}{$\forall$ \code{Symbol}} &%
Symbols for \gls{CLFV} observables: \local{contr} in \tabsref{tab_l3l-simple}{tab_llc-simple} \\
		& & Synonyms: in \file{\observable/FlexibleSUSY.m} \\ & & Connection to topologies: in \file{\observable/NPointFunctions.m}
		\\
		\option[j]{TopologyName} & $\forall$ \code{Symbol} & Definitions in \file{meta/\npf/Topologies.m}
		\\
		\hline
\end{tabularx}%

\caption{Syntax and options values for \code{topologies}.}
\label{tab_settings-topologies}
\end{table}
\tabref{tab_settings-topologies} shows the required command in the
file \file{\observable/NPointFunctions.m}. It connects each keyword
\option[i]{Contribution} with certain topology names.
The latter uniquely define topologies by connections to adjacency matrices, as shown in \figref{fig_topology-explanation}.
This explicit naming of topologies is useful, as they can also be used to
modify diagrams and amplitudes, as described in the following subsections.

The name of a topology can be any \wl symbol that has to be connected
with the actual adjacency matrix. All relevant topology names can be
defined in a separate file  \file{meta/\npf/Topologies.m}, as follows:
\begin{meta}{Example for definitions of topologies (in \file{meta/\npf/Topologies.m}).}
AllTopologies[{2, 2}] = {
	treeS   -> {1,0,1,0,0,1,0,1,0,1,0},
	treeT   -> {1,0,0,1,1,0,0,1,0,1,0},
	treeAll -> {treeS, treeT}
};
\end{meta}
Above, the \code{\{2, 2\}} argument represents the number of incoming and outgoing particles.
The definitions are stored in a list of key--value pairs of two kinds:
\begin{enumerate}
\item
	A connection between the user-given name of topology, e.g.~\code{treeS}, and \npf internal representation of this specific topology. It is obtained in a few steps, as follows from \figref{fig_topology-explanation}.
	One draws the \code{FeynArts} topology with explicit vertex numbers, creates an adjacency matrix, and eliminates all entries with redundant information: the zero top-left submatrix representing propagators between external particles, and entries below the diagonal as the adjacency matrix is symmetric. The rest is combined line by line into a one-dimensional list that is stored in \file{meta/\npf/Topologies.m}.
	These steps are done by the function \code{AdjaceTopology} defined there.
\item
	A connection between a set of simple topologies and their group name, like \code{treeAll}.
	This is useful to shorten the application of other settings.
\end{enumerate}
If the topology names are defined like this, an example content of
the file \file{\observable/NPointFunctions.m} might define the keyword
\code{TreeLevelSChannel} and 
associate it to the topology \code{treeS} as follows:
\begin{meta}{Example content of  \file{\observable/NPointFunctions.m}.}
topologies[0] = {
	TreeLevelSChannel -> treeS
};
\end{meta}

\subsubsection{\code{diagrams} (generic level modifications)}

The keywords \option[i]{Contribution} can also be associated with
generic-level field patterns (in \fa
nomenclature) to remove certain generic-level fields in
the Feynman diagrams.  
\tabref{tab_settings-diagrams} shows the required command in the
file \file{\observable/NPointFunctions.m} to define the associations
with field patterns.
\begin{table}[t]
\renewcommand*{\arraystretch}{1.2}
\setlength{\tabcolsep}{4pt}
\centering
\begin{tabular}{>{\centering\arraybackslash}p{\textwidth-2\tabcolsep}}
	\hline
	Usage
	\\\hline
	\cellcolor{color-listings-background}
	\placeintable
	\begin{meta}{}
diagrams[LoopNumber, WhenToApply] = { (* Or amplitudes[...] *)
	Contribution$_\optionindex{i}$ -> {
		TopologyName$_\optionindex{j}$ -> {Description$_\optionindex{a}$, Command$_\optionindex{b}$},
		...
	},
	...
};
	\end{meta}
	\\\hline
\end{tabular}
\begin{tabularx}{\textwidth}{>{\centering}p{2.8cm}>{\centering}p{3.5cm}X}
	\hline
	Abbreviation & Values & \multicolumn{1}{c}{Hints}
		\\\hline
		\option{LoopNumber} & \code{0} or \code{1} & As in \tabref{tab_settings-topologies}
		\\
		\option{WhenToApply} & \code{Present} or \code{Absent} & Apply \option[b]{Command} for \option[j]{TopologyName} if \option[i]{Contribution} is present (absent) in \code{KeepProcesses}, see \tabref{tab_simple-options}
		\\
		\option[i]{Contribution} & $\forall$ \code{Symbol} & As in \tabref{tab_settings-topologies}
		\\
		\option[j]{TopologyName} & $\forall$ \code{Symbol} & As in \tabref{tab_settings-topologies}
		\\
		\option[a]{Description} & $\forall$ \code{String} & Printed in the terminal during meta phase
		\\
		\option[b]{Command} & Special syntax & See the text description
		\\\hline
\end{tabularx}%
\caption{Syntax and options for \code{diagrams} and \code{amplitudes}.}
\label{tab_settings-diagrams}
\end{table}
The logic is the same as for \code{topologies}, but the main new
ingredient is the \option[b]{Command} option. It defines the kind of
generic fields that should be removed, and it should be a pure
  function which returns \code{True} or \code{False} and which takes
several arguments while traversing the tree of \fa
    insertions: \code{\context{FeynArts}TopologyList}, the current
    topology, and the current class level insertion. Apart from standard \wl expressions, the
following ingredients may be used to specify
\option[b]{Command}:\footnote{%
The syntax of \option[b]{Command} follows from the internal structure
of the diagram used by the \npf extension, and for more details we
refer to the file \file{meta/\npf/Settings.m} shipped with \fs.
}
\begin{itemize}
\item
	Fields appearing in tree-parts of diagrams are specified by \code{TreeFields}.
\item
	Fields appearing in loops are specified by \code{LoopFields}.
\item
	Any generic field of scalar type \code{\context{FeynArts}S}, fermion type \code{\context{FeynArts}F} or vector boson type \code{\context{FeynArts}V}.
\item
	A specific field, derived from external particles.
	For example, one can remove from the diagram fields that correspond to the external particles numbered 1 or 3 with the usage of the argument  \code{FieldPattern[\lsyntax{\#}, 1|3]} in the function \code{FreeQ} below.
\end{itemize}
Altogether, an example is given by the setting:
\begin{meta}{}
diagrams[1, Present] = {
	Vectors -> {penguinT -> {"no 1,3 particles", FreeQ[LoopFields[$\lsyntax{\#\#}$], FieldPattern[$\lsyntax{\#}$, 1|3]]$\lsyntax{\&}$}}
};
\end{meta}
Then,  the function call \code{NPointFunction[..., KeepProcesses -> \{Vectors\}]} deletes the diagrams with the topology \code{penguinT} when the external particles 1, 3 do appear inside the loop.
A second example,  demonstrating the \code{Absent} mode of
\option{WhenToApply}, is given by:
{\lstset{numbers=left}%
\begin{meta}{}
diagrams[1, Absent] = {
	Vectors -> {penguinT -> {"no tree vectors", FreeQ[TreeFields[$\lsyntax{\#\#}$], FeynArts`V]$\lsyntax{\&}$}},
	Scalars -> {penguinT -> {"no tree scalars", FreeQ[TreeFields[$\lsyntax{\#\#}$], FeynArts`S]$\lsyntax{\&}$}}
};
\end{meta}
}
\noindent
In this way, if \code{KeepProcesses -> \{Vectors\}} is specified (thus
\code{Scalars} is {\em not} specified), the third line of the listing
applies and eliminates diagrams of \code{penguinT} topology which
contain scalar particles in their tree-level parts.

\subsubsection{\code{amplitudes} (class level modifications)}

Often, modifications on the level of topologies and generic-level
amplitudes allowed by the \code{topologies} and \code{diagrams}
settings, together with other options of \tabref{tab_simple-options},
provide sufficient fine-tuning. 
Sometimes, however, removing amplitudes on the classes-level is
required. This is possible via specifying the
\code{amplitudes[\option{LoopNumber}, \option{WhenToApply}]}
setting. The syntax is identical to the one of \code{diagrams} and
given in  \tabref{tab_settings-diagrams}. As an example, amplitudes
with massless particles can require special treatment, and the
following setting
allows to remove them from the generated expressions (assuming they
will be  handled
properly elsewhere): 
\begin{meta}{}
amplitudes[1, Present] = {
	Vectors ->
		{penguinT -> {"no tree photons",	FreeQ[{$\lsyntax{\#\#}$}, InternalMass[FeynArts`V, 5] -> 0]$\lsyntax{\&}$}}
};
\end{meta}

\subsubsection{\code{order}, \code{chains} (Dirac algebra modifications)}\label{sec:chains}
\npf can be used to extract Wilson coefficients.
For observables with external fermions, there might exist a need to change or fine-tune fermion chains to achieve the desired operator structure.
This is done by the settings \code{order} and \code{chains}, where both the desired fermion order of external fermions as well as chains to be dropped are specified.

The syntax for \code{order} can be understood as follows.
Imagine that we would like to obtain Wilson coefficients corresponding
to the process $\mu^-\to e^-e^-e^+$, or more conveniently for the
crossed $2\to2$ process $\mu^-e^-\to e^-e^-$
instead, where the outgoing positron is replaced by an incoming
electron. That means we need to obtain coefficients of expressions
$\bar u(e^-)\Gamma_i u(\mu^-)\,\bar u(e^-)\Gamma_j u(e^-)$, where
$\Gamma$ represent structures involving Dirac matrices and momenta. 
For example, in the source code for the observable \code{BrLTo3L}, the \npf
extension is accordingly called as
\begin{meta}{}
NPointFunction[{lep, lep}, {lep, lep}, ...]
\end{meta}
where the variable \local{lep} will correspond to leptons during the
meta phase execution for concrete models. 
Interpreting the incoming particles as muon and electron, we specify
the required fermionic structure by the following line 
in the file \file{\npf.m}  (the field numbers are as in
\code{\context{FeynArts}InsertFields}): 
\begin{meta}{}
order[] = {3, 1, 4, 2};
\end{meta}

Once the fermionic order is fixed by \code{order}, and amplitudes are calculated,
one obtains multiple chains consisting of Dirac matrix products.
Often certain Dirac chains may be simplified or neglected, thanks to
the desired level of precision or the choice of basis 
of Wilson coefficients. 
Dropping specific types of Dirac chains can be implemented by using
the \code{chains} setting described in \tabref{tab_settings-chains}.
\begin{table}[t]
\renewcommand*{\arraystretch}{1.2}
\setlength{\tabcolsep}{4pt}
\centering
\begin{tabular}{>{\centering\arraybackslash}p{\textwidth-2\tabcolsep}}
	\hline
	Usage
	\\\hline
	\cellcolor{color-listings-background}
	\placeintable
	\begin{meta}{}
chains[LoopNumber] = {  
	Momenta$_\optionindex{i}$ -> {Rule$_\optionindex{j}$, ...}, 
	...
};
	\end{meta}
	\\\hline
\end{tabular}
\begin{tabularx}{\textwidth}{>{\centering}p{2.2cm}cX}
	\hline
	Abbreviation & Values & \multicolumn{1}{c}{Hints}
	\\\hline
	\option{LoopNumber} & \code{0} or \code{1} & As in \tabref{tab_settings-topologies}
	\\
	\option[i]{Momenta} & $\forall$ \code{Symbol} from \code{ZeroExternalMomenta} & See \tabref{tab_simple-options}
	\\
	\option[j]{Rule} & Special syntax & See \file{DiracChains.m}
	\\
	\hline
\end{tabularx}%
\caption{Syntax and options for \code{chains}. Files in column ``Hints'' are located in \dir{meta/\npf}.}\label{tab_settings-chains}
\end{table}
An example code which serves to 
neglect expressions proportional to the mass of the electron from the
example above is given by:
\begin{meta}{}
chains[1] = {
	ExceptLoops -> {1[k[4|2], ___] -> 0, 2[k[3|1], ___] -> 0}
};
\end{meta}

In general, the 
\option[j]{Rule} appearing on the right-hand side of this example (or
the general case in  \tabref{tab_settings-chains}) must be of the form
\begin{meta}{}
ChainNumber[Entry$_\optionindex{1}$, Entry$_\optionindex{2}$, ...] -> 0
\end{meta}%
The syntax can be described as follows (the source code which interprets these
expressions is in the file \file{meta/\npf/DiracChains.m}, and further
details can be found there):
\begin{description}
\item[\option{ChainNumber}] 
	is an integer which defines a chain number, e.g.~for the fermion order \code{\{3,1,4,2\}} the chain numbered \code{1} consists of particles \code{\{3,1\}} and chain numbered \code{2} is \code{\{4,2\}}.	
\item[{\option[1]{Entry}}] corresponds to a pattern for a Dirac chain
  which may be a non-commuting product of projection operators
  $P_{L,R}$, $\gamma^\mu$ matrices with open indices or $\gamma^\mu$
  matrices contracted with external momenta. We mimic \fc notation inside the Dirac chains so 
that the integer \code{6} represents $P_R$,  \code{7} is $P_L$,
a negative number leads to the antisymmetrized chain, \code{k[$i$]}
means $\gamma_\mu \code{k[$i$]}^\mu$ with $\code{k[$i$]}^\mu$ being
external momenta of $i$th particle. Further,  the short form
\code{l[$i$]} is converted to \code{Lor[$i$]} and represents a Lorentz
index connecting two Dirac chains. Accordingly, \option[1]{Entry}
  can take the following values: \code{6}, \code{-6}, \code{7},
  \code{-7}, \code{k[$i_1$]}, \code{l[$i_1$]} (\code{k} and \code{l}
  may have several integer arguments simultaneously such as
  \code{k[$i_1$|$i_2$|...]}, \code{l[$i_1$|$i_2$|...]}). 
In the case of \code{k[$i_1$|$i_2$|...]} and
\code{l[$i_1$|$i_2$|...]}, all projection operators are prepended 
automatically so that \code{2[k[3|4]]} becomes \code{2[6|-6|7|-7,
    k[3|4]]}; finally $i_j$ correspond to the integer numbers of
external particles. 
\item[{\option[2]{Entry}}] and further entries are similar but may
  take only the following values: \code{k[...]}, \code{l[...]},
  \code{\_}, \code{\_\_}, \code{\_\_\_}. 
\end{description}

\subsubsection{Other modifications}\label{sec:other-modifications}

Here we briefly describe and exemplify a set of settings which are
typically of minor 
importance.

In some cases, one should not sum over all generations for some generic field in the amplitude.
For example, in \gls{CLFV} observables, there often appear
penguin contributions with external self-energy-like corrections. In
such diagrams, the fermion in the tree-level propagator should differ from the external
fermion. This behavior can be defined via the setting \code{sum}:

\begin{meta}{}
sum[1] = {
	inSelfT -> {"skip initial lepton", {6, Field[$\lsyntax{\#}$, 1]$\lsyntax{\&}$}}
};
\end{meta}
This changes the expressions at the loop level \code{\option{LoopNumber} = 1} in the following way. For the topology identified as \code{inSelfT} one modifies the summation over generic fields in the propagator under \code{FeynArts} number \code{6}, so that the sum over the field being equal the first external particle is omitted on \cpp level.

The setting \code{momenta} can be used to eliminate certain external
momenta by using momentum conservation for a given topology,  for instance
\begin{meta}{}
momenta[1] = {
	penguinT -> 2
};
\end{meta}
This modifies the options for the function \code{\context{FormCalc}CalcFeynAmp} so that for \code{\option{LoopNumber} = 1} and topology \code{penguinT} the momenta of the second external particle is replaced by the momenta conservation expression.

The setting  \code{mass} allows to neglect masses, and improve the readability of the code, if desired:

\begin{meta}{}
mass[1] = {
	inSelfT -> {"explicit final lepton mass", InternalMass[FeynArts`F, 6] :> ExternalMass[3]},
	inSelfT -> {"keep initial lepton mass untouched", Hold :> ExternalMass[1]}
};
\end{meta}
The first example appends an additional replacement rule that allows the use of the explicit expression for the mass of the particle in the generic propagator. The second one prevents some simplification which would otherwise lead to incorrect amplitude expressions, see the explicit implementation for more details.

Finally, \code{regularization} overrides the option \code{\context{FormCalc}Dimension} for given topologies:
\begin{meta}{}
regularization[1] = {
	boxS -> D,
	boxU -> D
};
\end{meta}
This option might be useful to obtain the desired Wilson coefficients faster or more optimally: e.g.~sometimes the option \code{chains} might be skipped, as the default \fc setting has already produced the required expressions.

\subsubsection{Example~3: \examplethreename[(\code{Observable -> \observable{}[]})]}\label{sec:lepton-se-npf}

Let us exemplify how one enables the usage of the advanced settings
for \npf, by modifying Example~3 which calculates fermion self
energies. We continue from \secref{sec:lepton-se-simple} and
\secref{sec:ex3-cpp}, where the self-energy calculation was defined by
using only the simple mode of \npf. In order to use the advanced
settings, at first 
we need to apply the following modifications in the file \code{\fs.m}:
\begin{meta}{Modifications in \file{ExampleLeptonSE/\fs.m} for the option \code{Observable -> \observable{}[]}.}
NPointFunctions`Observable    -> obs[],
NPointFunctions`KeepProcesses -> If[Head[contr] === List, contr, {contr}]
...
name = "se_" <> CConversion`ToValidCSymbolString[contr];		
\end{meta}
Here the first lines are as explained in \secref{sec:advancedoverview}
and enable the advanced settings, including the variable \local{contr}
which allows users later to select different Feynman diagrams for the
evaluation of the self energies via
keywords corresponding to different \option[i]{Contribution}.
In the last line of the listing, the \local{name} of the function is
changed to reflect different possible values of the  
variable \local{contr}.

The goal in this example is to allow users to compute ``self
energies''  by  including either only one-particle irreducible diagrams, or
only diagrams with a tadpole part, or both. Hence we intend to use the
advanced settings to define
three keywords and associate them with the appropriate topologies via
the \code{topologies} setting described in
\tabref{tab_settings-topologies}. 

In	the following we describe a typical workflow how one might
interactively obtain the relevant information to create the
advanced settings
file \code{meta/Observables/\observable/\npf.m} which achieves
this. We begin with the definition of topologies we want to
enable/disable for the calculation. 

As the self-energy process is $1\to 1$, we start by looking for already defined topologies inside \file{meta/\npf/Topologies.m} in the form \code{AllTopologies[\{1, 1\}]}. 
Currently, this definition is missing, so its specification becomes our first task.
We can open a \code{Mathematica} notebook and evaluate the following code to figure out, which one-loop $1\to 1$ topologies exist in general and which ones are of interest to us:
{
\lstset{literate={`}{\textasciigrave}{1}}
\begin{lstlisting}[caption={Execute in a \code{Mathematica} notebook.}]
<<FeynArts`
(topologies = CreateTopologies[1, 1 -> 1]) // Paint
\end{lstlisting}
}
\begin{figure}[t!]
	\centering
\tikzset{node distance = 0.6cm}
	\begin{tikzpicture}
		\node (tadpole) 
		{\begin{fmffile}{example3-topologies-tadpole}
				\unitlength = 1mm
			\begin{fmfgraph*}(15, 15)
				\fmfpen{0.5}\fmfstraight\fmfset{arrow_len}{2.5mm}
				\fmfleft{v1}\fmfright{v2}\fmftop{t}
				\fmf{plain}{v1,v3}\fmf{plain}{v3,v2}
				\fmffreeze
				\fmf{plain,tension=1}{v3,v4}
				\fmf{plain,right,tension=0.2}{v4,t,v4}
				\fmfv{d.sh=circle,d.si=2pt,label=\tiny1,label.a=180, label.di=3}{v1}
				\fmfv{d.sh=circle,d.si=2pt,label=\tiny2,label.a=0, label.di=3}{v2}
				\fmfv{d.sh=circle,d.si=2pt,label=\tiny3,label.a=-90, label.di=3}{v3}
				\fmfv{d.sh=circle,d.si=2pt,label=\tiny4,label.a=90, label.di=3}{v4}
		\end{fmfgraph*}\end{fmffile}};
		\node[left=of tadpole] {T2:};
		\node[below=of tadpole] (sunset) 
		{\begin{fmffile}{example3-topologies-sunset}
				\unitlength = 1mm\begin{fmfgraph*}(15, 15)
				\fmfpen{0.5}\fmfstraight\fmfset{arrow_len}{2.5mm}
				\fmfleft{v1}\fmfright{v2}
				\fmf{plain,tension=1}{v1,v3}
				\fmf{plain,tension=1}{v4,v2}
				\fmf{plain,left,tension=0.3}{v3,v4,v3}
				\fmfv{d.sh=circle,d.si=2pt,label=\tiny1,label.a=180, label.di=3}{v1}
				\fmfv{d.sh=circle,d.si=2pt,label=\tiny2,label.a=0, label.di=3}{v2}
				\fmfv{d.sh=circle,d.si=2pt,label=\tiny3,label.a=0, label.di=3}{v3}
				\fmfv{d.sh=circle,d.si=2pt,label=\tiny4,label.a=180, label.di=3}{v4}
		\end{fmfgraph*}\end{fmffile}};
		\node[left=of sunset] {T3:};
		\node[right=1cm of tadpole, yshift=0.2cm] (tadadj) 
		{$\begin{array}{@{}c@{}c}
				& 
				\begin{array}{cccc}
					\helperintex{1}&
					\helperintex{2}&
					\helperintex{3}&
					\helperintex{4}\end{array} \\
				\begin{array}{cccc@{}}
					\text{\tiny1}\\
					\text{\tiny2}\\
					\text{\tiny3}\\
					\text{\tiny4} \end{array} &
				\left(
				\begin{array}{@{}c@{}c@{}}
					\mathbb 0_{2\times2} & \begin{array}{cc}
						1 & 0 \\
						1 & 0 \\
					\end{array} \\
					\begin{array}{cc}
						1 & 1 \\
						0 & 0
					\end{array} &
					\begin{array}{cc}
						0 & 1 \\
						1 & 1 \\
					\end{array}
				\end{array}
				\right)
			\end{array}
			$};
		\node[right=1cm of sunset, yshift=0.2cm] (sunadj) 
		{$\begin{array}{@{}c@{}c}
				& 
				\begin{array}{cccc}
					\helperintex{1}&
					\helperintex{2}&
					\helperintex{3}&
					\helperintex{4}\end{array} \\
				\begin{array}{cccc@{}}
					\text{\tiny1}\\
					\text{\tiny2}\\
					\text{\tiny3}\\
					\text{\tiny4} \end{array} &
				\left(
				\begin{array}{@{}c@{}c@{}}
					\mathbb 0_{2\times2} & \begin{array}{cc}
						1 & 0 \\
						0 & 1 \\
					\end{array} \\
					\begin{array}{cc}
						1 & 0 \\
						0 & 1
					\end{array} &
					\begin{array}{cc}
						0 & 2 \\
						2 & 0 \\
					\end{array}
				\end{array}
				\right)
			\end{array}
			$};
		\node[right=of tadadj, yshift=-0.2cm] {$\code{tadpole -> } 	
			\begin{array}{@{}c}
				\code{\{1, 0,} \\
				\code{\hphantom{\{}1, 0,} \\
				\code{\hphantom{\{}0, 1,} \\
				\code{\hphantom{\{0,} 1\}} \\
			\end{array}
			$};
		\node[right=of sunadj, yshift=-0.2cm] {$\code{\ sunset -> } 	
			\begin{array}{@{}c}
				\code{\{1, 0,} \\
				\code{\hphantom{\{}0, 1,} \\
				\code{\hphantom{\{}0, 2,} \\
				\code{\hphantom{\{0,} 0\}} \\
			\end{array}
			$};
	\end{tikzpicture}%
	\caption{Topologies, relevant for one-loop $1\to1$ processes, their \fa, mathematical and \npf representations, as in \figref{fig_topology-explanation}.}
	\label{fig_example3-topologies}
\end{figure}
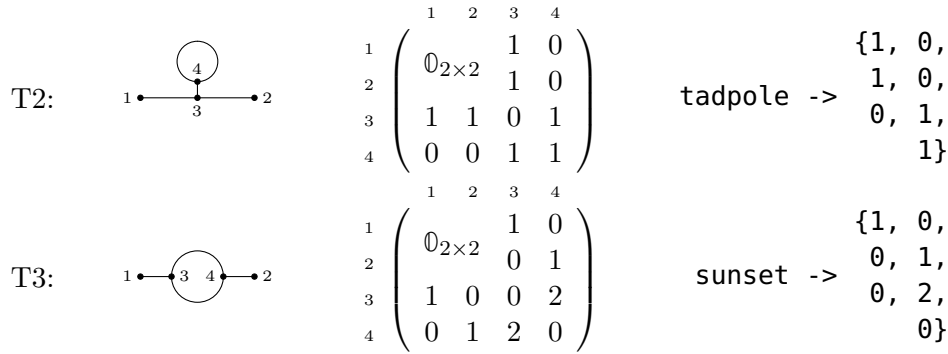

The printed output is similar to the content of the first column in \figref{fig_example3-topologies}. In this example, we are not interested in the first topology, while others represent what would like to include. 
We need to convert the \fa representation of the topology to the \npf one, this can be done as shown in \figref{fig_example3-topologies} or by execution of the code:\footnote{%
	The general definition of the function \code{AdjaceTopology} exists in \file{meta/\npf/Topologies.m}; one has to replace the specification of the pattern in the definition by \local{topology\_} and explicitly define \code{\local{external} = 2;} as this reflects the number of external vertices.
}
{
\lstset{literate={`}{\textasciigrave}{1}}
\begin{lstlisting}[caption={Execute in a \code{Mathematica} notebook.}]
AdjaceTopology[topologies[[2]]]
AdjaceTopology[topologies[[3]]]
\end{lstlisting}
}
\noindent Let us name the obtained topologies as in
\figref{fig_example3-topologies} and add their definition to the observable-independent file
\file{meta/\npf/Topologies.m} as:
\begin{meta}{In \file{meta/\npf/Topologies.m}.}
AllTopologies[{1, 1}] = {
	tadpole -> {1,0,1,0,0,1,1},
	sunset  -> {1,0,0,1,0,2,0},
	fermi   -> {tadpole, sunset}
};
\end{meta}
where the last line creates the synonym for both topologies combined.

Now we can finally define the content of the advanced settings file
for this observable:
\newpage 
\begin{meta}{Content of the file \file{ExampleLeptonSE/\npf.m}.}
topologies[1] = {
	Tadpoles -> tadpole,
	Sunsets  -> sunset,
	Fermi    -> fermi
};
\end{meta}
This achieves the goal. Now the observable \code{ExampleLeptonSE} has
an option \local{contr}, like all the examples in
\secref{sec:all-obs}. This option may be set to the values
\code{Tadpoles}, \code{Sunsets}, or \code{Fermi}. If some
model-specific configuration file 
\file{models/\model/\fs.m} contains the option \code{Tadpoles}, then
only the \code{tadpole} topology will be included, and similarly for
\code{Sunsets}. 
Once we specify \code{Fermi} (or \code{\{Tadpoles, Sunsets\}}), then
both topologies will be used to compute the (generalized) self-energy.
In this setup, \code{Sunsets} has the same effect as the
\code{Irreducible} setting from the simplified example. 

\subsection{Content of the optional file \file{\observable/FSMathLink.m}}

The spectrum generators created by \fs can be called from within a \wl
notebook or kernel. The necessary definitions to output the numerical
values of the generated observables at the \wl level are done in the
general file \file{meta/FSMathLink.m}. For a specific observable
$\observable$ one can create an optional file
\code{meta/Observables/\observable/FSMathLink.m} to specify the
desired interface via the function \code{PutObservable}. We refer the
reader to existing example files shipped with \fs for more details.

\subsection{New observables and how to calculate them with \fs}
\subsubsection*{General case}
To activate the \cpp code generation for a desired observable with \fs, one carries out the same steps as for predefined observables in \secref{sec:all-obs} (though now it is clear where all definitions come from).
For a user-selected model \model one modifies \file{models/\model/FlexibleSUSY.m} by changing \code{ExtraSLHAOutputBlocks}.
The pattern of observable \observable is defined inside the file called \file{meta/Observables/\observable/Observables.m} while the Les Houches blocks where this observable can be placed --- inside \file{meta/Observables/\observable/WriteOut.m}.
Finally, by execution of \code{make} one obtains the desired \cpp spectrum generator.

\subsubsection{Example 1: \exampleonename}\label{sec:ex1-final}
In summary,  to add the new observable corresponding to Example 1 to \fs, one needs to create several files:
\begin{enumerate}
\item 
	\file{\observable/Observables.m} with the \wl interface, the \cpp return type, and the \cpp prototype of the function that will be used to calculate the observable, see \secref{sec:constant-observable.m}.
\item 
	\file{\observable/WriteOut.m} with the connection of observable's return values and the \cpp spectrum generator output in Les Houches format, see \secref{sec:constant-writeout.m}.
\item 
	\file{\observable/\fs.m} which will fill the \cpp template files with the model-specific information, see \secref{sec:constant-flexiblesusy.m}.
\item 
	Two \cpp template files that will contain the \cpp code calculating the observable, see \secref{sec:constant-cpp}.
\end{enumerate}

After all this is done, the observable can be used in the same way as
the predefined observables discussed in \secref{sec:all-obs}, and we
need to proceed as described in that section.

We first need to choose a desired physical model
to perform the calculations (to continue this example we choose the \gls{SM}), create it via \code{./createmodel {-}{-}name=SM}, then configure \fs to make the \cpp spectrum generator for this model via \code{./configure {-}{-}with-models=SM}.

Then, two model-specific settings files have to be modified to enable the computation of the observable
\code{ExampleConstantObservable}.
To be specific, we modify the file with the meta-level model settings
\file{models/SM/\fs.m} to contain 
{\lstset{label={code:ex1-multiple-observables}}%
\begin{meta}{In \file{models/SM/\fs.m}.}
ExtraSLHAOutputBlocks = {
	{
		FlexibleSUSYLowEnergy,
		{
			{1, FlexibleSUSYObservable`ExampleConstantObservable[3]},
			{2, FlexibleSUSYObservable`ExampleConstantObservable[4]}
		}
	}
};
\end{meta}
}
\noindent In this way we specify that the observable is called twice, with two
different arguments (the arguments simply correspond to the numeric
constants which should be printed in the output), and that the output will be part of
the \code{Block FlexibleSUSYLowEnergy} with numbers 1 and 2,
respectively.
Finally, the calculation of all observables is enabled in the
runtime model-specific \gls{SLHA} input file: 
{
\lstset{backgroundcolor=\color{color-listings-background}}
\begin{lstlisting}[language=LesHouches,caption={In \file{models/SM/LesHouches.in.SM}.}]
Block FlexibleSUSY
	...
	15   1                    # calculate all observables
\end{lstlisting}
}
\noindent The execution of \code{make} command runs the meta phase and
compiles the final \cpp spectrum generator.
\fs provides shell scripts which merge these steps, as follows:
\begin{lstlisting}[language=sh, caption={Execute in terminal from the \fs directory.}]
./createmodel --name=SM
./configure --with-models=SM
./examples/new-observable/make-observable SM-example-1
make
./models/SM/run_SM.x --slha-input-file=models/SM/LesHouches.in.SM
\end{lstlisting}
to produce the following lines among the \gls{SLHA} output:
\begin{lstlisting}[language=LesHouches]
Block FlexibleSUSYLowEnergy Q= 1.73340000E+02
	1     3.00000000E+00   # exampleconstantobservable 3
	2     4.00000000E+00   # exampleconstantobservable 4
\end{lstlisting}
Here we see the appearance of the block numbers and the values of the
numeric constants in agreement with the definitions given above.

\subsubsection{Example 2: \exampletwoname}\label{sec:ex2-final}
In order to add the observable of Example~2 to \fs and	obtain a \gls{SM} spectrum generator including the output of this observable one needs to go through analogous steps. Again, \fs is shipped with a script which merges all steps:
\begin{lstlisting}[language=sh, caption={Execute in terminal from the \fs directory.}]
./createmodel --name=SM
./configure --with-models=SM
./examples/new-observable/make-observable SM-example-2
make
./models/SM/run_SM.x --slha-input-file=models/SM/LesHouches.in.SM
\end{lstlisting}
After executing these steps successfully, the \gls{SLHA} output will
contain the results corresponding to four instances of the observable
\code{ExampleLeptonMass} (where the fermion index 1 here means 2nd
generation due to the \cpp numbering convention):
\begin{lstlisting}[language=LesHouches]
Block FlexibleSUSYLowEnergy Q= 1.73340000E+02
	1     1.04187667E-01   # Fe[1] (lepton[1] if in Block ExampleLeptonMass) mass
	2     5.71258815E-02   # Fd[1] (lepton[1] if in Block ExampleLeptonMass) mass
Block ExampleLeptonMass Q= 1.73340000E+02
	1     1.05658357E-01   # Fe[1] (lepton[1] if in Block ExampleLeptonMass) mass
	2     1.05658357E-01   # Fd[1] (lepton[1] if in Block ExampleLeptonMass) mass
\end{lstlisting}
The script generates all the files discussed in the previous
Sections~\ref{sec:ex1-observables}, \ref{sec:ex2-writeout},
\ref{sec:ex2-fs}, and \ref{sec:fermion-mass-cpp}. Then it modifies the model-specific file \file{models/SM/\fs.m} as:
{\lstset{label={code:ex2-multiple-observables}}%
\begin{meta}{In \file{models/SM/\fs.m}.}
ExtraSLHAOutputBlocks = {
	{
		FlexibleSUSYLowEnergy,
		{
			{1, FlexibleSUSYObservable`ExampleFermionMass[Fe[2]]},
			{2, FlexibleSUSYObservable`ExampleFermionMass[Fd[2]]}
		}
	},
	{
		ExampleLeptonMass,
		{
			{1, FlexibleSUSYObservable`ExampleFermionMass[Fe[2]]},
			{2, FlexibleSUSYObservable`ExampleFermionMass[Fd[2]]}
		}
	}
};
\end{meta}
}
\noindent and assumes that the calculation of observables in the \gls{SLHA} input file should be already enabled:
{
\lstset{backgroundcolor=\color{color-listings-background}}
\begin{lstlisting}[language=LesHouches,caption={In \file{models/SM/LesHouches.in.SM}.}]
Block FlexibleSUSY
	...
	15   1                    # calculate all observables
\end{lstlisting}
}
Here the file \file{models/SM/\fs.m} defines the four instances of the
observable: they are distinguished by their argument (either
\code{Fe[2]} or \code{Fd[2]}) and by the Les Houches block (either
\code{FlexibleSUSYLowEnergy} or \code{ExampleLeptonMass}). We recall
that in the definition of the observable the running
\gls{MSBAR}/\gls{DRBAR} mass of the fermion in the \gls{BSM} model described
by the argument and the \gls{SM} lepton pole mass of the
specified generation (in
this case generation 2) are returned, see
\secref{sec:fermion-mass-cpp}. We also recall that the output depends
on the Les Houches block, see \secref{sec:ex2-writeout}. This explains
the output given above, where the \code{FlexibleSUSYLowEnergy} block
shows the \gls{MSBAR}/\gls{DRBAR} masses of the muon and the strange
quark in the BSM model, while the \code{ExampleLeptonMass} block shows
twice the muon pole mass from the \code{qedqcd} object.

\subsubsection{Example 3: \examplethreename}\label{sec:lepton-se-exe}
Technically, there are two versions of this example which differ by
\file{\observable/\fs.m}: using the simple mode of \npf  in
\secref{sec:lepton-se-simple} or using the advanced mode in \secref{sec:lepton-se-npf}.
To add the advanced version of the observable of Example~3 to \fs and obtain a \gls{SM} spectrum generator, one needs to go through steps analogous to previous examples. Again, \fs provides a script which combines all steps:
\begin{lstlisting}[language=sh, caption={Execute in terminal from the \fs directory.}]
./createmodel --name=SM
./configure --with-models=SM
./examples/new-observable/make-observable SM-example-3
make
./models/SM/run_SM.x --slha-input-file=models/SM/LesHouches.in.SM
\end{lstlisting}
to successfully see among the \gls{SLHA} output:
\begin{lstlisting}[language=LesHouches]
Block FWCOEF Q= 1.73340000E+02
	1111   31   00   2    -3.18762513E-05   # left
	1111   32   00   2    -3.18762513E-05   # right
\end{lstlisting}
Again, the script generates all files related to the definition of the
observable. Then it modifies the model-specific file
\file{models/SM/\fs.m} as 
\vspace{-0.5em} 
\begin{meta}{In \file{models/SM/\fs.m}.}
ExtraSLHAOutputBlocks = {
	{
		FWCOEF,
		{
			{1, FlexibleSUSYObservable`ExampleLeptonSE[Fe[1], {Sunsets}]}
		}
	}
};
\end{meta}
Note that the option \code{Sunsets} is used. It was defined via the
advanced settings of \npf in \secref{sec:lepton-se-npf}. It is further
assumed that the calculation of observables is enabled in the \gls{SLHA} input file:
\vspace{-0.5em} 
{
\lstset{backgroundcolor=\color{color-listings-background}}
\begin{lstlisting}[language=LesHouches,caption={In \file{models/SM/LesHouches.in.SM}.}]
Block FlexibleSUSY
	...
	15   1                    # calculate all observables
\end{lstlisting}
}

\section{Conclusions}
In this paper, we describe two novel essential features of \fs: a streamlined approach for incorporating new observables to \fs, and a simplified way to generate \cpp code for Feynman diagrams essential for the computation of these observables and other physical quantities.
To enhance the accessibility of this article  we have divided the content into two distinct parts.

The first part is concise, requires no detailed knowledge of the code
and is directed to readers interested  in the computation of
predefined \fs observables.
The currently predefined observables include \gls{CLFV} observables
such as \twodecays, \mec and \threedecays, whose physics definitions are
summarized in the appendix. Any user of \fs can now enable the
computation of these observables for any desired model by just adding
the appropriate flag in the model files.

The second part deals with the procedure of implementing new
observables. It is more extensive and technical and is of interest to
those seeking insights into the internal structure of the \npf code used for the
creation of new observables. In essence, the implementation of a new
observable requires five files (three \wl and two \cpp
template files) which essentially define in a model-independent way how the
observable is computed, how it is called and how its output is
organized.
The second part also illustrates the utilization of the \npf extension to streamline the generation of \cpp code for Feynman diagrams. 
This is achieved with the help of \fs-specialized wrappers designed for \fa, \fc, and \colormath packages.

Three fully worked out examples are provided. They correspond to a
minimal ``observable'' which simply outputs a single number, an
observable which outputs a set of particle masses, and an observable
which outputs the values of one-loop self energy diagrams. The code
snippets can be used as efficient starting points for future
implementations of further observables.
In the appendix additional features and details are discussed which may be
useful to accommodate special properties of new observables. They
correspond to actual use-cases motivated by implementing e.g.\ the
$b\to s\mu\mu$ or $h\to gg$ decays. 

The \fs extensions presented here have been thoroughly cross-checked
and used for concrete phenomenological applications in non-\gls{SUSY}
and in \gls{SUSY} models. Ref.\
\cite{Khasianevich:2023duu} uses and validates CLFV observables in
a leptoquark model,
where some observables arise at the one-loop level and some
observables arise at tree-level. Refs.~\cite{Dudenas:2022von,
  Dudenas:2022xnq} provide validations of loop-induced CLFV
observables in a model of neutrino masses where several neutrino
masses are themselves loop-induced. Finally, extensive validations
have been carried out in the context of a non-trivial \gls{SUSY}
realization by comparing with results of Ref.~\cite{Kotlarski:2019muo}
on CLFV phenomenology in the \gls{MRSSM}. These applications
demonstrate the reliability and
versatility of the code for observables defined via Feynman
diagrammatic calculations across a broad spectrum of models for wide variety of
observables. In the future, it is planned to add further observables to
the default distribution of \fs. In addition, users may add
observables individually. 
Finally, we introduce the possibility to request the implementation of
desired observables via
\href{https://github.com/FlexibleSUSY/FlexibleSUSY/issues}{github
  issues}, see the developer's repository.

\section*{Acknowledgements}
We thank the other \fs authors for helpful discussions. In particular, U.Kh.~thanks Jobst
Ziebell~\cite{Ziebell:2018th} and Kien Dang
Tran~\cite{DangTran:2019th} for the creation of an initial version of \npf.

We acknowledge support by the German Research Foundation (DFG) under grants STO 876/4 and STO 876/7.
W.K.~was supported by the National Science Centre (Poland) grant 2022/\allowbreak47/\allowbreak D/\allowbreak ST2/\allowbreak03087.
\appendix\gdef\thesection{\appendixname\Alph{section}}
\section{Available observables: physical details}\label{sec:physical-details}
In this section we discuss physics details of calculations for \gls{CLFV} observables implemented in \fs. 
Phenomenological applications connected to this section were presented
in Refs.~\cite{Kotlarski:2019muo, Dudenas:2022von, Dudenas:2022xnq,
  Khasianevich:2023duu}. In all expressions of this section, the masses are treated in the following way: in the amplitudes we use \gls{MSBAR}/\gls{DRBAR} values, everywhere else pole masses.

\subsection{Two-body CLFV decays (\twodecays)}\label{sec:two-decay-physics}
\begin{figure}[t!]
\input{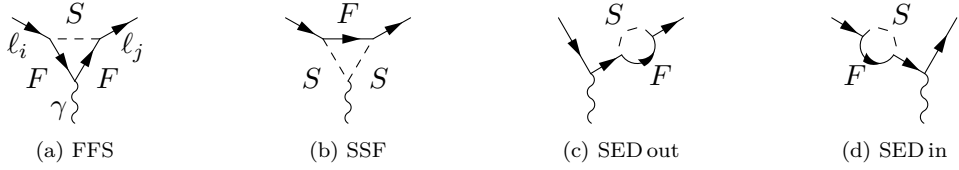}
\caption{
	Generic one-loop scalar-fermion Feynman diagrams contributing to \twodecays.
	Arrows show particle propagation.
	Diagrams of \gls{FFS} and \gls{SSF} types contribute to $A_2^X$, while
		all four are responsible for $A_1^X$.
}
\label{code:fig_generic-two-diagrams}
\end{figure}

The partial decay width is given by:
\begin{equation}\label{code:eq:two-decays-gamma}
	\Gamma_{\twodecays} = \frac{1}{4\pi}m_i^5\!
	\sum_{X=L,R}\norm{\coefficientEFT D X} ,
	\quad
	\coefficientEFT{D}{X} = -\frac{1}{2}A_2^X,
\end{equation}
where the minus sign corresponds to the outgoing photon momentum.
The form-factors $A_2^X$ are defined as
 (the contributions for both scalar and fermion insertions are shown in \figref{code:fig_generic-two-diagrams}):
\begin{equation}
	i\Gamma_{\bar \ell_j \ell_i \gamma}
	=
	i \bar{u}_{j}
	\Big[
	\left(q^2 \gamma^\mu
	-q^\mu\slashed{q}\right)\left(A_1^{L} P_L + A_1^{R}P_R
	\right)
	\\
	+im_{i}\sigma^{\mu\nu}q_\nu \left(
	A_2^{L}P_L+A_2^{R}P_R\right)
	\Big] u_i
	.
	\label{code:eq:2body-decay}
\end{equation}
It is convenient to uniquely separate contributions proportional to $f_\gamma$ and $s_\gamma$ ($f_\gamma = g^L_{[\bar F F \gamma]}/(32\pi^2 m_S^2)$ is proportional to fermion electric charge; $s_\gamma = g^L_{[S^*\!S\gamma]}/(32\pi^2 m_{S}^2)$ contains scalar electric charge):
\begin{align}\label{code:eq:generic-ffs}
	\begin{split}
		A_1^L &=
		- f_\gamma g^R_{[\bar F_j F S]} \hphantom{\Big(} g^L_{[\bar F F_i S^*]}\frac{F_D(x)}{18}
		,\\
		A_2^L &=
		- f_\gamma g^L_{[\bar F_j F S]}
		\left[
		g^R_{[\bar F F_i S^*]} \frac{F_E(x) }{12}
		+  g^L_{[\bar F F_i S^*]} \frac{m_{F}}{m_i} \frac{2F_F(x)}{3} \right]
		,
	\end{split}\\
  \label{code:eq:generic-ssf}
	\begin{split}
		A_1^L &= - s_\gamma g^R_{[\bar F_j F S]} \hphantom{\Big(}
		g^L_{[\bar F F_i S^*]}\frac{F_A(x)}{18}
		,\\
		A_2^L &= - s_\gamma g^L_{[\bar F_j F S]}
		\left[
		g^R_{[\bar F F_i S^*]}\frac{F_B(x)}{18}
		+
		g^L_{[\bar F F_i S^*]} \frac{m_F}{m_i} \frac{F_C(x)}{3}
		\right]
		.
	\end{split}
\end{align}
In both Eqs.~\eqref{code:eq:generic-ffs}--\eqref{code:eq:generic-ssf}
the argument of the loop functions is $x=m_F^2/m_S^2$ and the summation over repeated fields of model \model is assumed.
Also, the expressions for the right-handed amplitudes $A_{1,2}^R$ can be obtained from a substitution $[L\leftrightarrow R]$ in the generic couplings.
All $g^X_{[abc]}$ terms correspond to interaction vertices with removed imaginary prefactor.
Finally, $F_y(x)$ are loop functions, see the aforementioned file and Refs.~\cite{Hisano:1995cp, Ellis:2008zy}.

\subsection{Three-body CLFV decays (\threedecays)}\label{sec:three-decay-physics}
The partial decay width expressed in terms of Wilson coefficients has the form:
\begin{equation}\label{eq:gamma-mu3e}
	\begin{aligned}
		\Gamma_{\ell_i\to 3\ell_j}
		= \frac{m_i^5}{192\pi^3}\sum_{X=L,R}
		\Bigg\{&
		e^2 \norm{\coefficientEFT D X} \left(8 \ln\frac{m_i}{m_j} -11\right)
		+
		e \operatorname{Re}\left[\left(2\coefficientEFT[,j]{V}{XX} +\coefficientEFT[,j]{V}{X\bar X}\right)\coefficientEFT{D*}{\bar X}\right]
		\\
		&+
		\frac{1}{64}\left(
		\norm{\coefficientEFT[,j]{S}{XX}} + 16 \norm{\coefficientEFT[,j]{V}{XX}} + 8\norm{\coefficientEFT[,j]{V}{X \bar X}}
		\right)
		\Bigg\},
	\end{aligned}
\end{equation}
where $\bar X$ means complementary chirality, i.e.~$\bar L$ stands for $R$ (see the \code{width\_same} \cpp template in \file{templates/observables/br\_l\_to\_3l.cpp.in}).
In case of different final particles the expression becomes:
\begin{equation}
	\begin{alignedat}{4}\label{eq:three-different}
		\Gamma_{\threedecays}
		= \frac{m_i^5}{192\pi^3}\sum_{X=L,R}
		\!\Bigg\{
		&
		e^2 \norm{\coefficientEFT{D}{X}} \left(8\ln\frac{m_i}{m_k}-12\right)
		+
		e \operatorname{Re}\left[\left(\coefficientEFT[,k]{V}{XX}+\coefficientEFT[,k]{V}{X\bar X}\right)\coefficientEFT{D*}{\bar X}\right]
		\\&{+}
		\frac{1}{32} \left(\norm{\coefficientEFT[,k]{S}{XX}}{+}\norm{\coefficientEFT[,k]{S}{X\bar X}}\right)
		\!{+} \frac{1}{8}\left(\norm{\coefficientEFT[,k]{V}{XX}}{+}\norm{\coefficientEFT[,k]{V}{XX}}\right)
		\!{+} \frac{3}{2} \norm{\coefficientEFT[,k]{T}{XX}}
		\!\Bigg\},
	\end{alignedat}
\end{equation}
see the \code{width\_diff} \cpp template in \file{templates/observables/br\_l\_to\_3l.cpp.in}.

The Wilson coefficients are defined via the \gls{EFT} Lagrangian:
\begin{equation}\label{eq:low-lagrangian}
	\eftLagrangian=
	\Lagr_{\text{QED}}
	+
	\sum_{X,Y}
	\left(
	\coefficientEFT{D}{X}
	\operatorEFT{D}{X}
	+
	\sum_{I=\mathcal S,\mathcal V,\mathcal T}
	\sum_f
	\coefficientEFT[I,f]{}{XY}
	\operatorEFT[I,f]{}{XY}
	+\hc
	\right)
	,
	\quad
	\mathcal{D}_\mu=\partial_\mu + i e Q_f A_\mu.
\end{equation}
where the following set of operators is relevant for the considered \gls{CLFV} processes, see Refs.~\cite{Okada:1999zk, Crivellin:2017rmk, Kuno:1999jp}:%
\begin{equation}
	\begin{aligned}\label{eq:three-basis}
		\operatorEFT{D}{X} &= m_\mu\chain{\bar \ell_j\sigma^{\mu\nu}P_X\ell_i} F_{\mu\nu} , &
		\operatorEFT[,k]{S}{XY} &= \chain{\bar \ell_j P_X\ell_i}\chain{\bar\ell_k P_Y \ell_k} , &
		\\
		\operatorEFT[,k]{V}{XY} &= \chain{\bar\ell_j\gamma^\mu P_X\ell_i}\chain{\bar\ell_k \gamma_\mu P_Y \ell_k} , &
		\operatorEFT[,k]{T}{XX} &= \chain{\bar\ell_j\sigma^{\mu\nu} P_X\ell_i}\chain{\bar\ell_k\sigma_{\mu\nu} P_X \ell_k} ,
	\end{aligned}
\end{equation}
with the definitions of left-right chiral projectors $P_{L/R}=(1\mp\gamma_5)/2$; the square brackets emphasize different fermion chains with corresponding particles; \operatorEFT[,k]{T}{LR}, \operatorEFT[,k]{T}{RL} are zero.
The basis in Eq.~\eqref{eq:three-basis} is a redundant one:
scalar \operatorEFT[,k]{S}{LR}, \operatorEFT[,k]{S}{RL} and tensor \operatorEFT[,k]{T}{LL}, \operatorEFT[,k]{T}{RR} operators should be ignored because corresponding Wilson coefficients can be chosen to be zero.

Let us explain the amplitude matching procedure between \gls{EFT} and full \model-model Lagrangians specifically for the \meee process to simplify the notation and highlight the most essential details.
One needs to keep the same external fermion order for both theories.
It is enough to use the approximation of zero external momenta, and a convenient way to extract amplitudes is to consider the process $\mu^-_1 e^-_2\to e^-_3e^-_4$. Using the two Lagrangians \eftLagrangian and \fullLagrangian we get the following amplitudes (up to $\delta$-function and usual normalization factors; summation over repeated capital indices is assumed):
\begin{equation}\label{eq:mu3e-amplitude-matching}
  \bra{e^-_4e^-_3}T\exp\!\left(i\!\int\mathcal{L}\,\text{d}^4x\right)\!\ket{e^-_2\mu^-_1}
  =
  \begin{cases}
    -i\coefficientEFT[I,e]{}{XY}
    (\chain{\bar u_3 \Gamma^I_X u_1}\chain{\bar u_4\Gamma^I_Y u_2}
    - [3\leftrightarrow 4])
    &\text{for }\Lagr\!=\!\eftLagrangian,
    \\
    i \coefficientUV[I,e]{}{XY}
    \chain{\bar u_3 \Gamma^I_X u_1}\chain{\bar u_4\Gamma^I_Y u_2}
    &\text{for }\Lagr\!=\!\fullLagrangian.
  \end{cases}
\end{equation}
The first line contains the definition of the Wilson coefficients.
Note that there is no $[3\leftrightarrow4]$ for \coefficientUV[I,e]{}{XY} as we associate them with the results of \fc and the application of the setting \code{order} from \secref{sec:chains};
the derivation for coefficients  is explained further, right below the matching.

We can match \coefficientUV[I,e]{}{XY} onto low-energy coefficients \coefficientEFT[I,e]{}{XY}
by applying Fierz identities onto crossed $[3 \leftrightarrow 4]$ terms of
Eq.~\eqref{eq:mu3e-amplitude-matching}:
\begin{equation}\label{eq:mu3e-fierz-transformation}
  \begin{aligned}
    -\coefficientUV[,e]{S}{XX}
    &=\frac{1}{2}\coefficientEFT[,e]{S}{XX}-6\coefficientEFT[,e]{T}{XX},
    &
    -\coefficientUV[,e]{S}{X\bar X}&=\coefficientEFT[,e]{S}{X\bar X}-2\coefficientEFT[,e]{V}{X\bar X},
    \\
    -\coefficientUV[,e]{V}{XX}
    &=2\coefficientEFT[,e]{V}{XX},
    &
    -\coefficientUV[,e]{V}{X\bar X}&=\coefficientEFT[,e]{V}{X\bar X}-\frac{1}{2}\coefficientEFT[,e]{S}{X\bar X},
    \\
    -\coefficientUV[,e]{T}{XX}&=\frac{3}{2}\coefficientEFT[,e]{T}{XX}-\frac{1}{8}\coefficientEFT[,e]{S}{XX}.
  \end{aligned}
\end{equation}
To solve these equations one may set:
\begin{equation}\label{eq:m3e-coefficient-assertion}
	\coefficientEFT[,e]{S}{X\bar X} = 0, \qquad
	\coefficientEFT[,e]{T}{XX} = 0,
\end{equation}
then $\coefficientUV{S}{XX}/\coefficientUV{T}{XX}=-4$ and $\coefficientUV{S}{X\bar X}/\coefficientUV{V}{X\bar X}=-2$ should be fulfilled, which was checked numerically.

\begin{figure}[t!]
	\centering
\tikzset{node distance = 0.1cm}
\begin{tikzpicture}
\node (EQ) {$i\coefficientUV{}{}=$};
\node[right=of EQ] (inSED) 
{\begin{fmffile}{three-body-matching-inSED}\unitlength = 1mm\begin{fmfgraph*}(15, 15)
		\fmfpen{0.5}\fmfstraight\fmfset{arrow_len}{2.5mm}
		\fmfleft{v2,v1}\fmfright{v4,v3}
		\fmf{fermion}{v1,v5}
		\fmf{fermion}{v8,v7}
		\fmf{fermion,label=$e_3$,label.side=right,label.dist=2}{v7,v3}
		\fmfv{label=$e_2$,label.angle=70, label.dist=6}{v2}
		\fmf{fermion,label=,label.side=left,label.dist=3}{v2,v6}
		\fmf{fermion,label=$e_4$,label.side=left,label.dist=3}{v6,v4}
		\fmf{plain}{v6,v7}
		\fmf{plain,right,tension=0.01}{v5,v8}
		\fmf{plain,left}{v5,v8}
\end{fmfgraph*}\end{fmffile}};
\node[right=of inSED] (plus1) {$+$};
\node[right=of plus1] (outSED) 
{\begin{fmffile}{three-body-matching-outSED}\unitlength = 1mm\begin{fmfgraph*}(15, 15)
		\fmfpen{0.5}\fmfstraight\fmfset{arrow_len}{2.5mm}
		\fmfleft{v2,v1}\fmfright{v4,v3}
		\fmf{fermion}{v1,v5}
		\fmf{fermion}{v5,v8}
		\fmf{fermion}{v2,v6,v4}
		\fmf{fermion}{v7,v3}
		\fmf{plain}{v5,v6}
		\fmf{plain,right,tension=0.01}{v7,v8}
		\fmf{plain,left}{v7,v8}
\end{fmfgraph*}\end{fmffile}};
\node[right=of outSED] (plus2) {$+$};
\node[right=of plus2] (TRI) 
{\begin{fmffile}{three-body-matching-TRI}\unitlength = 1mm\begin{fmfgraph*}(15, 15)
		\fmfpen{0.5}\fmfstraight\fmfset{arrow_len}{2.5mm}
		\fmfleft{v2,v1}\fmfright{v4,v3}
		\fmf{fermion,tension=2}{v2,v6,v4}
		\fmf{fermion,tension=2}{v1,v5}
		\fmf{fermion,tension=2}{v7,v3}
		\fmf{plain,tension=2}{v6,v8}
		\fmf{plain}{v5,v7,v8,v5}
\end{fmfgraph*}\end{fmffile}};
\node[right=of TRI] (plus3) {$+$};
\node[right=of plus3] (UCH) 
{\rotatebox{45}{$u$-channel}};
\node[right=of UCH] (plus4) {$+$};
\node[right=of plus4] (SBOX) 
{\begin{fmffile}{three-body-matching-SBOX}\unitlength = 1mm\begin{fmfgraph*}(15, 15)
		\fmfpen{0.5}\fmfstraight\fmfset{arrow_len}{2.5mm}
		\fmfleft{v2,v1}\fmfright{v4,v3}
		\fmf{fermion,tension=3}{v1,v5}
		\fmf{fermion,tension=3}{v2,v6}
		\fmf{fermion,tension=3}{v7,v3}
		\fmf{fermion,tension=3}{v8,v4}
		\fmf{plain,tension=2}{v5,v6}
		\fmf{plain,tension=2}{v5,v7}
		\fmf{plain,tension=2}{v6,v8}
		\fmf{plain,tension=2}{v7,v8}
\end{fmfgraph*}\end{fmffile}};
\node[right=of SBOX] (plus5) {$+$};
\node[right=of plus5] {\rotatebox{45}{other boxes}};
\node[below=of inSED] {\code{inSelfT}};
\node[below=of outSED] {\code{outSelfT}};
\node[below=of TRI]{\code{triangleT}};
\node[below=of SBOX] {\code{boxS}};
\end{tikzpicture}%
	\caption{
		Calculation of four-lepton coefficients \coefficientUV{}{} in the full model \model.
		For self-energy and triangle topologies only $t$-channel is drawn.
		Decay \meee is considered so that $u$-channel for self-energy and penguin topologies should be added.
		All box channels are calculated directly.
		For convenience, \npf topology names are shown.
	}
	\label{fig_three-body-matching}
\end{figure}
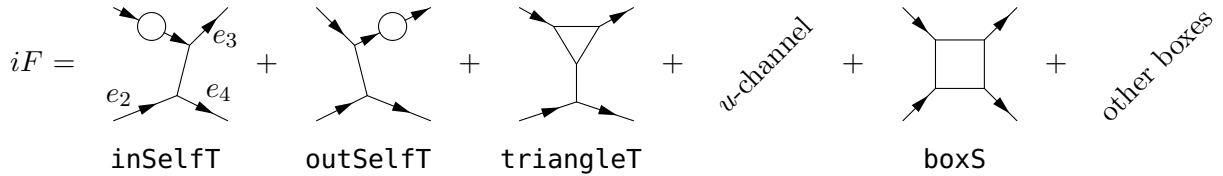

Now let us explain, how the \coefficientUV[I,e]{}{XY} are calculated in the full model \model via \npf.
As it is shown in \figref{fig_three-body-matching}, the \coefficientUV[I,e]{}{XY} coefficients take into account self-energy, penguin, and box diagrams.
If we consider non-box contributions (e.g.~for the \gls{MRSSM}~\cite{Kotlarski:2019muo} for convenience), then they can be categorized by the particle which is connected to the bottom chain of electrons to be the photon, $Z$- and Higgs-boson.
The photon contributes via dipole coefficients, and becomes a part of four-lepton vector operators as well, while the $Z$-boson contributes to the latter only, as follows from comparison with Ref.~\cite{Kotlarski:2019muo}:\footnote{%
This reference provides the calculation of the \gls{CLFV} processes in
the context of the \gls{MRSSM}, which represents a powerful way to
validate the code in a \gls{BSM} model for which no other public codes exist, in contrast to the \gls{MSSM}.  
}
\begin{equation}\label{eq:bosons-to-four-lepton}
		i\coefficientUV[,e]{V}{XY}\big|_{t\text{-channel for }\gamma} = -ieQ_eA_1^{X},
		\quad
		i\coefficientUV[,e]{V}{XY}\big|_{t\text{-channel for }Z} = - i\frac{g_Z^2}{m_Z^2}A_Z^X Z_e^Y,
\end{equation}
where the rhs.~incorporates an additional minus sign from embedding of $A^X_1$ coefficients into four-fermion amplitude.\footnote{%
	We use the left equation explicitly in the calculation of \threedecays with \fs. Instead of the right equation we use \npf directly and cross-checked it with unit test implemented in \fs.
}
Scalar and pseudoscalar Higgs bosons contribute purely to scalar
coefficients, and they are implemented and computed in all
models. 
Specifically for the
\gls{MRSSM}, they are much smaller than other diagrams,
\begin{equation}\label{eq:higgs-boson-to-four-fermion}
	\coefficientUV[,e]{S}{XY}\big|_{t\text{-channel for }h_i,A_i} \approx 0.
\end{equation}

Now, as one sees from \figref{fig_three-body-matching}, the expressions for crossed diagrams are required in addition to the mentioned $t$-channel self-energies and penguins. They can be calculated with the help of the same Fierz identities which were used for matching in Eq.~\eqref{eq:mu3e-fierz-transformation} because this channel differs only by the bispinor order.
One can take any equality from Eq.~\eqref{eq:mu3e-fierz-transformation} and remove the matching minus from the lhs.~and replace the \coefficientEFT[I,e]{}{XY} in the rhs.~by $t$-channel contributions of \coefficientUV[I,e]{}{XY}, e.g.~the first equality shown in Eq.~\eqref{eq:mu3e-fierz-transformation} transforms to the following one (see \code{fierz} \cpp template in \file{templates/observables/br\_l\_to\_3l.cpp.in}):
\begin{equation}\label{code:eq:example-tu-derivation}
	\begin{gathered}
		\coefficientUV[,e]{S}{XX}\big|_{t\text{- and }u\text{-channel}}
		=\frac{1}{2}\coefficientUV[,e]{S}{XX}\big|_{t\text{-channel}}
		-6\coefficientUV[,e]{T}{XX}\big|_{t\text{-channel}}.
	\end{gathered}
\end{equation}
Box diagrams contribute to all types of four-lepton coefficients.
Also, there is no sense to distinguish between different channels,
because we neglect external momenta so that all boxes directly
contribute to the complete \coefficientUV[I,e]{}{XY} coefficients. The
final contribution for \model being the \gls{MRSSM} can be written as
\begin{equation}\label{code:eq:three-adding boxes}
	\coefficientUV[I,e]{}{XY} = \coefficientUV[I,e]{}{XY}\big|_{t\text{- and }u\text{-channel for } \gamma,Z}+\coefficientUV[I,e]{}{XY}\big|_{\text{boxes}}.
\end{equation}
The numerical output of the code generated in this way for the
\gls{MRSSM} has been successfully validated against
Ref.~\cite{Kotlarski:2019muo}.

\subsection{Coherent conversion in nuclei (\mec)}\label{sec:mec-physics}
The conversion rate of the process is given by Refs.~\cite{Kotlarski:2019muo, Kitano:2002mt, Cirigliano:2009bz, Davidson:2017nrp, Davidson:2018kud}:

\begin{equation}\label{eq:gamma-conversion}
	\omega_{\mu-e}=4m_\mu^5\sum_{X=L,R}
	\norm{
		\frac{1}{4}D\coefficientEFT{D}{X}-\sum_{N=n,p}\left(S^{(N)}\cohFactor[,N]{S}{X}+V^{(N)}\cohFactor[,N]{V}{\bar X}\right)
	}
\end{equation}
with dimensionless integrals $D,S^{(N)},V^{(N)}$ defined in Ref.~\cite{Kitano:2002mt}; the minus sign reflects the different definitions of photon field, coming from a comparison of covariant derivatives (this observable is implemented in \file{templates/observables/l\_to\_l\_conversion.cpp.in}).

We use \gls{EFT} Lagrangian (with covariant derivative $\mathcal{D}_\mu=\partial_\mu + i e Q_f A_\mu$):
\begin{equation}\label{eq:mec-lagrangian}
	\eftLagrangian=
	\Lagr_{\text{QED}}
	+
	\Lagr_{\text{QCD}}
	+
	\sum_{X,Y}
	\left(
	\smashoperator[r]{\sum_{I=\mathcal D,\mathcal G}}
	\coefficientEFT{\mathnormal I}{X}
	\operatorEFT{\mathnormal{I}}{X}
	+
	\sum_{I=\mathcal S,\mathcal V,\mathcal T}
	\sum_f
	\coefficientEFT[I,f]{}{XY}
	\operatorEFT[I,f]{}{XY}
	+\hc
	\right)
\end{equation}
where the following set of operators is relevant, see Ref.~\cite{Crivellin:2017rmk}:%
\footnote{%
	Gluon dimension-7 operators \operatorEFT{G}{X} are important for \mec in general, see Ref.~\cite{Shifman:1978zn}.
	It was not necessary for our purposes to use $\muFermiFactor$ prefactor in the gluonic operator, as for considered models it was neglected, but we keep it to preserve the form of the operator in Ref.~\cite{Crivellin:2017rmk}.}%
\begin{equation}
	\begin{aligned}\label{eq:basis}
		\operatorEFT{D}{X} &= m_\mu\chain{\bar e\sigma^{\mu\nu}P_X\mu} F_{\mu\nu} ,
		&
		\operatorEFT{G}{X} &= \alpha_s\muFermiFactor\chain{\bar e P_X\mu}G_{\mu\nu}^aG^{a\mu\nu} ,
		\\
		\operatorEFT[,f]{S}{XY} &= \chain{\bar e P_X\mu}\chain{\bar{f} P_Yf} ,
		&
		\operatorEFT[,f]{V}{XY} &= \chain{\bar e\gamma^\mu P_X\mu}\chain{\bar{f}\gamma_\mu P_Yf} ,
		\\
		\operatorEFT[,f]{T}{XX} &= \chain{\bar e\sigma^{\mu\nu} P_X\mu}\chain{\bar{f}\sigma_{\mu\nu} P_Xf}
	\end{aligned}
\end{equation}
with minor changes:
\begin{equation}
	\begin{alignedat}{4}
		&\coefficientEFT[,q]{S}{X}
		&&=\frac{1}{2}
		(\coefficientEFT[,q]{S}{XL}+\coefficientEFT[,q]{S}{XR}),\qquad
		&&\operatorEFT[,q]{S}{X}
		&&=\chain{\bar e P_X\mu}\chain{\bar{q} q},
		\\
		&\coefficientEFT[,q]{V}{X}
		&&=\frac{1}{2}
		(\coefficientEFT[,q]{V}{XL}+\coefficientEFT[,q]{V}{XR}),\qquad
		&&\operatorEFT[,q]{V}{X}
		&&=\chain{\bar e\gamma^\mu P_X\mu}\chain{\bar{q}\gamma_\mu q},
		\\
		&\coefficientEFT[,q]{T}{X}
		&&=\coefficientEFT[,q]{T}{XX},\qquad
		&&\operatorEFT[,q]{T}{X}
		&&=\chain{\bar e\sigma^{\mu\nu} P_X\mu}\chain{\bar{q}\sigma_{\mu\nu} q}.
	\end{alignedat}
\end{equation}

\begin{figure}[t]
	\centering
\tikzset{node distance = 0.1cm}
	\begin{tikzpicture}
		\node (EQ) {$-i\coefficientEFT{}{}=$};
		\node[right=of EQ] (inSED) 
		{\begin{fmffile}{conversion-matching-inSED}\unitlength = 1mm\begin{fmfgraph*}(15, 15)
				\fmfpen{0.5}\fmfstraight\fmfset{arrow_len}{2.5mm}
				\fmfleft{v2,v1}\fmfright{v4,v3}
				\fmf{fermion}{v1,v5}
				\fmf{fermion}{v8,v7}
				\fmf{fermion,label=$e_3$,label.side=right,label.dist=2}{v7,v3}
				\fmf{fermion,label=,label.side=left,label.dist=3}{v2,v6}
				\fmfv{label=$q_2$,label.angle=70, label.dist=6}{v2}
				\fmf{fermion,label=$q_4$,label.side=left,label.dist=3}{v6,v4}
				\fmf{plain}{v6,v7}
				\fmf{plain,right,tension=0.01}{v5,v8}
				\fmf{plain,left}{v5,v8}
		\end{fmfgraph*}\end{fmffile}};
		\node[right=of inSED] (plus1) {$+$};
		\node[right=of plus1] (outSED) 
		{\begin{fmffile}{conversion-matching-outSED}\unitlength = 1mm\begin{fmfgraph*}(15, 15)
				\fmfpen{0.5}\fmfstraight\fmfset{arrow_len}{2.5mm}
				\fmfleft{v2,v1}\fmfright{v4,v3}
				\fmf{fermion}{v1,v5}
				\fmf{fermion}{v5,v8}
				\fmf{fermion}{v2,v6,v4}
				\fmf{fermion}{v7,v3}
				\fmf{plain}{v5,v6}
				\fmf{plain,right,tension=0.01}{v7,v8}
				\fmf{plain,left}{v7,v8}
		\end{fmfgraph*}\end{fmffile}};
		\node[right=of outSED] (plus2) {$+$};
		\node[right=of plus2] (TRI) 
		{\begin{fmffile}{conversion-matching-TRI}\unitlength = 1mm\begin{fmfgraph*}(15, 15)
				\fmfpen{0.5}\fmfstraight\fmfset{arrow_len}{2.5mm}
				\fmfleft{v2,v1}\fmfright{v4,v3}
				\fmf{fermion,tension=2}{v2,v6,v4}
				\fmf{fermion,tension=2}{v1,v5}
				\fmf{fermion,tension=2}{v7,v3}
				\fmf{plain,tension=2}{v6,v8}
				\fmf{plain}{v5,v7,v8,v5}
		\end{fmfgraph*}\end{fmffile}};
		\node[right=of TRI] (plus4) {$+$};
		\node[right=of plus4] (SBOX) 
		{\begin{fmffile}{conversion-matching-SBOX}\unitlength = 1mm\begin{fmfgraph*}(15, 15)
				\fmfpen{0.5}\fmfstraight\fmfset{arrow_len}{2.5mm}
				\fmfleft{v2,v1}\fmfright{v4,v3}
				\fmf{fermion,tension=3}{v1,v5}
				\fmf{fermion,tension=3}{v2,v6}
				\fmf{fermion,tension=3}{v7,v3}
				\fmf{fermion,tension=3}{v8,v4}
				\fmf{plain,tension=2}{v5,v6}
				\fmf{plain,tension=2}{v5,v7}
				\fmf{plain,tension=2}{v6,v8}
				\fmf{plain,tension=2}{v7,v8}
		\end{fmfgraph*}\end{fmffile}};
		\node[right=of SBOX] (plus5) {$+$};
		\node[right=of plus5] {\rotatebox{45}{other boxes}};
		\node[below=of inSED] {\code{inSelfT}};
		\node[below=of outSED] {\code{outSelfT}};
		\node[below=of TRI]{\code{triangleT}};
		\node[below=of SBOX] {\code{boxS}};
	\end{tikzpicture}%
	\caption{
		Amplitude matching $\eftLagrangian\leftrightarrow\fullLagrangian$ for quark coefficients.
		\mec is considered so that all existing self-energy and penguin topologies are in $t$-channel only and are shown explicitly.
		All channels for boxes are calculated directly.
		The sum over all represented diagrams can be matched to the corresponding \coefficientEFT[q]{}{X} coefficients without intermediate steps.
		For convenience, \npf topology names are shown.
	}
	\label{fig_t-matching-issues}
\end{figure}

In contrast to $\mu\to3e$, there are no $u$-channel penguin and self-energy diagrams and no coefficient vanishes due to Fierz identities.
So that $\eftLagrangian\leftrightarrow\fullLagrangian$ matching can be done directly by the consultation with \figref{fig_t-matching-issues}.
Let us again compare with the \gls{MRSSM} results of Ref.~\cite{Kotlarski:2019muo} for concreteness.
For four-fermion coefficients one, in general, adds self-energy, penguin, and box diagrams:
\begin{equation}\label{eq:bosons-to-four-fermion}
		-i\coefficientEFT[,q]{V}{X}\big|_\gamma = -ieQ_qA_1^{X},
		\quad
		-i\coefficientEFT[,q]{V}{X}\big|_Z = - i\frac{g_Z^2}{2m_Z^2}A_Z^X (Z_q^L+Z_q^R),
		\quad
		\coefficientEFT[,q]{S}{X}\big|_{h_i,A_i} \approx 0,
\end{equation}
where we justify the lhs.~by stressing that diagrams of \figref{fig_t-matching-issues} represent amplitudes. Note the additional minus sign from photon embedding, which corresponds to the chosen order of external fermions in $\mu^-_1 q^-_2\to e^-_3q^-_4$. Box diagrams contribute to all types of Wilson coefficients, so that:
\begin{equation}\label{four-fermion-matching}
	\coefficientEFT[I,q]{}{X} =
	\coefficientEFT[I,q]{}{X}\big|_{\gamma}+
	\coefficientEFT[I,q]{}{X}\big|_{Z}+\coefficientEFT[I,q]{}{X}\big|_{\text{boxes}}.
\end{equation}

Gluonic coefficients might be relevant in general, and are changed when heavy quarks $c,b$ are integrated out, as stated in Ref.~\cite{Shifman:1978zn}:
\begin{equation}\label{eq:shifman_change}
m_q\bar q q\to -\frac{\alpha_s}{12\pi}G_{\mu\nu}^aG^{a\mu\nu}
,\quad
\coefficientEFT{G}{X}\to
\gluonNew{G}{X}
=
\coefficientEFT{G}{X}-\frac{1}{12\pi}\sum_{q=c,b}\frac{1}{m_q\muFermiFactor}\coefficientEFT[,q]{S}{X}.
\end{equation}

To match \eftLagrangian onto \Lnucl, on-mass-shell condition is used for the same initial and final nucleon state~$\ket{N}$, see Refs.~\cite{Cirigliano:2009bz, Crivellin:2017rmk, Kosmas:2001mv, Kuno:1999jp, Kitano:2002mt, Kosmas:2001ia, Cirigliano:2017azj, Cirelli:2013ufw,
	Davidson:2017nrp,Davidson:2018kud}.
This is realized for a nucleon $N$, quark $q$, and coefficient type $I$ with a help of \nuclFactor[I,q]{}{N} form factors:
\begin{equation}
	\bra{N}\Lagr_{\text{EFT}}\ket{N} \approx \bra{N}\Lnucl\ket{N}
	\to%
	\begin{array}{c}
		\chain{\bar q \Gamma_I q} = \hspace{-2mm} \displaystyle\smashoperator[r]{\sum_{N=p,n}}\nuclFactor[I,q]{}{N}\chain{\bar N\Gamma_IN},
		\\
		\alpha_sG^a_{\mu\nu}G^{a\mu\nu} = \hspace{-2mm} \displaystyle\smashoperator[r]{\sum_{N=p,n}}m_N\nuclFactor{G}{N} \chain{\bar NN},
	\end{array}
\end{equation}
where gluonic form factors are obtained using a trace of the energy-momentum tensor, while others can be found in \tabref{tab_conversion-form-factors}:
\begin{equation}
	\nuclFactor{G}{N}
	=
	-\frac{8\pi}{9}\left(1-\sum_{q=u,d,s}\frac{m_q}{m_N}\nuclFactor[,q]{S}{N}\right).
\end{equation}

\begin{table}[t]
	\centering
\renewcommand*{\arraystretch}{1.5}
\begin{tabular}{|c | c |c |}
	\hline\rule{0pt}{2.6ex} 
	$\nuclFactor[,u]{S}{p}=0.021(2)\tfrac{m_p}{m_u}$
	& $\nuclFactor[,d]{S}{p}=0.041(3)\tfrac{m_p}{m_d}$
	& $\nuclFactor[,s]{S}{p}=0.043(11)\tfrac{m_p}{m_s}$
	\\\hline
	$\nuclFactor[,u]{S}{n}=0.019(2)\tfrac{m_n}{m_u}$
	& $\nuclFactor[,d]{S}{n}=0.045(3)\tfrac{m_n}{m_d}$
	& $\nuclFactor[,s]{S}{n}=0.043(11)\tfrac{m_n}{m_s}$
	\\\hline
	$\nuclFactor[,u]{V}{p}=\nuclFactor[,d]{V}{n}=2$
	& $\nuclFactor[,d]{V}{p}=\nuclFactor[,u]{V}{n}=1$
	& $\nuclFactor[,s]{V}{p}=\nuclFactor[,s]{V}{n}=0$
	\\\hline
	$\nuclFactor[\!,u]{T}{p}=\nuclFactor[\!,d]{T}{n}=0.77(7)$ 
	& $\nuclFactor[\!,d]{T}{p}=\nuclFactor[\!,u]{T}{n}=-0.23(3)$
	& $\nuclFactor[\!,s]{T}{p}=\nuclFactor[\!,s]{T}{n}=0.008(9)$
	\\\hline
\end{tabular}%
	\caption{
		Matching (\Lnucl$\leftrightarrow\Lagr_{\text{EFT}}$) form factors, see \ref{sec:extra-mec-input}. Scalar form factors measure the contribution of the quark condensate to the mass of nucleon. They are determined from pion-nucleon $\sigma_{\pi N}$ term for $u,d$-quarks \cite{Crivellin:2013ipa, Hoferichter:2015dsa} and from the lattice calculation for $s$-quark.
		Vector form factors do not suffer from theoretical uncertainty during the matching and are derived from the conservation of vector current (counting of valence quarks).
		For tensor form factors the values calculated in lattice \gls{QCD} at $2~\text{GeV}$ \cite{Bhattacharya:2015esa, Cirigliano:2017azj} are taken.
		See both \file{src/observables/l\_to\_l\_conversion/settings.cpp} used for the storage of form factors and \file{templates/observables/l\_to\_l\_conversion.cpp.in} for observable implementation.
	}
	\label{tab_conversion-form-factors}
\end{table}

At finite recoil, tensor operator contributes to scalar ones \cite{Cirigliano:2017azj,Cirelli:2013ufw}, which is expressed via the replacement:
\begin{equation}\label{eq:mec-tensor-contribution}
	\chain{\bar e\sigma^{\mu\nu}P_X\mu}\chain{\bar N\sigma_{\mu\nu}N}
	\rightarrow
	\tensorRecoil\frac{m_\mu}{m_N}
	\chain{\bar eP_X\mu}\chain{\bar NN}.
\end{equation}

The relevant for $\mu-e$ conversion Lagrangian takes the following form:
\begin{equation}
	\Lnucl =
	\sum_{X=L,R}
	\left[
	\coefficientEFT{D}{X}\operatorEFT{D}{X}
	+
	\sum_{N=n,p}
	\left(
	\cohFactor[,N]{S}{X}\operatorEFT[,N]{S}{X}
	+
	\cohFactor[,N]{V}{X}\operatorEFT[,N]{V}{X}
	\right)
	+\hc
	\right],
\end{equation}
where new coefficients and operators are introduced (\nuclFactor{V}{N,s} are zero and may be omitted):
\begin{equation}
	\begin{gathered}
		\cohFactor[,N]{S}{X}=
		m_N\muFermiFactor\nuclFactor{G}{N}\gluonNew{G}{X}
		+
		\sum_{q=u,d,s}
		\left(\nuclFactor[,q]{S}{N}\coefficientEFT[,q]{S}{X}
		+\tensorRecoil\frac{m_\mu}{m_N}\nuclFactor[\!,q]{T}{N}\coefficientEFT[\!,q]{T}{X}
		\right),
		\quad
		\operatorEFT[,N]{S}{X}=\chain{\bar eP_X\mu}\chain{\bar NN},
		\\
		\cohFactor[,N]{V}{X}=
		\sum_{q=u,d,s}\nuclFactor[,q]{V}{N}\coefficientEFT[,q]{V}{X},
		\quad
		\operatorEFT[,N]{S}{X}=\chain{\bar e\gamma^\mu P_X\mu}\chain{\bar N\gamma_\mu N}.
	\end{gathered}
\end{equation}

\section{Additional features and examples}\label{sec:additional-features}
In this section we show several advanced ways to use
\fs in order to obtain more freedom in both input and output formats, as well as to perform non-conventional intermediate calculations.
Though it is currently
possible to use these features in the way described below, they might be automatized in future.
\subsection{Example 4: post-processing in $h\to gg$}
Let us demonstrate a possible way to do post-processing when additional user input is required after \npf calculations.
When all external particles are fermions, one can consult the implementation of \threedecays and \mec. In this example, we consider external bosons which implies several changes to aforementioned observables.

The code template of this section can be integrated into \fs by the execution of the following command:
\begin{lstlisting}[language=sh, caption={}]
./examples/new-observable/make-observable example-4
\end{lstlisting}
The observable defined in this example calculates amplitudes required for the $h\to gg$ process (but not the branching ratio itself) with the help of \npf.
After \npf calculations are done, one is left with amplitudes containing many abbreviations.
From physics we know that the computed amplitudes must contain several
structures of covariants. For $h\to gg$, the possible covariants are
\begin{equation}
\epsilon_2^\mu \epsilon^{\vphantom{\mu}}_{3\mu}
,\quad
\epsilon_2^\mu p^{\vphantom{\mu}}_{3\mu}
,\quad
\epsilon_3^\mu p^{\vphantom{\mu}}_{2\mu}
,\quad
\varepsilon^{\alpha\beta\mu\nu} \epsilon^{\vphantom{\mu}}_{2\alpha} \epsilon^{\vphantom{\mu}}_{3\beta} p^{\vphantom{\mu}}_{2\mu} p^{\vphantom{\mu}}_{3\nu},
\end{equation}
where $\epsilon_i$ is the polarization vector of the corresponding
gluon and $\varepsilon$ is the Levi-Civita tensor.

It is the coefficients of these structures which are relevant for the calculation of the branching ratio. 
Hence we need to extract the prefactors of these structures. 
This can be performed as shown in \lstref{code:gg}. There, we apply all remaining sub-expressions in \lineref{line:gg-apply}, then abbreviate all basis structures in \linesref{line:gg-abbrev-start}{line:gg-abbrev-end}, and extract the coefficients that come as prefactors of the second argument of \code{InterfaceToMatching} in line~\ref{line:gg-matching-interface} (similarly to line~\ref{line:ex3-simple-fs-interface} in Listing~\ref{code:ex3-simple-fs}).  
Finally, the \cpp code definitions are generated by the function \code{NPFDefinitions} in \lineref{line:gg-cpp-generation} (similarly to \lineref{line:ex3-simple-fs-create} in \lstref{code:ex3-simple-fs}; note, that \code{NPFDefinitions} may accept strings that are different to the last argument of \code{InterfaceToMatching}):
{%
\lstset{numbers=left,label={code:gg}}
\begin{meta}{Content of \file{HiggsTo2Gluons/\fs.m}.}
npf = NPointFunctions`ApplySubexpressions[npf];$\label{line:gg-apply}$
npf = npf /. {$\label{line:gg-abbrev-start}$
	SARAH`sum[__, FormCalc`ec[2, l_] FormCalc`ec[3, l_] SARAH`g[__]] :>         "e2e3",
	SARAH`sum[__, FormCalc`ec[2, l_] SARAH`Mom[3, l_] SARAH`g[__]]   :>         "e2m3",
	SARAH`sum[__, FormCalc`ec[3, l_] SARAH`Mom[2, l_] SARAH`g[__]]   :>         "e3m2",
	FormCalc`Eps[FormCalc`ec[2], FormCalc`ec[3], SARAH`Mom[2], SARAH`Mom[3]] :> "eps",
	SARAH`sum[__, SARAH`Mom[2, l_] SARAH`Mom[3, l_] SARAH`g[__]] :> SARAH`Mass[higgs]^2/2
};$\label{line:gg-abbrev-end}$
npf = WilsonCoeffs`InterfaceToMatching[npf, {"eps", "e2e3", "e2m3" "e3m2"}];$\label{line:gg-matching-interface}$
	
...
	
AppendTo[npfDefinitions,
	NPointFunctions`NPFDefinitions[npf, "cpp_name", SARAH`Delta, {"eps", "e2e3", "e2m3_e3m2"}]$\label{line:gg-cpp-generation}$
];
\end{meta}
}
\noindent
In principle, one can apply the routines from the code snippet above for the processes with external fermions as well. This might be relevant, in particular, if one prefers to ignore the setting \code{chains} from \secref{sec:chains} and deal with Dirac chains in some other ways.

\subsection{Example 5: \code{flavio} output in $b\to s\mu\mu$}
One might want to generate the output of Wilson coefficients to be used later via other programs, like \code{flavio}~\cite{straub2018flavio} that requires a \file{.json} file with  specific content.
The code template of this section can be integrated into \fs by the execution of the following command:
\begin{lstlisting}[language=sh, caption={}]
./examples/new-observable/make-observable example-5
\end{lstlisting}
Currently, the way to fill a \file{.json} file can be demonstrated by the operators required for $b\to s\mu\mu$ using \gls{JSON} library \cite{nlohmann} distributed with \fs:
\newpage 
\begin{cppcode}{Content of \file{templates/observables/br\_d\_l\_to\_d\_l.cpp.in}.}
#include <fstream>
#include <iomanip>
#include "json.hpp"
// Definition of C9_bsmumu and other Wilson coefficients
nlohmann::json j;
j["eft"]    = "WET";
j["basis"]  = "flavio";
j["scale"]  = dynamic_cast<@ModelName@_mass_eigenstates const&>(context.model).get_scale();
j["values"] = {
	{"C9_bsmumu",   {{"Re", Re(C9_bsmumu)},   {"Im", Im(C9_bsmumu)}}},
	// Other Wilson coefficients
};
std::ofstream wc_json("WET_bsmumu.json");
wc_json << std::setw(4) << j << std::endl;
wc_json.close();
\end{cppcode}

The code from the listing above will generate the file \file{WET\_bsmumu.json} filled with the numerical values obtained by \fs and, in principle, ready to be used for the program \code{flavio}.
It is clear, that one is generally required to calculate the complete set of operators closed under the \gls{RGE} running and only then use it for phenomenological studies.
In particular for this example, one is expected currently to generate several \file{.json} files and appropriately merge them before the execution of \code{flavio}, which is planned to be improved.

\subsection{Example 6: additional LH input blocks in \mec}\label{sec:extra-mec-input}
The process of \mec depends on several parameters, see \tabref{tab_conversion-form-factors}, and contains multiple factors that might not be relevant for a given model, see \ref{sec:mec-physics}.
There is a way to modify these settings during the runtime of \cpp
spectrum generator via an additional Les Houches input block that is automatically added if \fs is configured to calculate \mec.

Let us provide more information about how this functionality works and can be used for other observables as well if they require additional configuration options.

First of all, multiple changes within \cpp template files should be made in \fs globally. 
This is realized in the function \code{WriteClass} via its additional
output value, called \str{"C++ replacements"}, see
\secref{sec:content-fs-file} and the \file{LToLConversion/\fs.m} file.

On the \cpp side, one defines files to store new values in the
\dir{src/obsevables/\fntoken} directory (see the definition for
\code{descriptions} and their default values in the function
\code{reset}) and lets \fs know about the new Les Houches block name
in \file{src/slha\_io.*} files (in this case, the name is
\str{"LToLConversion"}). 

For example, if we want to enable the tensor contributions from Eq.~\eqref{eq:mec-tensor-contribution}, then we need to include the following line into the input Les Houches file:
{
\lstset{backgroundcolor=\color{color-listings-background}}
\begin{lstlisting}[language=LesHouches,caption={In \file{models/\model/LesHouches.in.\model}, if \fs is configured to calculate \mec.}]
Block LToLConversion
    0   1   # include tensor contribution
\end{lstlisting}
}
\noindent Similarly, one can change numerical values of all coefficients from \tabref{tab_conversion-form-factors}.
\bibliographystyle{elsarticle-num}
\biboptions{sort&compress}
\bibliography{bibliography}

\begin{thebibliography}{10}
\expandafter\ifx\csname url\endcsname\relax
  \def\url#1{\texttt{#1}}\fi
\expandafter\ifx\csname urlprefix\endcsname\relax\def\urlprefix{URL }\fi
\expandafter\ifx\csname href\endcsname\relax
  \def\href#1#2{#2} \def\path#1{#1}\fi

\bibitem{Athron:2014yba}
P.~Athron, J.-h. Park, D.~St\"ockinger, A.~Voigt, {FlexibleSUSY \textemdash{} A
  spectrum generator generator for supersymmetric models}, Comput. Phys.
  Commun. 190 (2015) 139--172.
\newblock \href {http://arxiv.org/abs/1406.2319} {\path{arXiv:1406.2319}},
  \href {https://doi.org/10.1016/j.cpc.2014.12.020}
  {\path{doi:10.1016/j.cpc.2014.12.020}}.

\bibitem{Athron:2017fvs}
P.~Athron, M.~Bach, D.~Harries, T.~Kwasnitza, J.-h. Park, D.~St\"ockinger,
  A.~Voigt, J.~Ziebell, {FlexibleSUSY 2.0: Extensions to investigate the
  phenomenology of SUSY and non-SUSY models}, Comput. Phys. Commun. 230 (2018)
  145--217.
\newblock \href {http://arxiv.org/abs/1710.03760} {\path{arXiv:1710.03760}},
  \href {https://doi.org/10.1016/j.cpc.2018.04.016}
  {\path{doi:10.1016/j.cpc.2018.04.016}}.

\bibitem{Staub:2013tta}
F.~Staub, {SARAH 4: A tool for (not only SUSY) model builders}, Comput. Phys.
  Commun. 185 (2014) 1773--1790.
\newblock \href {http://arxiv.org/abs/1309.7223} {\path{arXiv:1309.7223}},
  \href {https://doi.org/10.1016/j.cpc.2014.02.018}
  {\path{doi:10.1016/j.cpc.2014.02.018}}.

\bibitem{Porod:2003um}
W.~Porod, {SPheno, a program for calculating supersymmetric spectra, SUSY
  particle decays and SUSY particle production at \ensuremath{e^+e^-}
  colliders}, Comput. Phys. Commun. 153 (2003) 275--315.
\newblock \href {http://arxiv.org/abs/hep-ph/0301101}
  {\path{arXiv:hep-ph/0301101}}, \href
  {https://doi.org/10.1016/S0010-4655(03)00222-4}
  {\path{doi:10.1016/S0010-4655(03)00222-4}}.

\bibitem{Porod:2011nf}
W.~Porod, F.~Staub, {SPheno 3.1: Extensions including flavour, CP-phases and
  models beyond the MSSM}, Comput. Phys. Commun. 183 (2012) 2458--2469.
\newblock \href {http://arxiv.org/abs/1104.1573} {\path{arXiv:1104.1573}},
  \href {https://doi.org/10.1016/j.cpc.2012.05.021}
  {\path{doi:10.1016/j.cpc.2012.05.021}}.

\bibitem{Porod:2014xia}
W.~Porod, F.~Staub, A.~Vicente, {A Flavor Kit for BSM models}, Eur. Phys. J. C
  74~(8) (2014) 2992.
\newblock \href {http://arxiv.org/abs/1405.1434} {\path{arXiv:1405.1434}},
  \href {https://doi.org/10.1140/epjc/s10052-014-2992-2}
  {\path{doi:10.1140/epjc/s10052-014-2992-2}}.

\bibitem{Mathematica}
{\relax Wolfram Research, Inc.},
  \href{https://www.wolfram.com/language}{Wolfram {L}anguage, {V}ersion 14.0},
  {C}hampaign, {I}llinois, 2024.
\newline\urlprefix\url{https://www.wolfram.com/language}

\bibitem{Staub:2009bi}
F.~Staub, {From Superpotential to Model Files for FeynArts and
  CalcHep/CompHep}, Comput. Phys. Commun. 181 (2010) 1077--1086.
\newblock \href {http://arxiv.org/abs/0909.2863} {\path{arXiv:0909.2863}},
  \href {https://doi.org/10.1016/j.cpc.2010.01.011}
  {\path{doi:10.1016/j.cpc.2010.01.011}}.

\bibitem{Staub:2010jh}
F.~Staub, {Automatic Calculation of supersymmetric Renormalization Group
  Equations and Self Energies}, Comput. Phys. Commun. 182 (2011) 808--833.
\newblock \href {http://arxiv.org/abs/1002.0840} {\path{arXiv:1002.0840}},
  \href {https://doi.org/10.1016/j.cpc.2010.11.030}
  {\path{doi:10.1016/j.cpc.2010.11.030}}.

\bibitem{Staub:2012pb}
F.~Staub, {SARAH 3.2: Dirac Gauginos, UFO output, and more}, Comput. Phys.
  Commun. 184 (2013) 1792--1809.
\newblock \href {http://arxiv.org/abs/1207.0906} {\path{arXiv:1207.0906}},
  \href {https://doi.org/10.1016/j.cpc.2013.02.019}
  {\path{doi:10.1016/j.cpc.2013.02.019}}.

\bibitem{Allanach:2001kg}
B.~C. Allanach, {SOFTSUSY: a program for calculating supersymmetric spectra},
  Comput. Phys. Commun. 143 (2002) 305--331.
\newblock \href {http://arxiv.org/abs/hep-ph/0104145}
  {\path{arXiv:hep-ph/0104145}}, \href
  {https://doi.org/10.1016/S0010-4655(01)00460-X}
  {\path{doi:10.1016/S0010-4655(01)00460-X}}.

\bibitem{Allanach:2013kza}
B.~C. Allanach, P.~Athron, L.~C. Tunstall, A.~Voigt, A.~G. Williams,
  {Next-to-Minimal SOFTSUSY}, Comput. Phys. Commun. 185 (2014) 2322--2339,
  [Erratum: Comput.Phys.Commun. 250, 107044 (2020)].
\newblock \href {http://arxiv.org/abs/1311.7659} {\path{arXiv:1311.7659}},
  \href {https://doi.org/10.1016/j.cpc.2014.04.015}
  {\path{doi:10.1016/j.cpc.2014.04.015}}.

\bibitem{Athron:2022isz}
P.~Athron, M.~Bach, D.~H.~J. Jacob, W.~Kotlarski, D.~St\"ockinger, A.~Voigt,
  {Precise calculation of the W boson pole mass beyond the standard model with
  FlexibleSUSY}, Phys. Rev. D 106~(9) (2022) 095023.
\newblock \href {http://arxiv.org/abs/2204.05285} {\path{arXiv:2204.05285}},
  \href {https://doi.org/10.1103/PhysRevD.106.095023}
  {\path{doi:10.1103/PhysRevD.106.095023}}.

\bibitem{Athron:2021kve}
P.~Athron, A.~B\"uchner, D.~Harries, W.~Kotlarski, D.~St\"ockinger, A.~Voigt,
  {FlexibleDecay: An automated calculator of scalar decay widths}, Comput.
  Phys. Commun. 283 (2023) 108584.
\newblock \href {http://arxiv.org/abs/2106.05038} {\path{arXiv:2106.05038}},
  \href {https://doi.org/10.1016/j.cpc.2022.108584}
  {\path{doi:10.1016/j.cpc.2022.108584}}.

\bibitem{Khasianevich:2022ess}
U.~Khasianevich, W.~Kotlarski, D.~St\"ockinger, {{NPointFunctions}: a
  calculator of amplitudes and observables in {FlexibleSUSY}}, PoS
  CompTools2021 (2022) 036.
\newblock \href {http://arxiv.org/abs/2206.00745} {\path{arXiv:2206.00745}},
  \href {https://doi.org/10.22323/1.409.0036} {\path{doi:10.22323/1.409.0036}}.

\bibitem{Hahn:2000kx}
T.~Hahn, {Generating Feynman diagrams and amplitudes with FeynArts 3}, Comput.
  Phys. Commun. 140 (2001) 418--431.
\newblock \href {http://arxiv.org/abs/hep-ph/0012260}
  {\path{arXiv:hep-ph/0012260}}, \href
  {https://doi.org/10.1016/S0010-4655(01)00290-9}
  {\path{doi:10.1016/S0010-4655(01)00290-9}}.

\bibitem{Hahn:1998yk}
T.~Hahn, M.~Perez-Victoria, {Automatized one loop calculations in
  four-dimensions and D-dimensions}, Comput. Phys. Commun. 118 (1999) 153--165.
\newblock \href {http://arxiv.org/abs/hep-ph/9807565}
  {\path{arXiv:hep-ph/9807565}}, \href
  {https://doi.org/10.1016/S0010-4655(98)00173-8}
  {\path{doi:10.1016/S0010-4655(98)00173-8}}.

\bibitem{Sjodahl:2012nk}
M.~Sj\"odahl, {ColorMath \textemdash{} A package for color summed calculations
  in \ensuremath{SU(N_c)}}, Eur. Phys. J. C 73~(2) (2013) 2310.
\newblock \href {http://arxiv.org/abs/1211.2099} {\path{arXiv:1211.2099}},
  \href {https://doi.org/10.1140/epjc/s10052-013-2310-4}
  {\path{doi:10.1140/epjc/s10052-013-2310-4}}.

\bibitem{Skands:2003cj}
P.~Z. Skands, et~al., {SUSY Les Houches accord: Interfacing SUSY spectrum
  calculators, decay packages, and event generators}, JHEP 07 (2004) 036.
\newblock \href {http://arxiv.org/abs/hep-ph/0311123}
  {\path{arXiv:hep-ph/0311123}}, \href
  {https://doi.org/10.1088/1126-6708/2004/07/036}
  {\path{doi:10.1088/1126-6708/2004/07/036}}.

\bibitem{Allanach:2008qq}
B.~C. Allanach, et~al., {SUSY Les Houches Accord 2}, Comput. Phys. Commun. 180
  (2009) 8--25.
\newblock \href {http://arxiv.org/abs/0801.0045} {\path{arXiv:0801.0045}},
  \href {https://doi.org/10.1016/j.cpc.2008.08.004}
  {\path{doi:10.1016/j.cpc.2008.08.004}}.

\bibitem{Mahmoudi:2010iz}
F.~Mahmoudi, et~al., {Flavour Les Houches Accord: Interfacing Flavour related
  Codes}, Comput. Phys. Commun. 183 (2012) 285--298.
\newblock \href {http://arxiv.org/abs/1008.0762} {\path{arXiv:1008.0762}},
  \href {https://doi.org/10.1016/j.cpc.2011.10.006}
  {\path{doi:10.1016/j.cpc.2011.10.006}}.

\bibitem{Kribs:2007ac}
G.~D. Kribs, E.~Poppitz, N.~Weiner, {Flavor in supersymmetry with an extended
  R-symmetry}, Phys. Rev. D 78 (2008) 055010.
\newblock \href {http://arxiv.org/abs/0712.2039} {\path{arXiv:0712.2039}},
  \href {https://doi.org/10.1103/PhysRevD.78.055010}
  {\path{doi:10.1103/PhysRevD.78.055010}}.

\bibitem{Khasianevich:2023duu}
U.~Khasianevich, D.~St\"ockinger, H.~St\"ockinger-Kim, J.~W\"unsche,
  {Constraint on scalar leptoquark from low-energy leptonic observables}, Phys.
  Rev. D 108~(9) (2023) 095027.
\newblock \href {http://arxiv.org/abs/2305.05016} {\path{arXiv:2305.05016}},
  \href {https://doi.org/10.1103/PhysRevD.108.095027}
  {\path{doi:10.1103/PhysRevD.108.095027}}.

\bibitem{Dudenas:2022von}
V.~D\={u}d\.{e}nas, T.~Gajdosik, U.~Khasianevich, W.~Kotlarski,
  D.~St\"ockinger, {Charged lepton flavor violating processes in the
  Grimus-Neufeld model}, JHEP 09 (2022) 174.
\newblock \href {http://arxiv.org/abs/2206.00661} {\path{arXiv:2206.00661}},
  \href {https://doi.org/10.1007/JHEP09(2022)174}
  {\path{doi:10.1007/JHEP09(2022)174}}.

\bibitem{Dudenas:2022xnq}
V.~D\={u}d\.{e}nas, T.~Gajdosik, U.~Khasianevich, W.~Kotlarski,
  D.~St\"ockinger, {Box-enhanced charged lepton flavor violation in the
  Grimus-Neufeld model}, Phys. Rev. D 107~(5) (2023) 055027.
\newblock \href {http://arxiv.org/abs/2211.14384} {\path{arXiv:2211.14384}},
  \href {https://doi.org/10.1103/PhysRevD.107.055027}
  {\path{doi:10.1103/PhysRevD.107.055027}}.

\bibitem{Kotlarski:2019muo}
W.~Kotlarski, D.~St\"ockinger, H.~St\"ockinger-Kim, {Low-energy lepton physics
  in the MRSSM: $(g-2)_\mu$, $\mu \to e\gamma$ and $\mu\to e$ conversion}, JHEP
  08 (2019) 082.
\newblock \href {http://arxiv.org/abs/1902.06650} {\path{arXiv:1902.06650}},
  \href {https://doi.org/10.1007/JHEP08(2019)082}
  {\path{doi:10.1007/JHEP08(2019)082}}.

\bibitem{Ziebell:2018th}
J.~Ziebell, {Precise Higgs boson mass calculation in the MSSM with one-loop
  pole mass matching to the THDM}, Master's thesis, TU Dresden (2018).

\bibitem{DangTran:2019th}
K.~D. Tran, {B physics in the Minimal R-symmetric Supersymmetric Standard
  Model}, Master's thesis, TU Dresden (2019).

\bibitem{Hisano:1995cp}
J.~Hisano, T.~Moroi, K.~Tobe, M.~Yamaguchi, {Lepton flavor violation via
  right-handed neutrino Yukawa couplings in supersymmetric standard model},
  Phys. Rev. D 53 (1996) 2442--2459.
\newblock \href {http://arxiv.org/abs/hep-ph/9510309}
  {\path{arXiv:hep-ph/9510309}}, \href
  {https://doi.org/10.1103/PhysRevD.53.2442}
  {\path{doi:10.1103/PhysRevD.53.2442}}.

\bibitem{Ellis:2008zy}
J.~R. Ellis, J.~S. Lee, A.~Pilaftsis, {Electric Dipole Moments in the MSSM
  Reloaded}, JHEP 10 (2008) 049.
\newblock \href {http://arxiv.org/abs/0808.1819} {\path{arXiv:0808.1819}},
  \href {https://doi.org/10.1088/1126-6708/2008/10/049}
  {\path{doi:10.1088/1126-6708/2008/10/049}}.

\bibitem{Okada:1999zk}
Y.~Okada, K.-i. Okumura, Y.~Shimizu, {\ensuremath{\mu\to e\gamma} and
  \ensuremath{\mu\to3e} processes with polarized muons and supersymmetric grand
  unified theories}, Phys. Rev. D 61 (2000) 094001.
\newblock \href {http://arxiv.org/abs/hep-ph/9906446}
  {\path{arXiv:hep-ph/9906446}}, \href
  {https://doi.org/10.1103/PhysRevD.61.094001}
  {\path{doi:10.1103/PhysRevD.61.094001}}.

\bibitem{Crivellin:2017rmk}
A.~Crivellin, S.~Davidson, G.~M. Pruna, A.~Signer, {Renormalisation-group
  improved analysis of $\mu\to e$ processes in a systematic
  effective-field-theory approach}, JHEP 05 (2017) 117.
\newblock \href {http://arxiv.org/abs/1702.03020} {\path{arXiv:1702.03020}},
  \href {https://doi.org/10.1007/JHEP05(2017)117}
  {\path{doi:10.1007/JHEP05(2017)117}}.

\bibitem{Kuno:1999jp}
Y.~Kuno, Y.~Okada, {Muon decay and physics beyond the standard model}, Rev.
  Mod. Phys. 73 (2001) 151--202.
\newblock \href {http://arxiv.org/abs/hep-ph/9909265}
  {\path{arXiv:hep-ph/9909265}}, \href
  {https://doi.org/10.1103/RevModPhys.73.151}
  {\path{doi:10.1103/RevModPhys.73.151}}.

\bibitem{Kitano:2002mt}
R.~Kitano, M.~Koike, Y.~Okada, {Detailed calculation of lepton flavor violating
  muon electron conversion rate for various nuclei}, Phys. Rev. D 66 (2002)
  096002, [Erratum: Phys.Rev.D 76, 059902 (2007)].
\newblock \href {http://arxiv.org/abs/hep-ph/0203110}
  {\path{arXiv:hep-ph/0203110}}, \href
  {https://doi.org/10.1103/PhysRevD.76.059902}
  {\path{doi:10.1103/PhysRevD.76.059902}}.

\bibitem{Cirigliano:2009bz}
V.~Cirigliano, R.~Kitano, Y.~Okada, P.~Tuzon, {On the model discriminating
  power of \ensuremath{\mu\to e} conversion in nuclei}, Phys. Rev. D 80 (2009)
  013002.
\newblock \href {http://arxiv.org/abs/0904.0957} {\path{arXiv:0904.0957}},
  \href {https://doi.org/10.1103/PhysRevD.80.013002}
  {\path{doi:10.1103/PhysRevD.80.013002}}.

\bibitem{Davidson:2017nrp}
S.~Davidson, Y.~Kuno, A.~Saporta,
  {\textquotedblleft{}Spin-dependent\textquotedblright{} ${\mu \rightarrow e}$
  conversion on light nuclei}, Eur. Phys. J. C 78~(2) (2018) 109.
\newblock \href {http://arxiv.org/abs/1710.06787} {\path{arXiv:1710.06787}},
  \href {https://doi.org/10.1140/epjc/s10052-018-5584-8}
  {\path{doi:10.1140/epjc/s10052-018-5584-8}}.

\bibitem{Davidson:2018kud}
S.~Davidson, Y.~Kuno, M.~Yamanaka, {Selecting $\mu \to e$ conversion targets to
  distinguish lepton flavour-changing operators}, Phys. Lett. B 790 (2019)
  380--388.
\newblock \href {http://arxiv.org/abs/1810.01884} {\path{arXiv:1810.01884}},
  \href {https://doi.org/10.1016/j.physletb.2019.01.042}
  {\path{doi:10.1016/j.physletb.2019.01.042}}.

\bibitem{Shifman:1978zn}
M.~A. Shifman, A.~I. Vainshtein, V.~I. Zakharov, {Remarks on Higgs Boson
  Interactions with Nucleons}, Phys. Lett. B 78 (1978) 443--446.
\newblock \href {https://doi.org/10.1016/0370-2693(78)90481-1}
  {\path{doi:10.1016/0370-2693(78)90481-1}}.

\bibitem{Kosmas:2001mv}
T.~S. Kosmas, S.~Kovalenko, I.~Schmidt, {Nuclear muon- e- conversion in strange
  quark sea}, Phys. Lett. B 511 (2001) 203.
\newblock \href {http://arxiv.org/abs/hep-ph/0102101}
  {\path{arXiv:hep-ph/0102101}}, \href
  {https://doi.org/10.1016/S0370-2693(01)00657-8}
  {\path{doi:10.1016/S0370-2693(01)00657-8}}.

\bibitem{Kosmas:2001ia}
T.~S. Kosmas, {Exotic \ensuremath{\mu\to e} conversion in nuclei: Energy
  moments of the transition strength and average energy of the outgoing $e^-$},
  Nucl. Phys. A 683 (2001) 443--462.
\newblock \href {https://doi.org/10.1016/S0375-9474(00)00471-1}
  {\path{doi:10.1016/S0375-9474(00)00471-1}}.

\bibitem{Cirigliano:2017azj}
V.~Cirigliano, S.~Davidson, Y.~Kuno, {Spin-dependent $\mu \to e$ conversion},
  Phys. Lett. B 771 (2017) 242--246.
\newblock \href {http://arxiv.org/abs/1703.02057} {\path{arXiv:1703.02057}},
  \href {https://doi.org/10.1016/j.physletb.2017.05.053}
  {\path{doi:10.1016/j.physletb.2017.05.053}}.

\bibitem{Cirelli:2013ufw}
M.~Cirelli, E.~Del~Nobile, P.~Panci, {Tools for model-independent bounds in
  direct dark matter searches}, JCAP 10 (2013) 019.
\newblock \href {http://arxiv.org/abs/1307.5955} {\path{arXiv:1307.5955}},
  \href {https://doi.org/10.1088/1475-7516/2013/10/019}
  {\path{doi:10.1088/1475-7516/2013/10/019}}.

\bibitem{Crivellin:2013ipa}
A.~Crivellin, M.~Hoferichter, M.~Procura, {Accurate evaluation of hadronic
  uncertainties in spin-independent WIMP-nucleon scattering: Disentangling two-
  and three-flavor effects}, Phys. Rev. D 89 (2014) 054021.
\newblock \href {http://arxiv.org/abs/1312.4951} {\path{arXiv:1312.4951}},
  \href {https://doi.org/10.1103/PhysRevD.89.054021}
  {\path{doi:10.1103/PhysRevD.89.054021}}.

\bibitem{Hoferichter:2015dsa}
M.~Hoferichter, J.~Ruiz~de Elvira, B.~Kubis, U.-G. Mei\ss{}ner, {High-Precision
  Determination of the Pion-Nucleon \ensuremath{\sigma} Term from Roy-Steiner
  Equations}, Phys. Rev. Lett. 115 (2015) 092301.
\newblock \href {http://arxiv.org/abs/1506.04142} {\path{arXiv:1506.04142}},
  \href {https://doi.org/10.1103/PhysRevLett.115.092301}
  {\path{doi:10.1103/PhysRevLett.115.092301}}.

\bibitem{Bhattacharya:2015esa}
T.~Bhattacharya, V.~Cirigliano, R.~Gupta, H.-W. Lin, B.~Yoon, {Neutron Electric
  Dipole Moment and Tensor Charges from Lattice QCD}, Phys. Rev. Lett. 115~(21)
  (2015) 212002.
\newblock \href {http://arxiv.org/abs/1506.04196} {\path{arXiv:1506.04196}},
  \href {https://doi.org/10.1103/PhysRevLett.115.212002}
  {\path{doi:10.1103/PhysRevLett.115.212002}}.

\bibitem{straub2018flavio}
D.~M. Straub, flavio: a python package for flavour and precision phenomenology
  in the standard model and beyond (2018).
\newblock \href {http://arxiv.org/abs/1810.08132} {\path{arXiv:1810.08132}}.

\bibitem{nlohmann}
N.~Lohmann, \href{https://json.nlohmann.me}{{JSON} for modern {C++}} (11 2023).
\newline\urlprefix\url{https://json.nlohmann.me}

\end{thebibliography}
\end{document}